\documentclass[aps,prb,twocolumn,showpacs,groupedaddress ]{revtex4-2}
\usepackage{graphicx}  % needed for figures
\usepackage{ulem}
\usepackage{dcolumn}   % needed for some tables
\usepackage{bm}        % for math
\usepackage{amssymb}   % for math
\usepackage{epstopdf}
\usepackage{bbm}
\usepackage{array}
\usepackage{subcaption}
\usepackage{amsmath}
\usepackage{graphicx}
\usepackage{float}
\usepackage{subcaption}
\usepackage{tikz}
\usepackage[colorlinks,linkcolor=blue,citecolor=blue,hyperindex]{hyperref}
\usepackage{color}

\begin{document}
\title{Schwinger boson symmetric spin liquids of Shastry-Sutherland model}
\author{Ke Liu}
\affiliation{International Center for Quantum Materials, School of Physics, Peking University, Beijing 100871, China}

\author{Fa Wang}
\email{wangfa@pku.edu.cn}
\affiliation{International Center for Quantum Materials, School of Physics, Peking University, Beijing 100871, China}
\affiliation{Collaborative Innovation Center of Quantum Matter, Beijing 100871, China}

\newcommand{\su}{\vphantom{-1}}

\newcommand{\TX}{T_X^{\vphantom{-1}}}
\newcommand{\TY}{T_Y^{\vphantom{-1}}}
\newcommand{\Cf}{C_4^{\vphantom{-1}}}
\newcommand{\Mm}{\sigma^{\vphantom{-1}}}

\newcommand{\TXi}{T_X^{-1}}
\newcommand{\TYi}{T_Y^{-1}}
\newcommand{\Cfi}{C_4^{-1}}
\newcommand{\Mmi}{\sigma^{-1}}

\newcommand{\TXn}[1]{T_X^{#1}}
\newcommand{\TYn}[1]{T_Y^{#1}}
\newcommand{\Cfn}[1]{C_4^{#1}}
\newcommand{\Mmn}[1]{\sigma^{#1}}

\newcommand{\Ncell}{N_c}

\date{\today}
\begin{abstract}
    Motivated by recent experimental and numerical evidences of
    deconfined quantum critical point and
    quantum spin-liquid states
    in spin-$1/2$ Heisenberg model on Shastry-Sutherland lattice,
    we studied possible symmetric spin liquid states and their proximate ordered states
    under Schwinger boson formalism.
    We found a symmetric gapped $Z_2$ spin-liquid state for intermediate model
    parameter $0.66<J_1/J_2<0.71$ under mean-field approximation.
    The Schwinger boson  mean-field
    %phase diagram 
    picture
    is partially supported by exact-diagonalization and
    self-consistent spin wave theory results.
\end{abstract}
\maketitle
\section{Introduction}
\begin{figure}[ht]
    \centering{\begin{subfigure}[ht]{0.45\textwidth}
            \includegraphics[width=\textwidth]{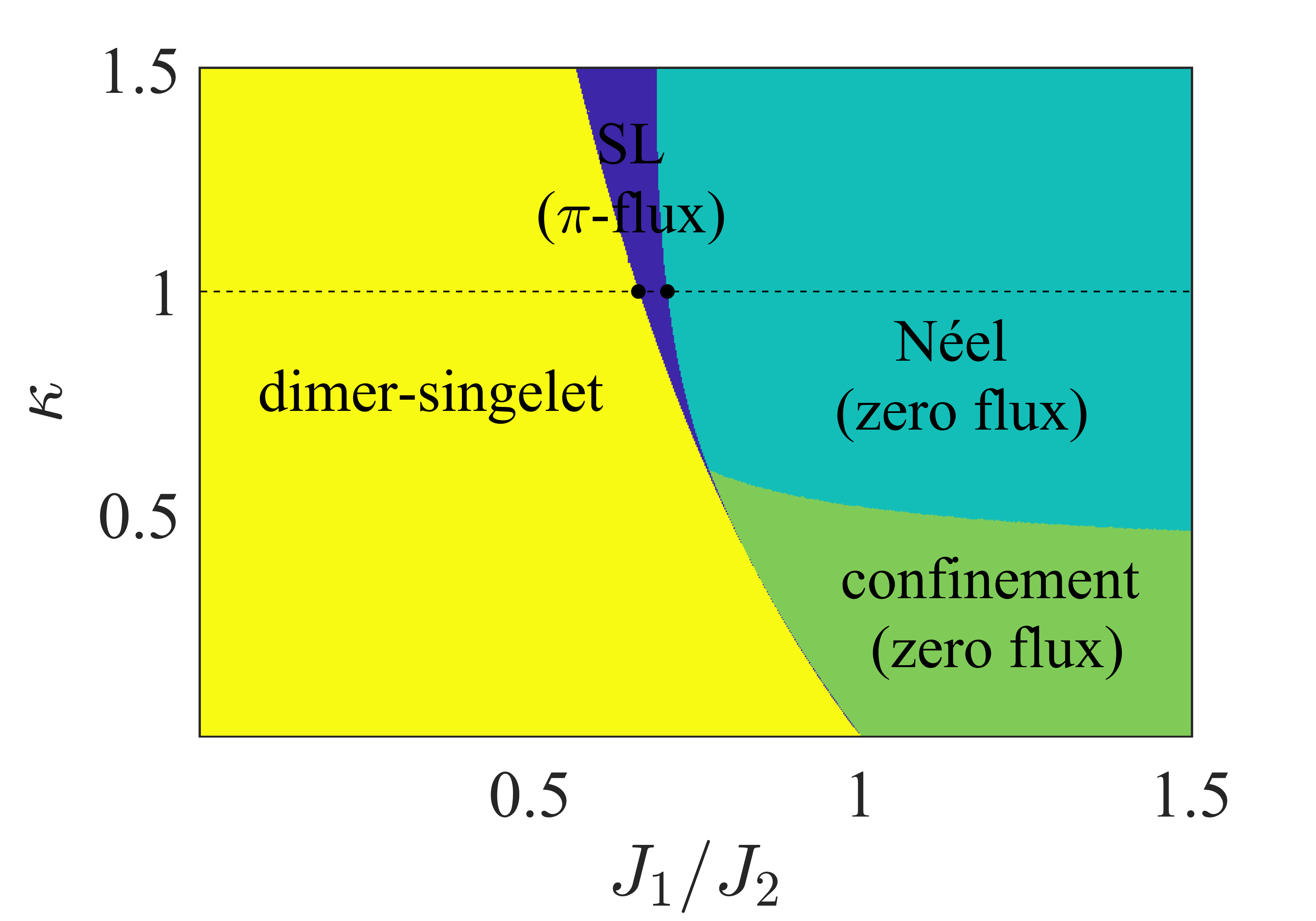}
            %\label{fig:a}
        \end{subfigure}}
    \caption{ The phase diagram of Schwinger boson mean-field theory
        for Shastry-Sutherland model.
        The phase boundaries at physical condition $\kappa=2S=1$ are $J_1/J_2=0.66$ and $J_1/J_2=0.71$.
        In the mean-field level,
        the phase transition from DS to SL is continuous while the phase transition from SL to N\'eel is first order.
    }
    \label{phase diagram}
\end{figure}

Recently the quasi-two-dimensional materials $S=1/2$ quantum magnets $\rm SrCu_2(BO_3)_2$\cite{DQCPwithChangeOffield,PS&DQCP-E2017,PSExperimentThermodynamics1,PSExperimentThermodynamics2}
attracted a lot of research interest,
as they are one of the most promising realizations of deconfined quantum critical points (DQCP)\cite{DQCP1,DQCP2,DQCP3,DQCP4}
or spin-liquid states\cite{QSL}.
The in-plane antiferromagnetic (AFM) Heisenberg interactions
%of 
between
the copper atoms of $\rm SrCu_2(BO_3)_2$ makes it
%a very faithful
%an accurate
a potential
realization of the Shastry-Sutherland model\cite{Shastry-Sutherland} shown in Fig.~\ref{phase diagram} (a).
% A N\'eel state forms when the nearest neighbor coupling $J_1$ is large and a dimer valence bond solid forms when second neighbor coupling $J_2$ is large\cite{Shastry-Sutherland}.
Depending on the values of the nearest neighbor couplings $J_1$ and the second neighbor couplings $J_2$,
the Shastry-Sutherland model can host a dimer valence bond solid ground state in the $J_2/J_1\gg 1$ limit, or a N\'eel state in the $J_1/J_2\gg 1$ limit.
The issue is the intermediate phase
% for $0.67\lesssim J_1/J_2\lesssim 0.76$
\cite{EDforPhaseBoundaryOfDSandNeel,PhaseTransitionDS-PS-Neel1,DQCPandEmergentO4inPSandNeel,PSbyTensorNetwork}
between the N\'eel phase and dimer-singlet (DS) phase [also called orthogonal dimer (OD) phase in some literature].
There are a variety of prediction of the intermediate phase by difference theories and numerics:
a direct transition from dimer-singlet phase to N\'eel phase\cite{PhaseTransitionDS-Neel1,PhaseTransitionDS-PS-Neel2},
a helical order\cite{PhaseTransitionDS-Helix-Neel,SCBOTheoryHelixAndPS},
columnar dimers\cite{DS-ColumnarDimer-Neel},
or plaquette-singlet\cite{PhaseTransitionDS-PS-Neel1,PhaseTransitionDS-PS-Neel2,SCBOTheoryHelixAndPS} intermediate phase.
Recently there are experimental\cite{PS&DQCP-E2017,PSExperimentThermodynamics1,PSExperimentThermodynamics2}
and numerical\cite{PS&DQCPbyIDMRG2019,PSbyTensorNetwork,PSIndeucedByDifferentPlaqutte,PSIndeucedByDifferentPlaqutte2,wang2023plaquette} evidences of the existence of the plaquette-singlet (PS) phase [also called plaquette singlet solid (PSS) or plaquette valence bond solid (PVBS) in some literature].
The phase transition between the N\'eel and plaquette-singlet phase may be described by a DQCP with emergent $O(4)$ symmetry\cite{DQCPandEmergentO4inPSandNeel,PS&DQCPbyIDMRG2019}
and deconfined spinon excitations.
% In the phase transformation from the N\'eel state to the VBS, 
% an intermediate spin-liquid phase is also possible\cite{SLbyDMRG,SLbyED}.
And evidences of a proximate DQCP were found on the boundary of plaquette-singlet
phase and N\'eel AFM phase in a NMR study of \(\rm SrCu_2(BO_3)_2\) under external magnetic field\cite{DQCPwithChangeOffield}.

% Most recently, a proximate DQCP was found in the boundary of plaquette-singlet (PS)
% phase and a N\'eel AFM phase in the NMR study of $\rm SrCu_2(BO_3)_2$ with the change of magnetic field\cite{DQCPwithChangeOffield}.
% Moreover, recent results of density-matrix renormalization group (DMRG)\cite{SLbyDMRG}
% and the exact diagonalization method\cite{SLbyED} show that there is a spin liquid phase in a narrow range of coupling parameter $J_1/J_2$ between the PS phase and N\'eel phase without magnetic field.
% It also argued that the narrow spin-liquid states shrink to a DQCP as the magnetic field increases.
Some recent numerical studies including density-matrix renormalization group (DMRG)\cite{SLbyDMRG}, exact diagonalization(ED)\cite{SLbyED},
and pseudofermion functional renormalization group\cite{PhysRevB.105.L041115}
also suggest the existence of a spin liquid (SL) phase in a narrow range of coupling parameter \(J_1/J_2\) between the PS phase and N\'eel phase without magnetic field.
This intriguing possibility has not been thoroughly studied theoretically, especially using the traditional slave particle language for spin liquids\cite{RevModPhys.89.025003}.
With this motivation
we study the possible symmetric spin-liquid states and their proximate ordered states on Shastry-Sutherland lattice under Schwinger boson formalism\cite{spin-liquid-boson} in this paper.
The goal of our work is not to accurately determine the phase diagram of the Shastry-Sutherland model,
but to explore the possibilities of quantum spin liquids compatible with this lattice,
which might be realized in related models and materials.

%Motivated by these experimental and numerical evidences of DQCP and quantum spin-liquid states
%in the spin-1/2 Heisenberg model on Shastry-Sutherland lattice,
%we study the possible symmetric spin-liquid states and their proximate ordered states under Schwinger boson formalism
%in this paper.
%Generally, the excitations in DQCP and spin-liquid states are fractional spinons
%and emergent gauge fields\cite{DQCP1,DQCP2,DQCP3,spin-liquid-boson,spin-liquid-fermion}.
%These spinon may come form the anyon excitations of $Z_2$ topological order\cite{Z2}.
In this paper, we focus on the Schwinger boson\cite{PhysRevB.38.316} description of quantum spin liquids.
This formalism and its large-$N$ generalization\cite{PhysRevLett.66.1773} are convenient to describe the transition between gapped $Z_2$ spin-liquid phases and magnetic ordered phases\cite{spin-liquid-boson}, and have been successful in the studies of several quantum magnets\cite{Ch14-IntroFrusMag}.
% which is convenient to describe the transition between gapped $Z_2$ spin-liquid phases and magnetic ordered phases\cite{SB}.
By projective symmetry group (PSG)\cite{PSG-fermion,PSG-boson},
we find 6 possible algebraic $Z_2$ PSG solutions
and 4 gauge inequivalent ansatz.
Comparing the mean-field energies of these ansatz,
we get the mean-field phase diagram, which is shown in Fig.~\ref{phase diagram}.
We find a symmetric gapped $Z_2$ spin liquid for the intermediate model parameter
$0.66<J_1/J_2<0.71$ under mean-field approximation.
A dimer-singlet phase forms in $J_1/J_2<0.66$ while a N\'eel AFM state forms in $J_1/J_2>0.71$,
where the Schwinger boson condensation happens.
To further investigate the PS state,
we also study ansatz with PS order, and find that these PS ansatz have higher ground state energy compared with the symmetric spin liquids in the mean-field level.
% However, the energy differences of the plaquette ansatz and the other two 
% algebraic $Z_2$ PSG solutions are small in the intermediate model parameter,
% and the PS phase may also exist by considering the gauge fluctuation
% and projecting the mean-field wave function to physical conditions,
% which is left for further investigation.

We also study the spin correlations of ground state wave function and structure factor of the spin-liquid states and N\'eel state by the Schwinger boson mean-field theory(SBMFT),
and compare them with the results of exact-diagonalization and spin-wave theory.
We find that the spin correlations in SBMFT have similar behavior compared with the result of the exact diagonalization method.
The structure factor in the N\'eel phase can also be calculated by the spin wave theory.
However, the linear spin wave theory breaks down near $J_1/J_2\sim 1$,
which is far from the N\'eel phase boundary.
To investigate the N\'eel phase in $J_1/J_2<1$ region,
% model in this region,
we use a self-consistent spin wave theory,
which pushes the N\'eel phase boundary down to $J_1/J_2\approx 0.65$.
With the self-consistent spin wave theory,
we get qualitatively consistent dynamical spin correlations with Schwinger boson mean-field theory.

This paper is organized as follows.
In Sec.~\ref{PSG},
we introduce the Shastry-Sutherland model and Schwinger boson mean-field theory.
We then introduce the PSG classification and show the results of algebraic PSG.
In Sec.~\ref{MF results},
we show the mean field results and properties of the 4 gauge inequivalent symmetric spin liquid states,
and briefly discuss the two PS states.
% we present the main results of the Schwinger boson mean-field theory.
In Sec.~\ref{MF phase diagram},
we present
the mean-field phase diagram with a gapped $Z_2$ spin liquid state
in the intermediate parameter $0.66<J_1/J_2<0.71$.
In Sec.~\ref{compare},
we compare the structure factors for the spin liquid and N\'eel states by SBMFT and self-consistent spin wave theory,
and compare some of the ground state properties by SBMFT and exact diagonalization.
Section~\ref{Conclusion} contains further discussion and a summary of results.
The technical and numerical details are presented in the Appendixes.

\section{Projective symmetry group of Shastry-Sutherland lattice Schwinger boson states}\label{PSG}
\begin{figure}[ht]
    \centering{\begin{subfigure}[ht]{0.45\textwidth}
            \includegraphics[width=\textwidth]{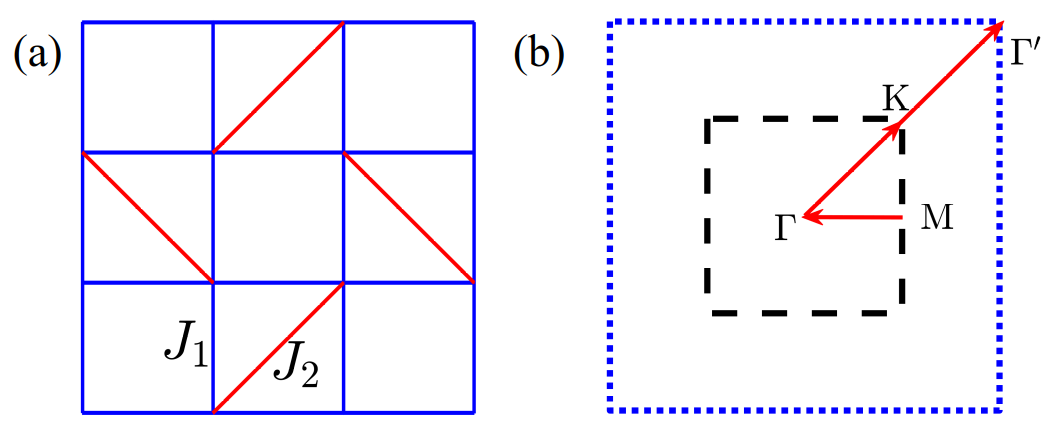}
            %\label{fig:a}
        \end{subfigure}}
    \caption{(a) is the Shastry-Sutherland lattice. The Heisenberg couplings in blue and red bonds are $J_1$ and $J_2$ respectively.
        The black dashed line in (b) is the first Brillouin zone of Shastry-Sutherland lattice,
        and the blue dotted line is the first Brillouin zone of square lattice without $J_2$ bonds.
        The red arrows are the plot paths ($\mathrm{K}\rightarrow\mathrm{M}\rightarrow\mathrm{\Gamma}\rightarrow\mathrm{K}\rightarrow\mathrm{\Gamma^\prime}$) in this paper.
    }
    \label{lattice-BZ}
\end{figure}
The Shastry-Sutherland model Hamiltonian is
\begin{equation}\label{Eq:SS Hamiltonian}
    H=J_1 \sum_{\langle i j\rangle} \mathbf{S}_i \cdot \mathbf{S}_j+J_2 \sum_{\langle\langle i j\rangle\rangle^{\prime}} \mathbf{S}_i \cdot \mathbf{S}_j,
\end{equation}
where $\mathbf{S}_i$ are $S=1/2$ spin operators and
$\langle i j\rangle$ are the nearest-neighbor(n.n.)[blue in Fig.~\ref{lattice-BZ}(a)] bonds, while
$\langle\langle i j\rangle\rangle^{\prime}$ are some of the next-nearest-neighbor(n.n.n.)[red in Fig.~\ref{lattice-BZ}(a)] bonds.
Here we study this Hamiltonian by the Schwinger boson mean-field theory.
The spin operator is expressed by the Schwinger bosons as
\begin{eqnarray}
    \bm{S}_i=\frac{1}{2}\sum_{\alpha, \beta=\uparrow, \downarrow} b_{i \alpha}^{\dagger} \bm{\sigma}_{\alpha \beta} b_{i \beta},
\end{eqnarray}
with the constraints at every site
\begin{eqnarray}
    \sum_\sigma b_{i \sigma}^{\dagger} b_{i \sigma}=\kappa=2S, \label{numConstrain}
\end{eqnarray}
for a spin system with spin $S$.
For the convenience of analysis,
$\kappa$ is usually regarded as a continuous parameter.
Using the Schwinger boson representation,
the Heisenberg interaction can be rewritten as
\begin{eqnarray}
    \bm{S}_i \cdot \bm{S}_j=: \hat{B}_{i j}^{\dagger} \hat{B}_{i j}:-\hat{A}_{i j}^{\dagger} \hat{A}_{i j}
\end{eqnarray}
where $::$ is normal ordering and boson pairing operator $\hat{A}$ and hopping operator $\hat{B}$ are defined as
\begin{eqnarray}
    \hat{B}_{i j}&=&\frac{1}{2} \sum_\sigma b_{i \sigma}^{\dagger} b_{j \sigma}, \\
    \hat{A}_{i j}&=&\frac{1}{2} \sum_{\sigma, \sigma^{\prime}} \epsilon_{\sigma \sigma^{\prime}} b_{i \sigma} b_{j \sigma^{\prime}}.
\end{eqnarray}
After decoupling the quartic terms by the Hubbard-Stratonovich transformation,
we get the mean-field Hamiltonian:
\begin{eqnarray}\label{Eq:MF Hamiltonian}
    H_{\mathrm{MF}}&= &\sum_{i j}J_{ij}\left(-A_{i j}^* \hat{A}_{i j}+B_{i j}^* \hat{B}_{i j}+\text { H.c. }\right)      \nonumber                              \\
    &+&\sum_{i j}J_{ij}\left(\left|A_{i j}\right|^2-\left|B_{i j}\right|^2\right)-\mu_i \sum_i\left(\hat{n}_i-\kappa\right),
\end{eqnarray}
where the complex numbers $A_{i j}=\langle \hat{A}_{ij}\rangle=-A_{j i}$ and $B_{ij}=\langle \hat{B}_{ij}\rangle=B_{ji}^*$ are the
mean-field ansatz.
$\mu_i$ are the real Lagrangian multipliers to enforce the constraints of Eq.~(\ref{numConstrain}),
which are usually site independent, namely $\mu_i=\mu$.
% In the mean-field level, we assume $\mu_i$ is site independent, namely $\mu_i=\mu$.
% Because of the existence of emergent $U(1)$ gauge redundancy,
This mean-field description has emergent $U(1)$ gauge redundancy, namely that the following $U(1)$ gauge transformation,
\begin{subequations}\label{eq:U1 gauge transformation}
    \begin{eqnarray}
        b_{j \sigma} & \rightarrow &e^{i \phi(j)} b_{j \sigma},\\
        A_{i j} & \rightarrow &e^{i[\phi(i)+\phi(j)]} A_{i j}, \\
        B_{i j} & \rightarrow& e^{i[-\phi(i)+\phi(j)]} B_{i j},
    \end{eqnarray} 
\end{subequations}
will not change the physical spin states.
Therefore
the mean-field solutions should be classified by projective symmetry group(PSG)\cite{PSG-fermion,PSG-boson}.
In the remainder of this section, we briefly show the details of PSG classification of the symmetric spin liquid
on Shastry-Sutherland lattice.

As mentioned above,
the Schwinger boson mean-field theory has a emergent $U(1)$ gauge symmetry.
After the local gauge transformation of Eq.~(\ref{eq:U1 gauge transformation}),
% \begin{eqnarray}
%     b_{j \sigma} & \rightarrow &e^{i \phi(j)} b_{j \sigma},\\
%     A_{i j} & \rightarrow &e^{i[\phi(i)+\phi(j)]} A_{i j}, \\
%     B_{i j} & \rightarrow& e^{i[-\phi(i)+\phi(j)]} B_{i j},
% \end{eqnarray}
the mean-field Hamiltonian is invariant and all physical observables are unchanged,
as the wave function is same after projected to physical condition.
Because of the existence of emergent gauge symmetry,
for different spin liquids with same symmetry,
the ansatz are invariant under symmetry transformations followed by gauge transformations,
$\hat{b}_{\boldsymbol{r},s}\to
    \exp[\mathbbm{i} \phi_g(g\boldsymbol{r})] \hat{b}_{g\boldsymbol{r},s}$, for space group element $g$.
Therefore, the spin-liquid states should be classified by the projective representation
of the space group.
The ansatz are invariant under operations of projective symmetry group (PSG).
Different PSGs characterize different kinds of spin-liquid states with the same symmetries.

\begin{figure}
    \centering
    \begin{subfigure}[h]{0.3\textwidth}
        \includegraphics[width=\textwidth]{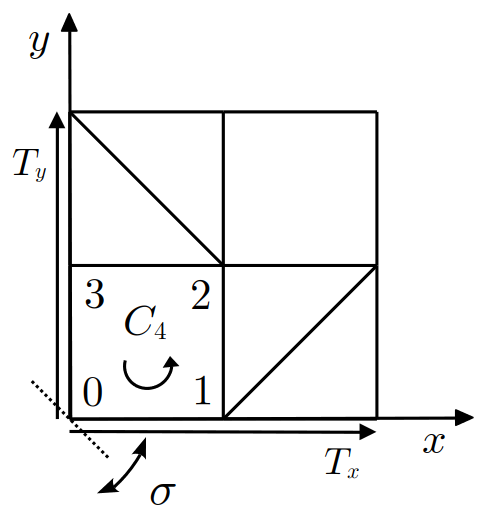}
        %\label{fig:a}
    \end{subfigure}
    \caption{
        The coordinate system and space group generators $T_x$, $T_y$, $C_4$, $\sigma$ of the Shastry-Sutherland lattice.
        These generators are translation $T_x$ along $\hat{e}_x$ by 2 units,
        translation $T_y$ along $\hat{e}_y$ by 2 units,
        a reflection $\sigma$ about $x=-y$ and the $90^\circ$ rotation $C_4$ around $(1/2,1/2)$.
    }
    \label{sslatice}
\end{figure}

We set up a Cartesian coordinate system and represent the site coordinate by $(x,y)=x\hat{e}_x+y\hat{e}_y $ with $x,y\in \mathbb{Z} $.
For the discussion convenience, the site coordinate can also be expressed by cell-sublattice index,
$(X,Y,s)$, where $X,Y\in\mathbb{Z}$ and $s\in \mathbb{Z}_4$ ($s=0,1,2,3$),
which means Cartesian $(2X+x_s,2Y+y_s)$ with
$(x_0,y_0)=(0,0)$, $(x_1,y_1)=(1,0)$, $(x_2,y_2)=(1,1)$, $(x_3,y_3)=(0,1)$.
The sublattice labeling is shown in Fig.~\ref{sslatice}.
With these two coordinate systems,
the nearest-neighbor(n.n.) bonds are $(x,y)-(x+1,y)$ and $(x,y)-(x,y+1)$ while the
next-nearest-neighbor(n.n.n.) bonds are $(X,Y,0)-(X,Y-1,2)$ and $(X,Y,1)-(X+1,Y,3)$.

The space group of the square lattice is generated by translation $T_x$ along $\hat{e}_x$ by 2 units,
translation $T_y$ along $\hat{e}_y$ by 2 units,
a reflection $\sigma$ along $x=-y$ and the $90^\circ$ rotation $C_4$ around $(1/2,1/2)$, which are also shown in Fig.~\ref{sslatice}.
The action of these generators on the Shastry-Sutherland lattice reads
\begin{subequations}\label{eq:generators}
    \begin{eqnarray}
        \TX&:&\quad (x,y)\mapsto (x+2,y),\\
        \TY&:&\quad (x,y)\mapsto (x,y+2),\\
        \Cf&:&\quad (x,y)\mapsto (-y+1,x),\\
        \Mm&:&\quad (x,y)\mapsto (-y,-x).
    \end{eqnarray}
\end{subequations}
Note that the glide-reflection generators (\(G_x\) and \(G_y\)) can be generated by
these four generators, which can be written as
\begin{eqnarray}
    G_x&=&C_4\sigma:\quad (x,y)\mapsto (x+1,-y),\\
    G_y&=&T_y\sigma C_4:\quad (x,y)\mapsto (-x,y+1),
\end{eqnarray}
and are not used in solving PSGs.

These 4 generators in Eq.~(\ref{eq:generators}) have the following commutative relations,
\begin{eqnarray}
    \TXi \TY \TX \TYi&=&\mathbbm{1},\label{re1}\\
    \TYi\TXi\Cf\TX\Cfi&=&\mathbbm{1},\label{re2}\\
    \TXi\Cf\TYi\Cfi&=&\mathbbm{1},\label{re3}\\
    \Cfn{4} &=&\mathbbm{1},\label{re4}\\
    \TXi\Mm\TYi\Mmi&=&\mathbbm{1},\label{re5}\\
    \TYi\Mm\TXi\Mmi&=&\mathbbm{1},\label{re6}\\
    \Mmn{2} &=&\mathbbm{1},\label{re7}\\
    \TXi \Cf\Mm\Cf\Mmi&=&\mathbbm{1}\label{re8}.
\end{eqnarray}

We have solved the algebraic PSG in Appendix~\ref{App:1} and we only show the results here,
\begin{eqnarray}
    \phi_{\TX}(X,Y,s) & = & 0,\\
    \phi_{\TY}(X,Y,s) & = & 0 ,\\
    \phi_{\Cf}(X,Y,s) & = & p_2\pi\cdot Y+p_4\pi\cdot \delta_{s,0} ,\\
    \phi_{\Mm}(X,Y,s) & = & \frac{p_7\pi}{2}+p_2\pi\cdot x_s y_s +p_4\pi \cdot \delta_{s,0},
\end{eqnarray}
with three remaining free $Z_2$ integer parameters $p_2,p_4,p_7=0$ or $1$ $(\mod 2)$.
Therefore, there are at most 8 kinds of PSGs.

Then we need to consider the constraints on PSG by ansatz.
The nn bond poses no constraint, because there is no nontrivial space group element that maps one nn bond to itself or its reverse.
For the nnn bond, if $A_{\mathrm{n.n.n.}}\neq 0$,
consider $(0,0,0)-(0,-1,2)$, which is invariant under $\Mm$,
then $\phi_{\Mm}(0,0,0)+\phi_{\Mm}(0,-1,2)=p_7\pi+p_2\pi+p_4\pi=0$,
namely $p_2+p_4+p_7=0$;
if $B_{\mathrm{n.n.n.}}\neq 0$,
consider $(0,0,0)-(0,-1,2)$, which is invariant under $\Mm$,
then $\phi_{\Mm}(0,0,0)-\phi_{\Mm}(0,-1,2)=-p_2\pi-p_4\pi=0$,
namely $p_2+p_4=0$, this is incompatible with $A_{\mathrm{n.n.n.}}\neq 0$;
consider $(-1,0,1)-(0,0,3)$, which is reverted by $\Mm$,
then $\phi_{\Mm}(-1,0,1)-\phi_{\Mm}(0,0,3)=0$,
then if $B_{\mathrm{n.n.n.}}\neq 0$, it must be real.
If we only consider the condition where at least one of next-nearest-neighbor ansatz $A_{\mathrm{n.n.n.}}$ and $B_{\mathrm{n.n.n.}}$ is not zero,
there are at most 6 kinds of PSGs with these constraints.
If we assume the nearest-neighbor ansatz $A_1$ is nonzero,
these 6 states can be classified by two gauge invariant phase $\Phi_1$ and $\Phi_2$,
which are defined on empty square plaquettes and $J_2$ square plaquettes respectively,
\begin{eqnarray}
    A_{i j}\left(-A_{j k}^*\right) A_{k l}\left(-A_{l i}^*\right)=\left|A_1\right|^4 e^{i \Phi}.
\end{eqnarray}
The gauge invariant ``flux'' values $\Phi_1$ and $\Phi_2$ in
the empty squares and $J_2$ squares respectively, are defined as the complex phase of
the product of nearest-neighbor boson pairing ansatz $A_{i j}(-A_{j k}^*) A_{k l}(-A_{l i}^*)$ around a plaquette\cite{TchernyshyovEPL06}.
Due to time reversal symmetry $\Phi_1$ and $\Phi_2$ can only be $0$ or $\pi$,
and the four different combinations of $(\Phi_1,\Phi_2)$ corresponds to the 4 gauge inequivalent ansatz solved by PSG.
Therefore, the 6 PSG states only have 4 gauge inequivalent ansatz which are named as $(\pi,\pi)$, $(0,\pi)$, $(\pi,0)$, and $(0,0)$-flux states according to their gauge flux distribution.
The configuration details of these states are shown in Appendix~\ref{App:1}.
\section{Mean-field states}\label{MF results}
In this section we will show the properties of these four gauge inequivalent spin liquid states,
including the ansatz amplitudes, spinon dispersions and the static and dynamic spin structure factors.

In the Schwinger boson mean field theory,
the structure factor can be expressed by the imaginary part of ``bubble'' Feynman diagrams.
Note that the anomalous green's function of the spinons also takes important parts.
The static and dynamic spin structure factor can be measured experimentally by neutron scattering.
In the following we will show the mean field results of the four gauge inequivalent symmetric spin liquid states.
To consider the existence of PS phase, we also briefly discuss the PS states in open square and \(J_2\)-square,
which are not included in the symmetric spin liquid states because they break the glide symmetry generated by \(G_x\) and \(G_y\).

\begin{widetext}
    All these mean-field ansatz have the four-site unit cell depicted in Fig.~\ref{sslatice}
    and the ansatz configurations are shown in the figures in Appendix~\ref{App:1}.
    After Fourier transformation
    %$b_{\mathbf{r}}=\frac{1}{\sqrt{\Ncell}} \sum_{\mathbf{r}} e^{-i \mathbf{k} \cdot \mathbf{r}} b_{\mathbf{k}}$,
    $b_{s\mathbf{k}}=\frac{1}{\sqrt{\Ncell}} \sum_{\mathbf{r}} e^{-i \mathbf{k} \cdot \mathbf{r}} b_{\mathbf{r}s}$,
    where $\mathbf{r}=(X,Y)$ labels unit cell and $\Ncell$ is the number of unit cells,
    the mean-field Hamiltonian in Eq.~(\ref{Eq:MF Hamiltonian}) can be formally written as
    \begin{eqnarray}
        H_{\mathrm{MF}}&=&\sum_{\mathbf{k}} \Psi_{\mathbf{k}}^{\dagger} D_{\mathbf{k}} \Psi_{\mathbf{k}} +\Ncell\left[\mu+\mu \kappa+8J_1(\left|A_1\right|^2-\left|B_1\right|^2)+2J_2(\left|A_2\right|^2-\left|B_2\right|^2)\right],
    \end{eqnarray}
    where we have used the Nambu spinor $\Psi_{\mathbf{k}}=(b_{0\mathbf{k}\uparrow},b_{1\mathbf{k}\uparrow},b_{2\mathbf{k}\uparrow},b_{3\mathbf{k}\uparrow},b_{0\mathbf{-k}\downarrow}^\dagger,b_{1\mathbf{-k}\downarrow}^\dagger,b_{2\mathbf{-k}\downarrow}^\dagger,b_{3\mathbf{-k}\downarrow}^\dagger)^T$.
    \(A_1\) and \(B_1\) are boson pairing and hopping ansatz amplitudes on n.n. bonds respectively,
    and \(A_2\) and \(B_2\) are ansatz on n.n.n. bonds.
    The $8\times8$ matrix $D_{\mathbf{k}}$ satisfies
    \begin{eqnarray}\label{eq:Hk}
        D_{\mathbf{k}}&=&-\mu\mathbf{1}+\left(\begin{array}{cc}
                \boldsymbol{B}_{\mathbf{k}}           & \boldsymbol{A}_{\mathbf{k}}      \\
                \boldsymbol{A}_{\mathbf{k}}^{\dagger} & \boldsymbol{B}_{-\mathbf{k}}^{T}
            \end{array}\right),
    \end{eqnarray}
    where $\mathbf{1}$ is the $8\times 8$ identity matrix.
    The $4\times 4$ matrices $\boldsymbol{A}_{\mathbf{k}}$ and $\boldsymbol{B}_{\mathbf{k}}$
    have different expression in the 4 gauge inequivalent states.
    After a  Bogoliubov transformation,
    the mean-field Hamiltonian can be diagonalized as
    \begin{eqnarray}
        H_{\mathrm{MF}}=\sum_{\mathbf{k}s}\omega_{s\mathbf{k}}(\gamma_{s\mathbf{k}\uparrow}^\dagger\gamma_{s\mathbf{k}\uparrow}+\gamma_{s\mathbf{k}\downarrow}^\dagger\gamma_{s\mathbf{k}\downarrow}+1) +\Ncell\left[\mu+\mu \kappa+8J_1(\left|A_1\right|^2-\left|B_1\right|^2)+2J_2(\left|A_2\right|^2-\left|B_2\right|^2)\right],
    \end{eqnarray}
    where the $\omega_{s\mathbf{k}}$ is the spinon dispersions, and $s=0,1,2,3$.
    The self-consistent equations for symmetric spin liquid states are
    \begin{subequations}\label{eq:self-consistent}
        \begin{eqnarray}
            16 J_1 A_1&=&-\sum_s\int_{B Z} \frac{\partial \omega_{s\mathbf{k}}}{\partial A_1} \mathbf{d}^2 k, \mathrm{\ if\ }A_1\neq0, \\
            4 J_2 A_2&=-&\sum_s\int_{B Z} \frac{\partial \omega_{s\mathbf{k}}}{\partial A_2} \mathbf{d}^2 k,\mathrm{\ if\ }A_2\neq0,\\
            16 J_1 B_1&=&\sum_s\int_{B Z} \frac{\partial \omega_{s\mathbf{k}}}{\partial B_1} \mathbf{d}^2 k,\mathrm{\ if\ }B_1\neq0,\\
            4 J_2 B_2&=&\sum_s\int_{B Z} \frac{\partial \omega_{s\mathbf{k}}}{\partial B_2} \mathbf{d}^2 k,\mathrm{\ if\ }B_2\neq0, \\
            1+\kappa&=&-\sum_s\int_{B Z} \frac{\partial \omega_{s\mathbf{k}}}{\partial \mu} \mathbf{d}^2 k,
        \end{eqnarray}
    \end{subequations}
    where the integral is over the first Brillouin zone (BZ).
    With these self-consistent equations,
    the mean-field ansatz can be solved.
    \subsection{$(0,0)$-flux state}
    In the $(0,0)$-flux state, only \(A_1\) and \(B_2\) are nonzero and the configuration is shown in Fig.~\ref{00} in Appendix~\ref{App:1B}.
    The $\boldsymbol{A}_{\mathbf{k}}$ and $\boldsymbol{B}_{\mathbf{k}}$ of this state are shown in Appendix~\ref{App:3 A}.
    After a  Bogoliubov transformation,
    we get the spinon dispersion $\omega_{s\mathbf{k}}$.
    Then the mean-field ansatz is solve with the self-consistent equations in Eq.~(\ref{eq:self-consistent}).
    The spinon dispersion at \(\kappa_c\) and the mean-field ansatz amplitudes are shown in Fig.~\ref{00 ansatz}.
    \begin{figure*}[ht]
        \centering
        \begin{subfigure}[h]{0.3\textwidth}
            \includegraphics[width=\textwidth]{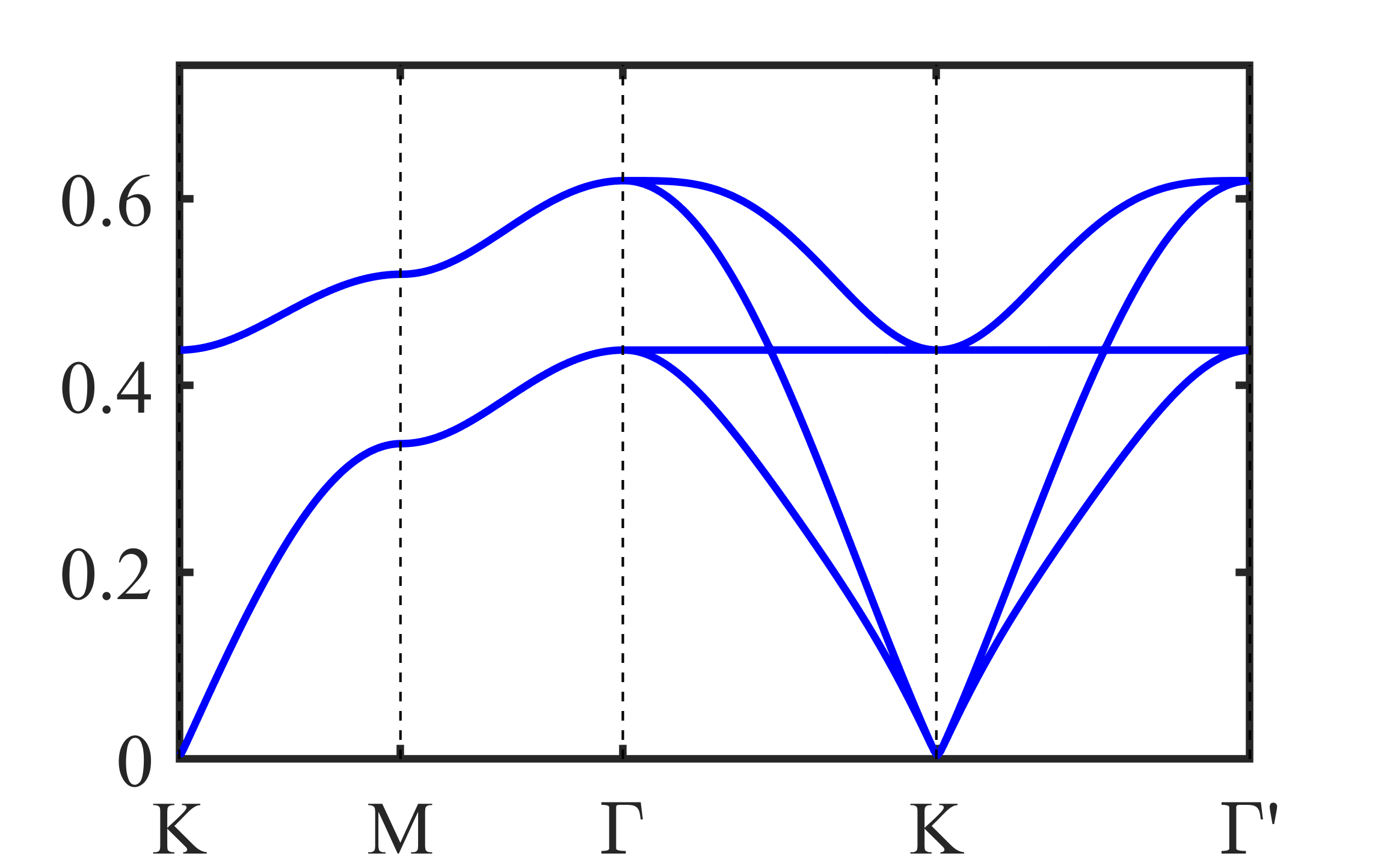}
            %\label{fig:a}
            \caption{Spinon dispersion at $\kappa_c$}
        \end{subfigure}
        \begin{subfigure}[h]{0.3\textwidth}
            \includegraphics[width=\textwidth]{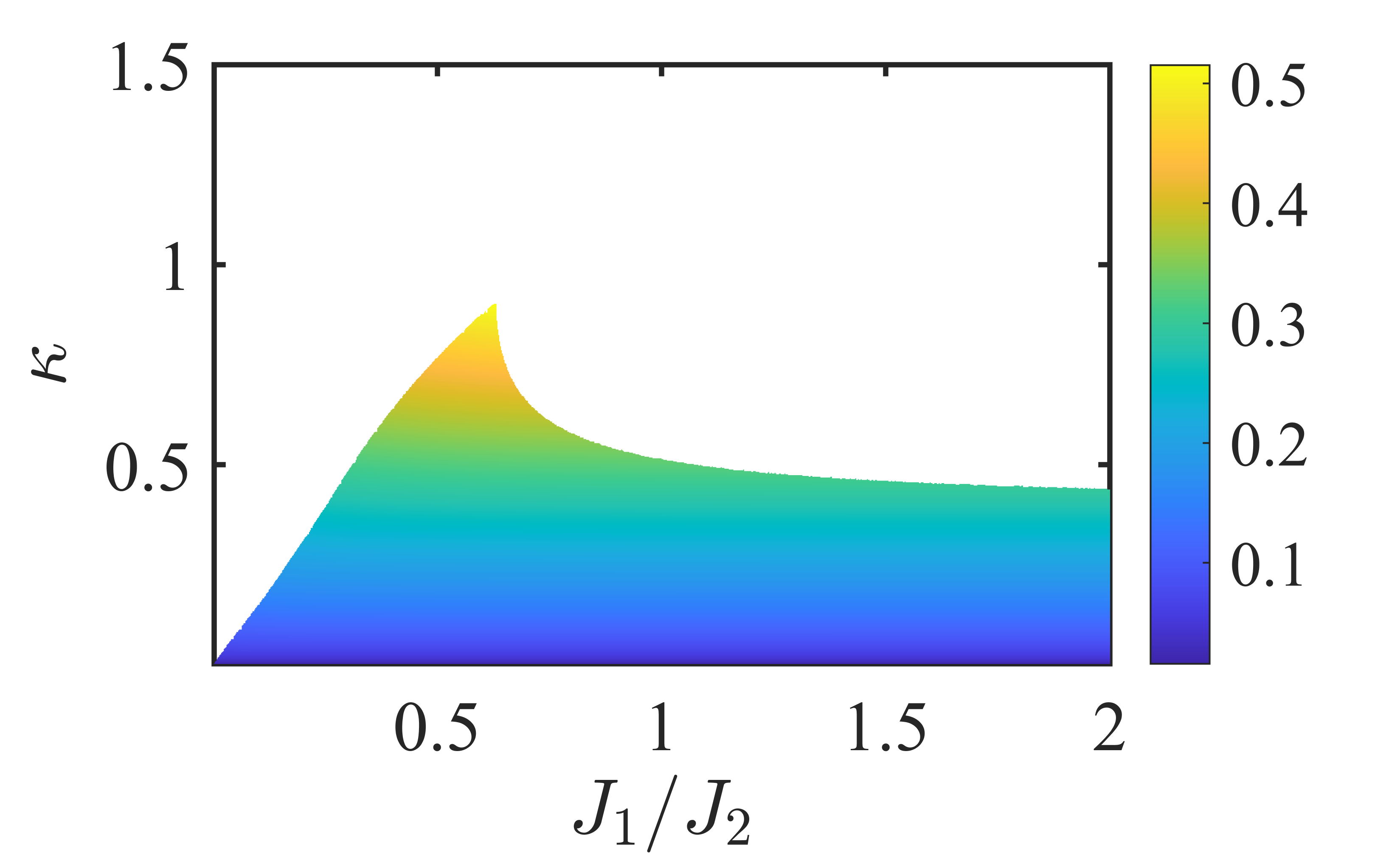}
            \caption{$A_1$}
        \end{subfigure}
        \begin{subfigure}[h]{0.3\textwidth}
            \includegraphics[width=\textwidth]{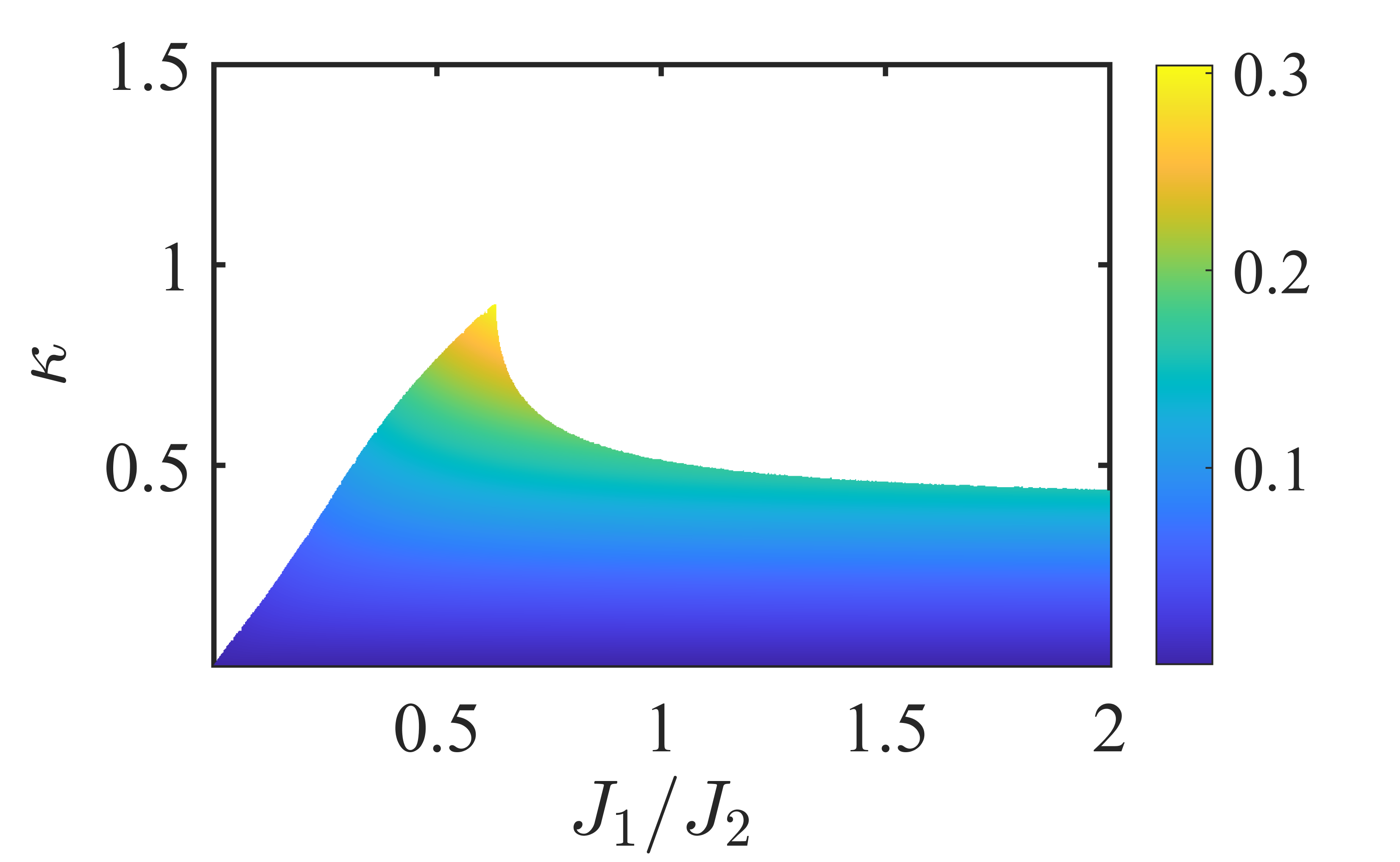}
            \caption{$B_2$}
        \end{subfigure}
        \caption{(a) is the spinon dispersion at $\kappa=\kappa_c$ and $J_1=J_2=1$ of (0,0)-flux state.
            The plot path is shown in Fig.~\ref{lattice-BZ} (b).
            (b) and (c) are the ansatz $A_1$ and $B_2$ value respectively.
            $A_2$ and $B_1$ are always zero in these two conditions.
            The spinon condenses at the white area in (b) and (c).
        }
        \label{00 ansatz}
    \end{figure*}

    As shown in Fig.~\ref{00 ansatz} (a),
    the minima of the spinon dispersion is located at $\mathbf{Q}=(\pi,\pi)$,
    and the gap vanishes at $\kappa=\kappa_c$.
    When $\kappa$ is larger than $\kappa_c$,
    the spinon will condense and form a N\'eel magnetic order.
    The details of the formation of the magnetic order are discussed in Appendix \ref{App:2}.
    The spinon condenses at the white area in Fig.~\ref{00 ansatz} (b) and (c),
    and the contour of the ansatz amplitudes indicate the critical $\kappa_c$.
    We find that $\kappa_c$ is always smaller than 1 with the change of \(J_1/J_2\).
    Therefore,
    the (0,0)-flux state contributes to the magnetic ordered state in the physical condition.
    In the condition $\kappa<\kappa_c$,
    the IGG of this state is $U(1)$,
    and this state can not exist stably and always confines.

    \subsection{$(0,\pi)$-flux state}
    In the $(0,\pi)$-flux state, the ansatz amplitudes \(A_1\), \(A_2\) and \(B_1\) are nonzero.
    The configuration of this state is shown in Fig.~\ref{0-Pi} in Appendix~\ref{App:1A}.
    and the $\boldsymbol{A}_{\mathbf{k}}$ and $\boldsymbol{B}_{\mathbf{k}}$ are show in Appendix~\ref{App:3 B}.
    Diagonalizing the mean-field Hamiltonian by Bogoliubov transformation and
    solving these self-consistent equations in Eq.~(\ref{eq:self-consistent}) we have get the ansatz amplitudes,
    which are shown in Fig.~\ref{zero-pi ansatz}.
    \begin{figure*}[h]
        \centering
        \begin{subfigure}[h]{0.3\textwidth}
            \includegraphics[width=\textwidth]{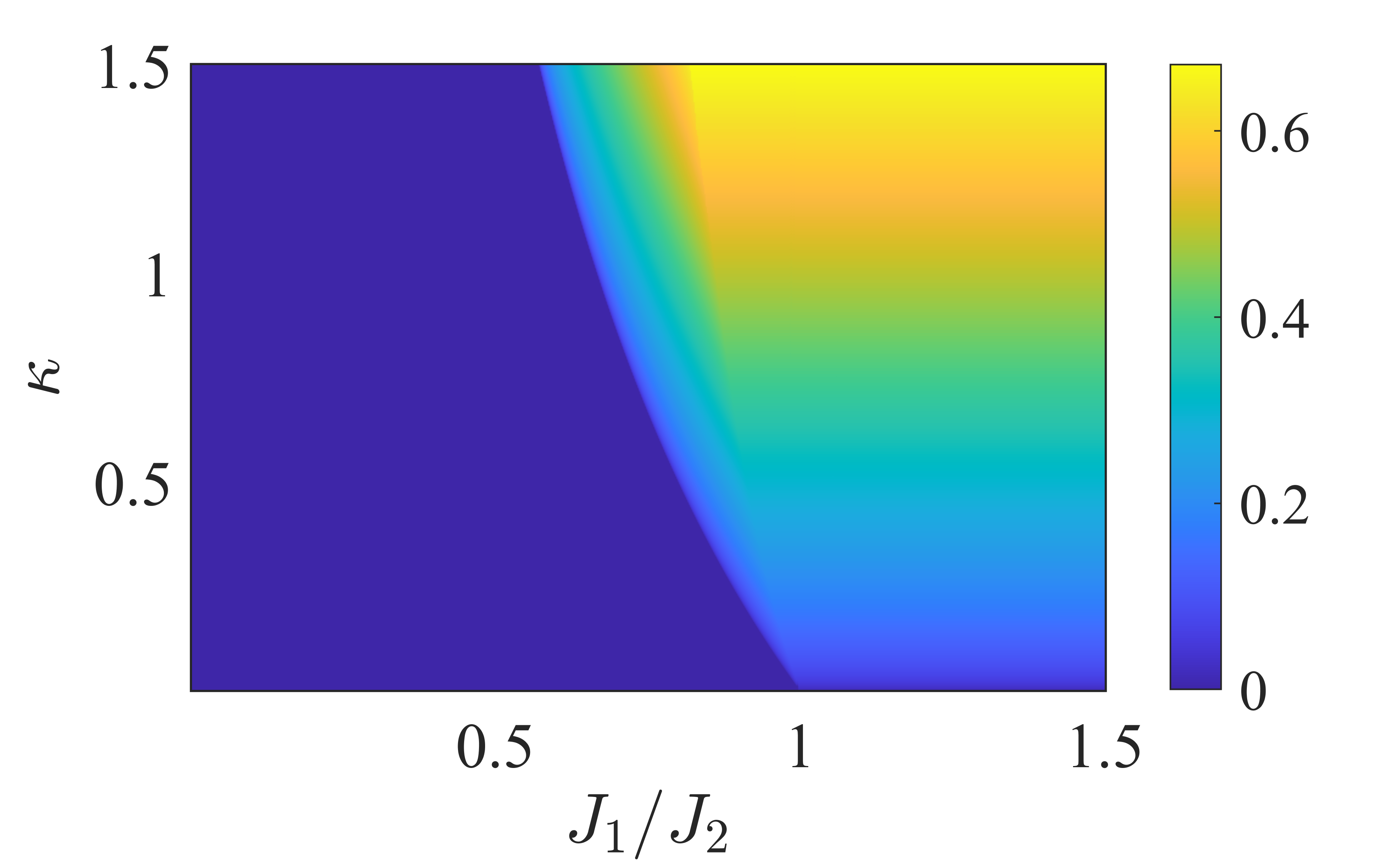}
            %\label{fig:a}
            \caption{$A_1$}
        \end{subfigure}
        \begin{subfigure}[h]{0.3\textwidth}
            \includegraphics[width=\textwidth]{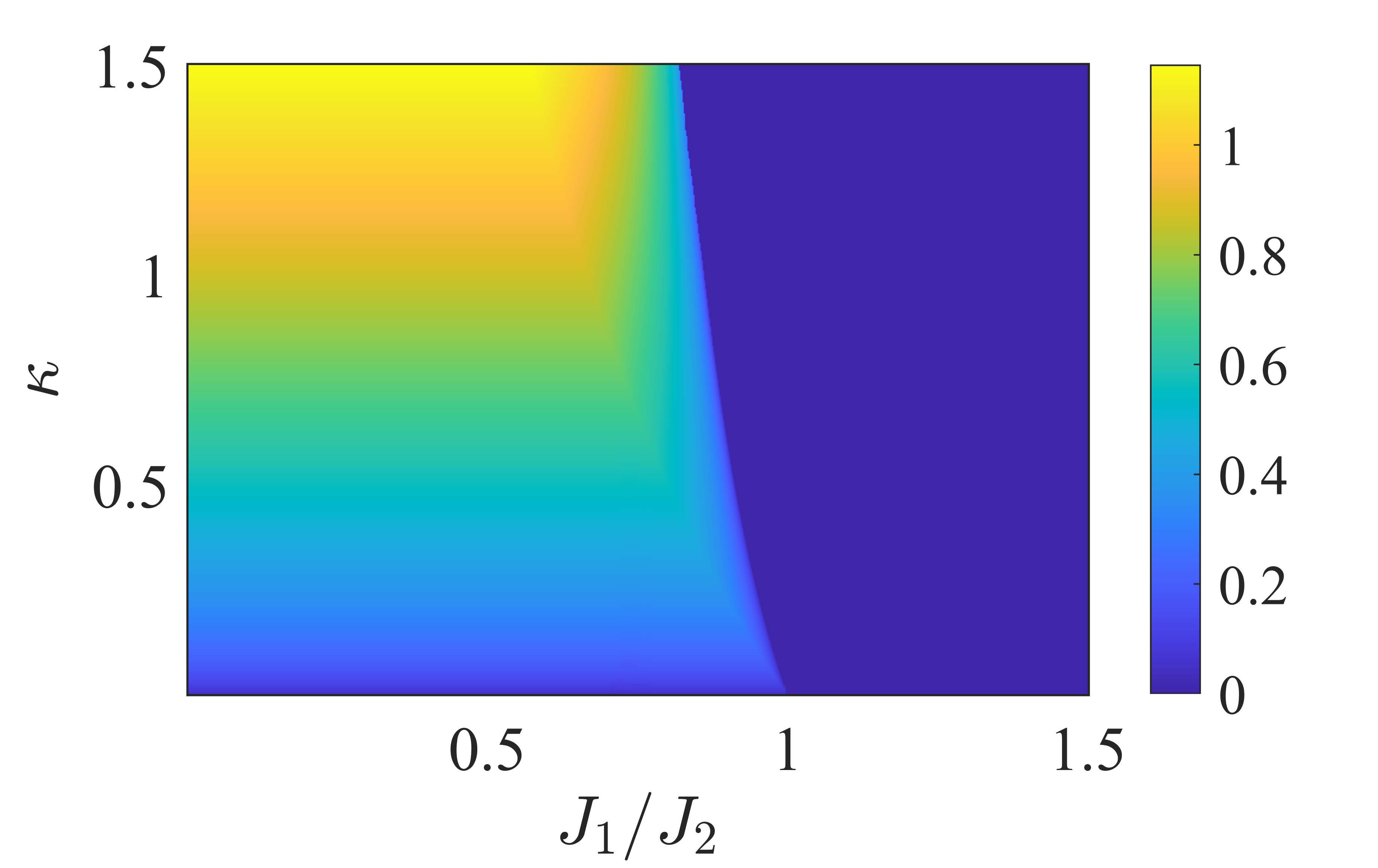}
            \caption{$A_2$}
        \end{subfigure}
        \begin{subfigure}[h]{0.3\textwidth}
            \includegraphics[width=\textwidth]{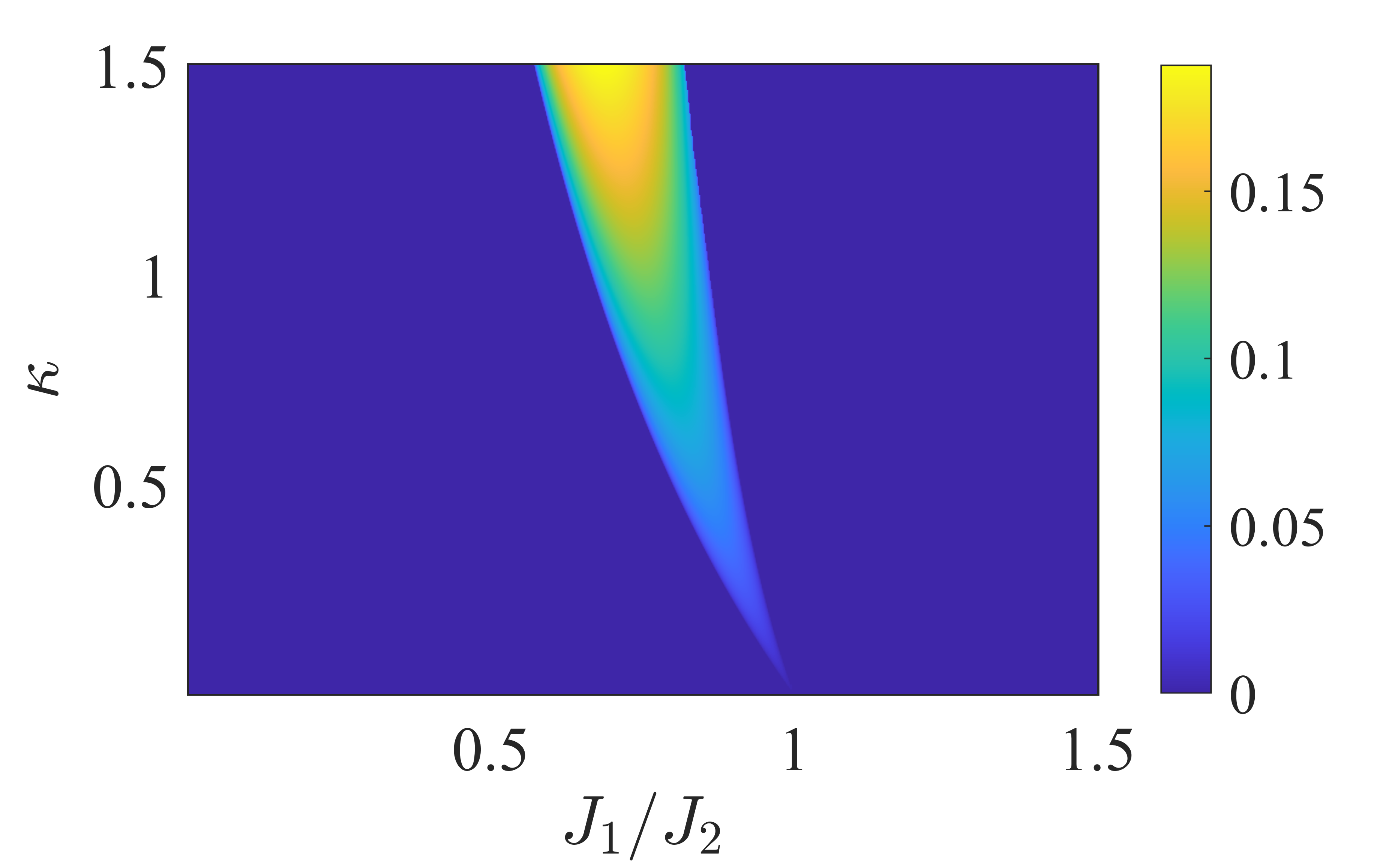}
            \caption{$B_1$}
        \end{subfigure}
        \caption{The ansatz value (a) $A_1$, (b) $A_2$ and (c) $B_1$ of $(0,\pi)$-flux.
            $A_1$ is finite at $J_1$ is large and $A_2$ is large at $J_1$ is large.
            $B_1$ is finite only when $A_1$ and $A_2$ are both finite.
            $B_2$ is always zero in this condition.
        }
        \label{zero-pi ansatz}
    \end{figure*}
    With these ansatz values,
    the spinon dispersion and the structure factor can be obtained,
    which are shown in Fig.~\ref{0-pi structure}.
    We find the bottom branch of the spinon dispersion is flat,
    therefore,
    the critical spinon density $\kappa_c$ is large in this state.
    \begin{figure*}[ht]
        \centering
        \begin{subfigure}[h]{0.3\textwidth}
            \includegraphics[width=\textwidth]{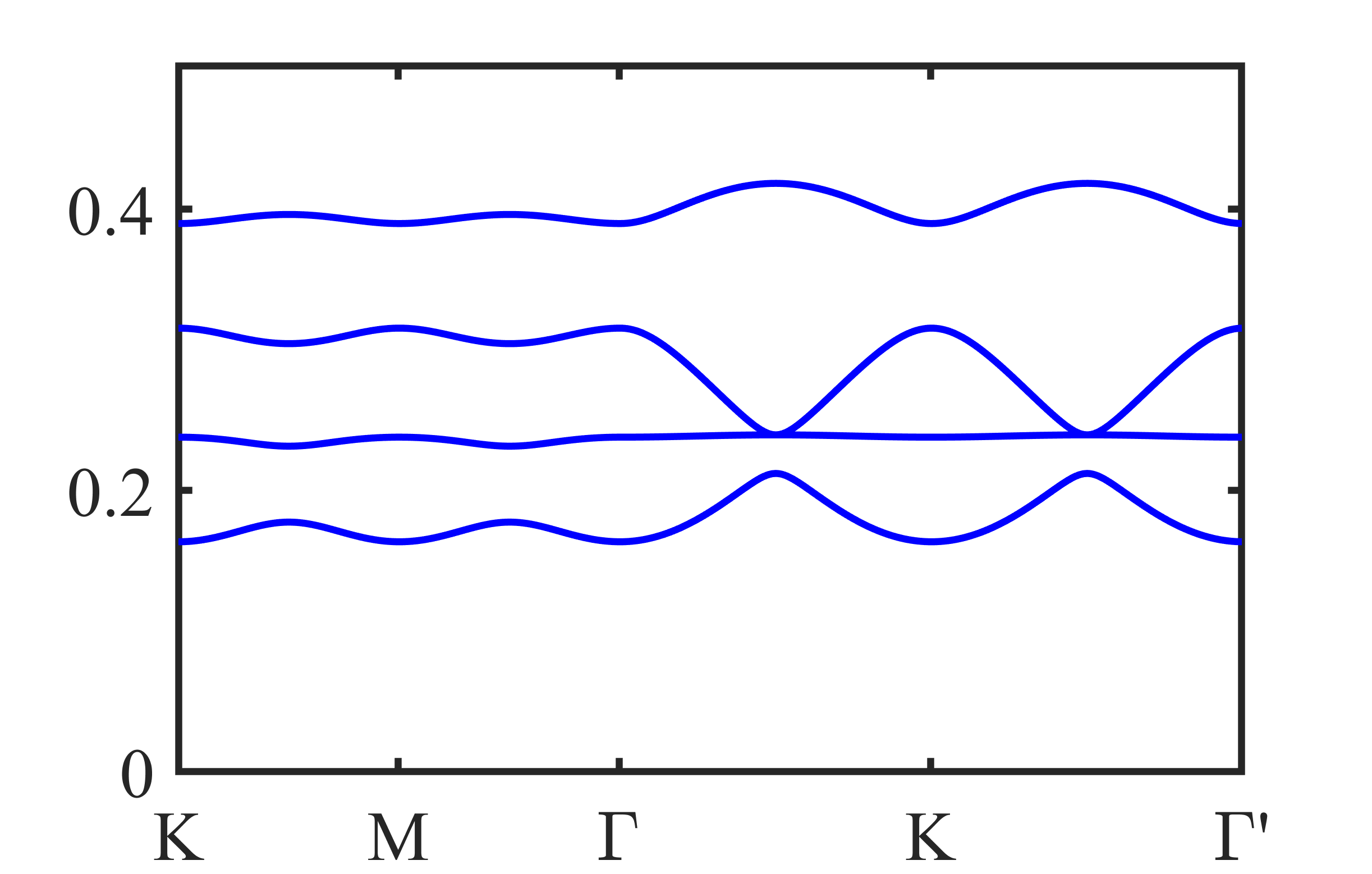}
            %\label{fig:a}
            \caption{Spinon dispersion}
        \end{subfigure}
        \begin{subfigure}[h]{0.3\textwidth}
            \includegraphics[width=\textwidth]{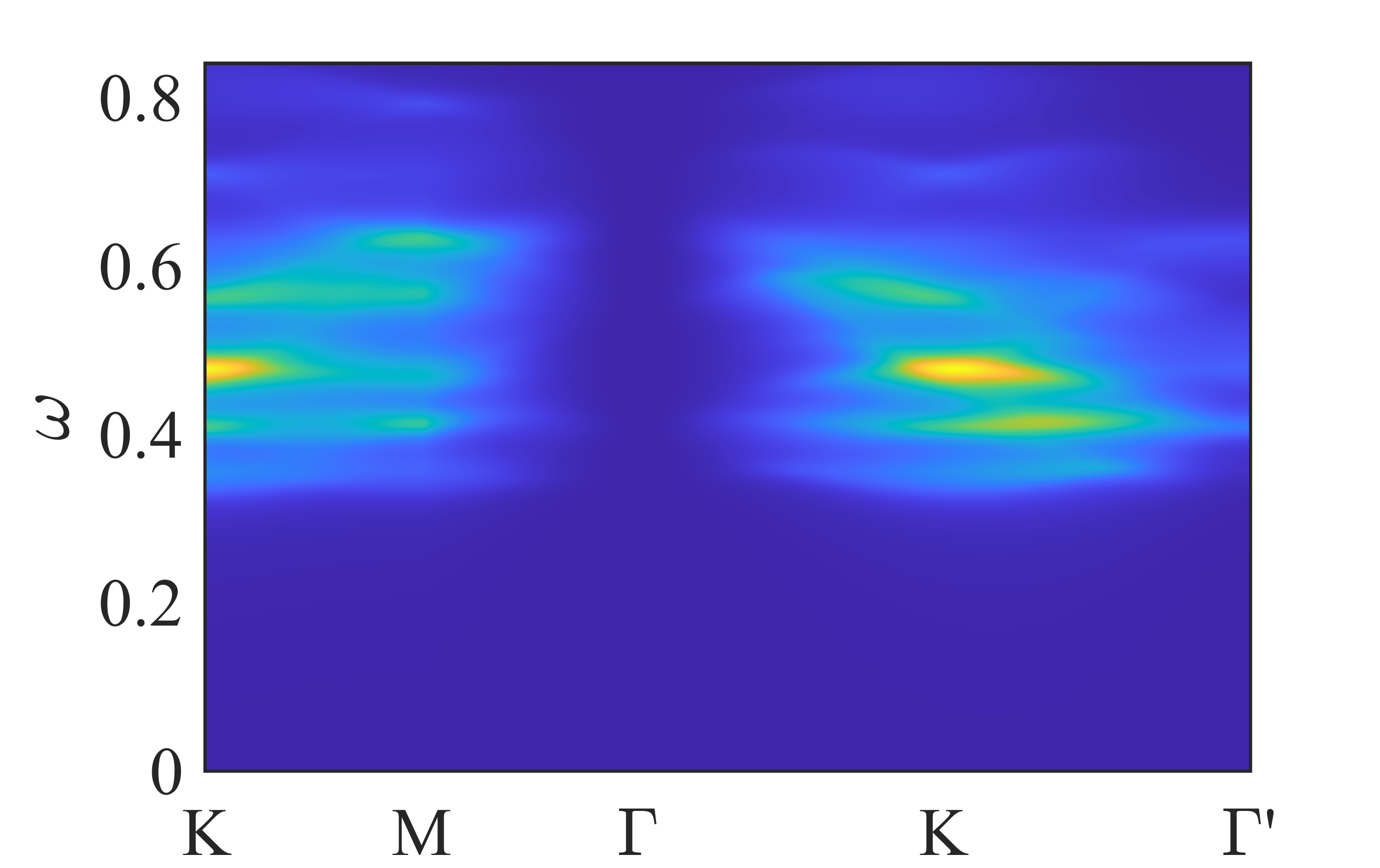}
            \caption{dynamic structure factor}
        \end{subfigure}
        \begin{subfigure}[h]{0.3\textwidth}
            \includegraphics[width=\textwidth]{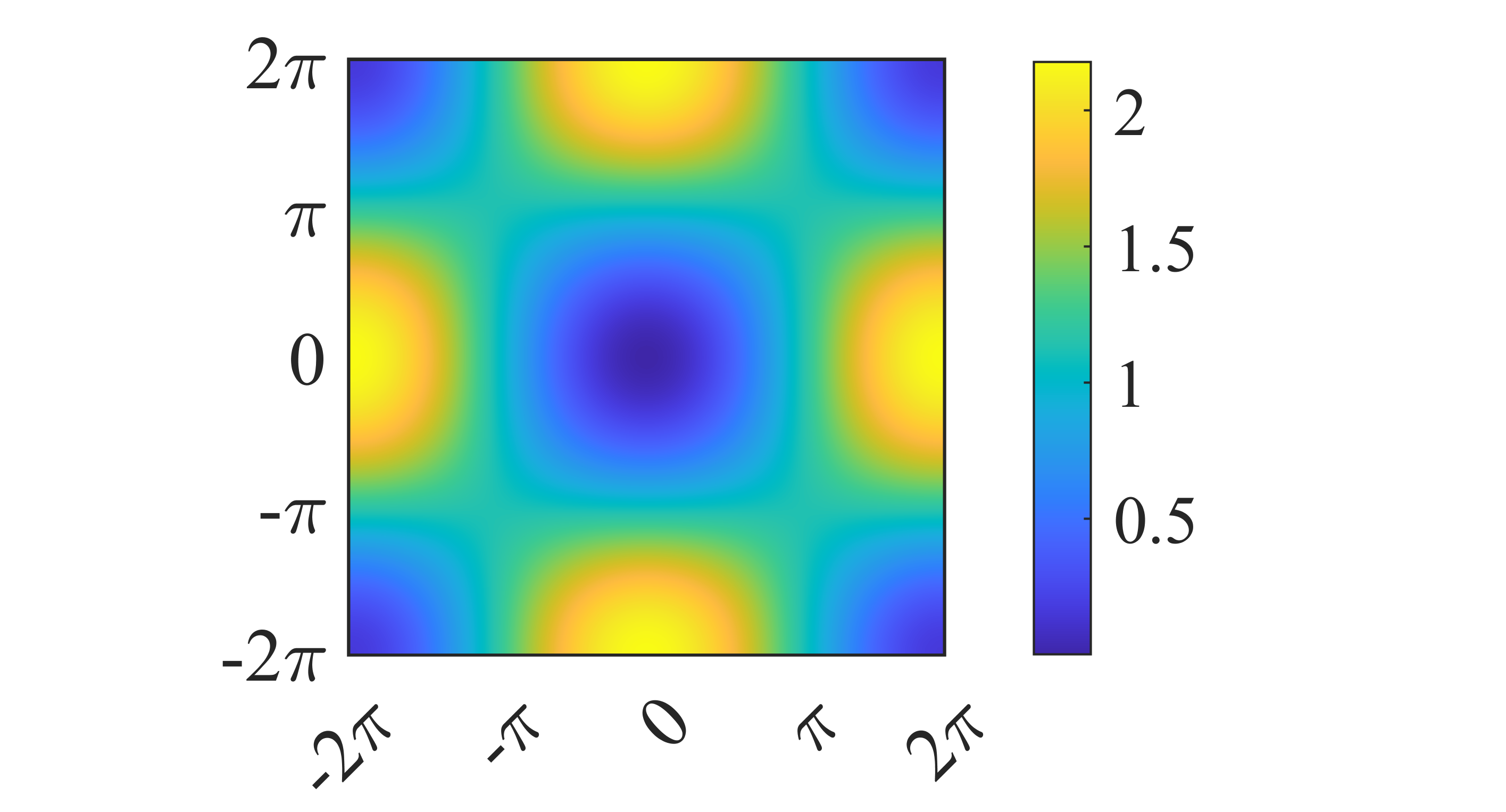}
            \caption{static structure factor}
        \end{subfigure}
        \caption{(a) is the spinon dispersion,
            (b) is the dynamic structure factor, and (c) is the static structure factor of $(0,\pi)$-flux state ($J_1/J_2=0.7$).
            The plot path of (a) and (b) is shown in Fig.~\ref{lattice-BZ} (b).
        }
        \label{0-pi structure}
    \end{figure*}
    % The ground state energy of this state is slightly higher than the energy minimum of $\pi$-flux ($(\pi,\pi)$-flux) and zero-flux ((0,0)-flux) state as shown in Fig.~(2)(c) in Main Text,
    % therefore,
    % this state is not shown in the phase diagram in Fig.~1 (c) in Main Text.
    % However,
    % the energy difference of $(\pi,\pi)$ and $(0,\pi)$-flux state is very small.
    % Therefore,
    % this state may also exist beyond the mean-field level.
    \subsection{$(\pi,0)$-flux state}
    In the $(\pi,0)$-flux state,
    only the ansatz amplitudes \(A_1\) and \(B_2\) are nonzero,
    and the configuration is shown in Fig.~\ref{pi-0} in Appendix~\ref{App:1B}.
    and the $\boldsymbol{A}_{\mathbf{k}}$ and $\boldsymbol{B}_{\mathbf{k}}$ are shown in Appendix.~\ref{App:3 C}.
    With these self-consistent equations in Eq.~(\ref{eq:self-consistent}),
    the mean-field ansatz can be solved.
    The spinon dispersion at physical condition $\kappa=1$ and the ansatz amplitudes and are shown in Fig.~\ref{Pi0 ansatz}.
    \begin{figure}[H]
        \centering
        \begin{subfigure}[h]{0.3\textwidth}
            \includegraphics[width=\textwidth]{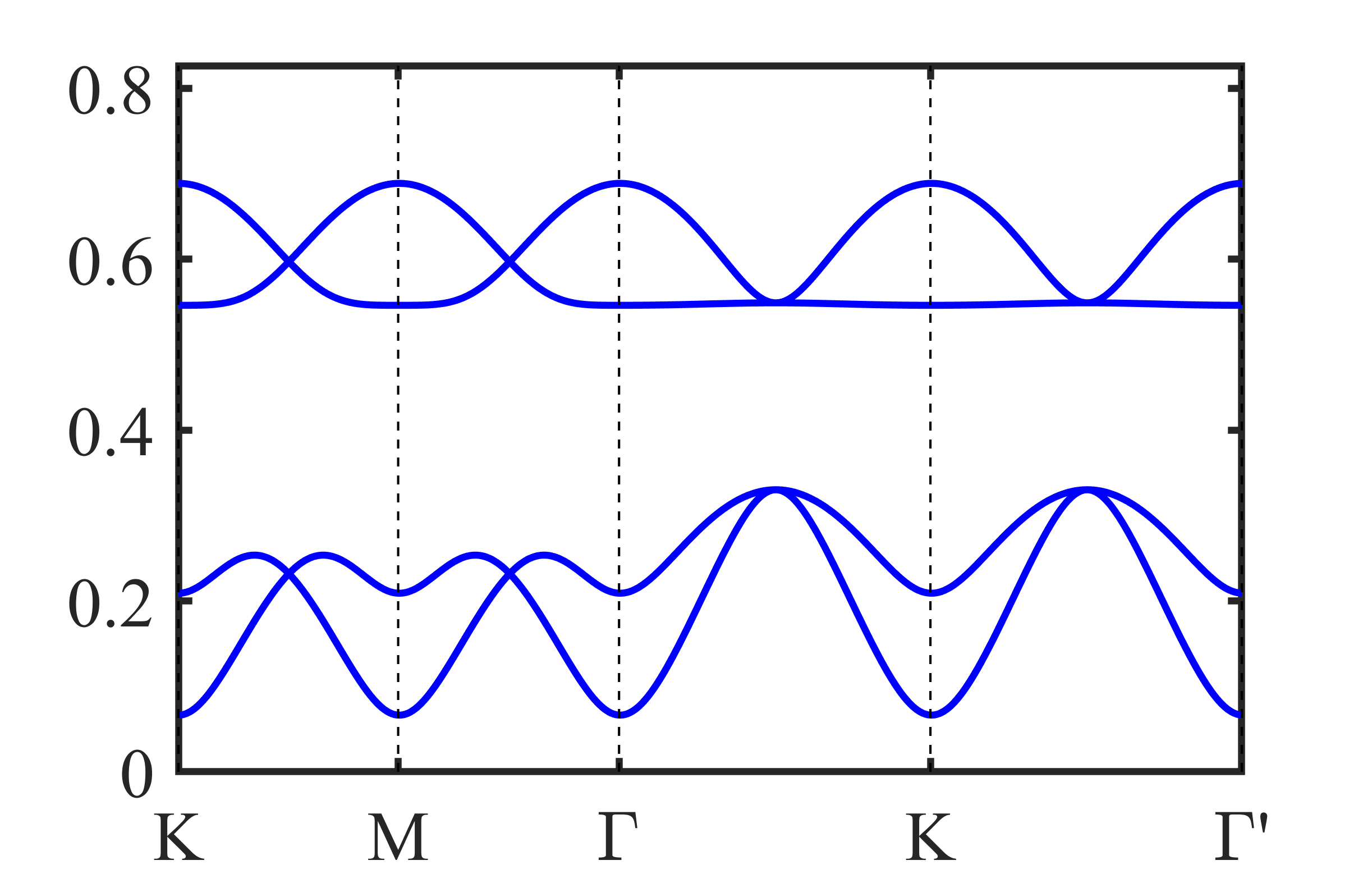}
            %\label{fig:a}
            \caption{Spinon dispersion at $\kappa=1$}
        \end{subfigure}
        \begin{subfigure}[h]{0.3\textwidth}
            \includegraphics[width=\textwidth]{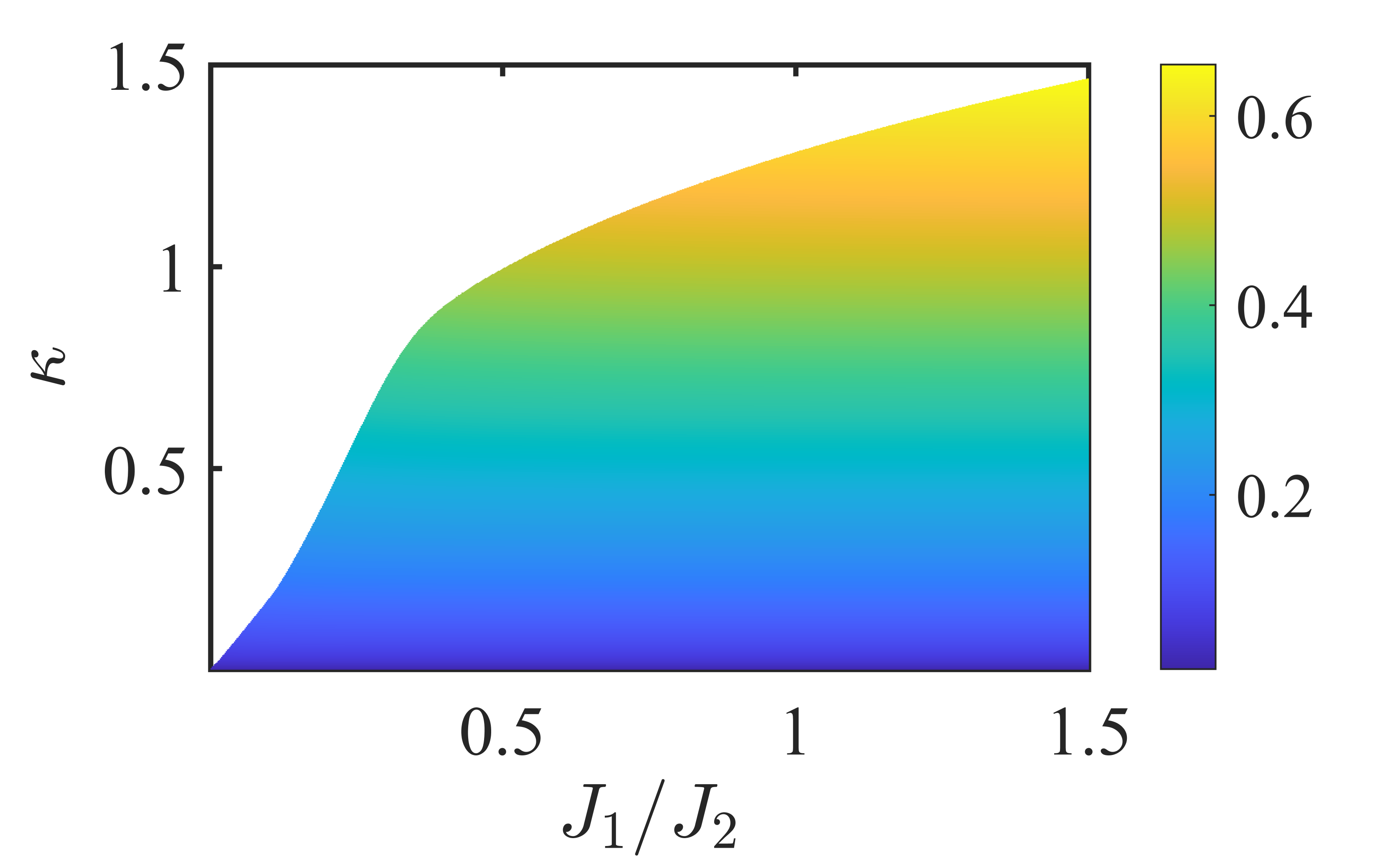}
            \caption{$A_1$}
        \end{subfigure}
        \begin{subfigure}[h]{0.3\textwidth}
            \includegraphics[width=\textwidth]{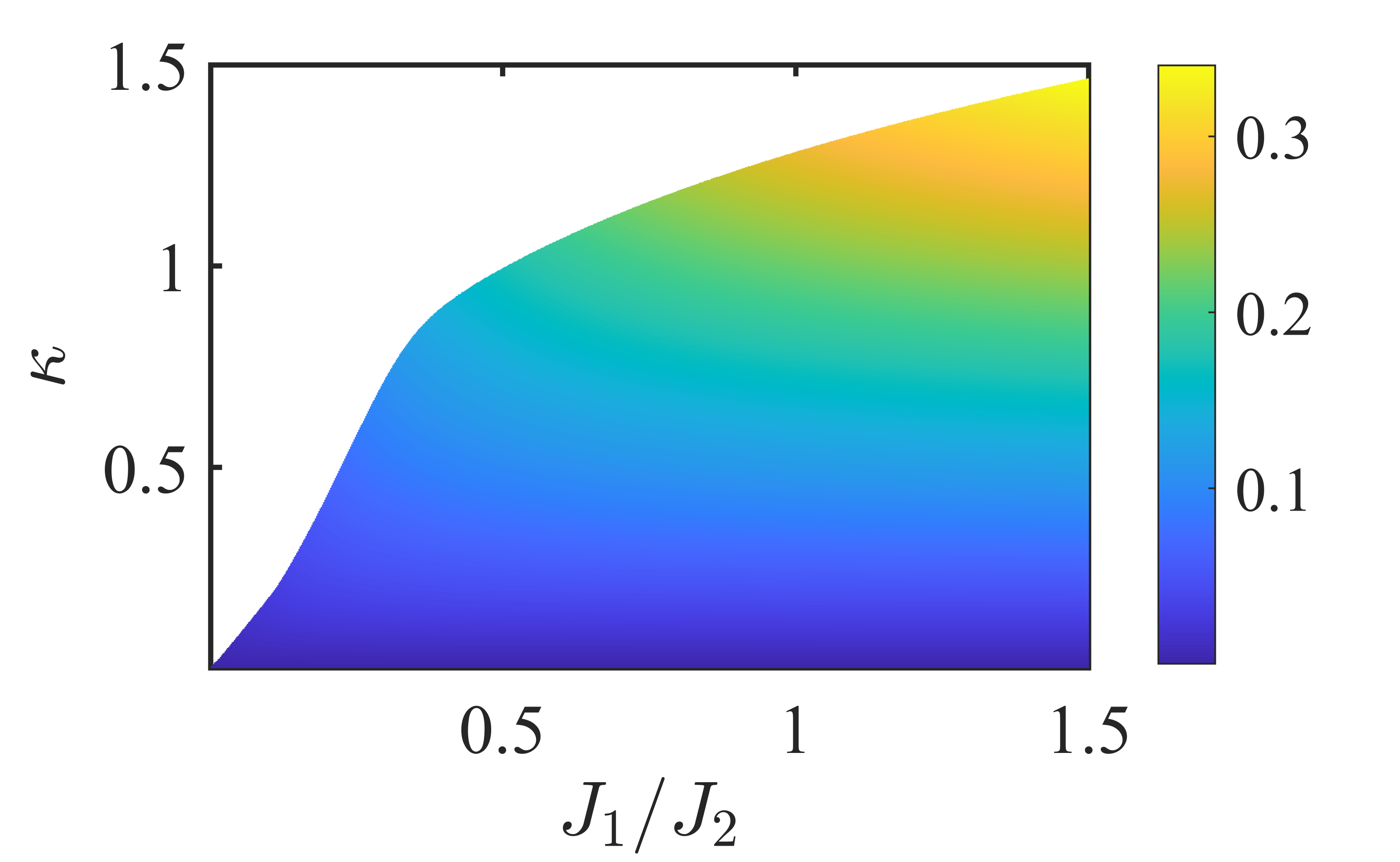}
            \caption{$B_2$}
        \end{subfigure}
        \caption{(a) is the spinon dispersion at $\kappa=\kappa_c$ and $J_1=J_2=1$ of ($\pi,0$)-flux.
            The plot path is shown in Fig.~\ref{lattice-BZ} (b) .
            (b) and (c) are the ansatz $A_1$ and $B_2$ value respectively.
            $A_2$ and $B_1$ are always zero in this condition.
        }
        \label{Pi0 ansatz}
    \end{figure}
    \subsection{$(\pi,\pi)$-flux state}
    In the $(\pi,\pi)$-flux state,
    the ansatz amplitudes \(A_1\), \(A_2\) and \(B_1\) are nonzero.
    The configuration of this state is shown in Fig.~\ref{pi-pi} in Appendix~\ref{App:1A}.
    and the $\boldsymbol{A}_{\mathbf{k}}$ and $\boldsymbol{B}_{\mathbf{k}}$ are shown in Appendix~\ref{App:3 D}.
    Solving the self-consistent equations in Eq.~(\ref{eq:self-consistent}) we have get the ansatz amplitudes,
    which are shown in Fig.~\ref{pi-pi ansatz}.
    \begin{figure}[H]
        \centering
        \begin{subfigure}[h]{0.3\textwidth}
            \includegraphics[width=\textwidth]{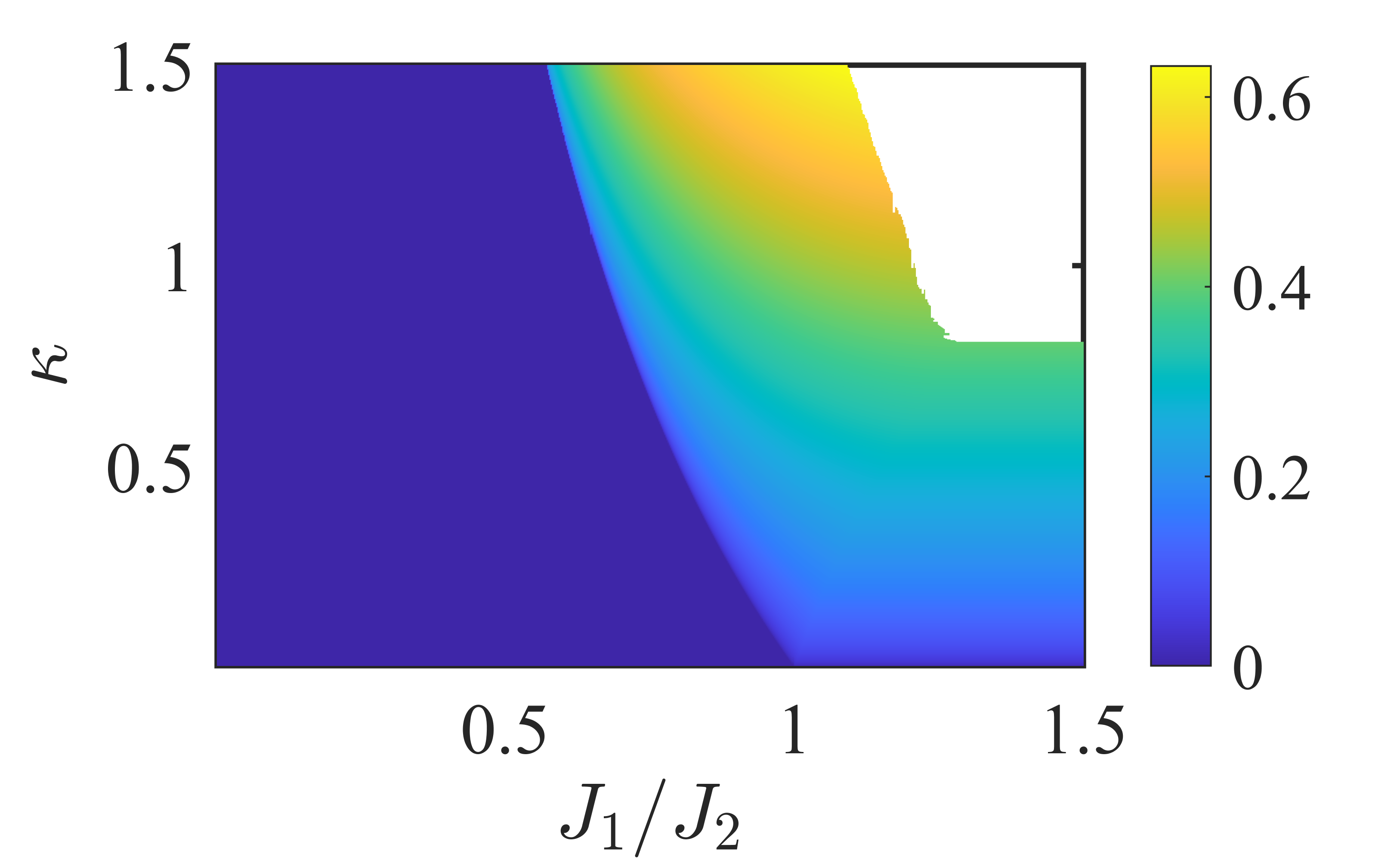}
            %\label{fig:a}
            \caption{$A_1$}
        \end{subfigure}
        \begin{subfigure}[h]{0.3\textwidth}
            \includegraphics[width=\textwidth]{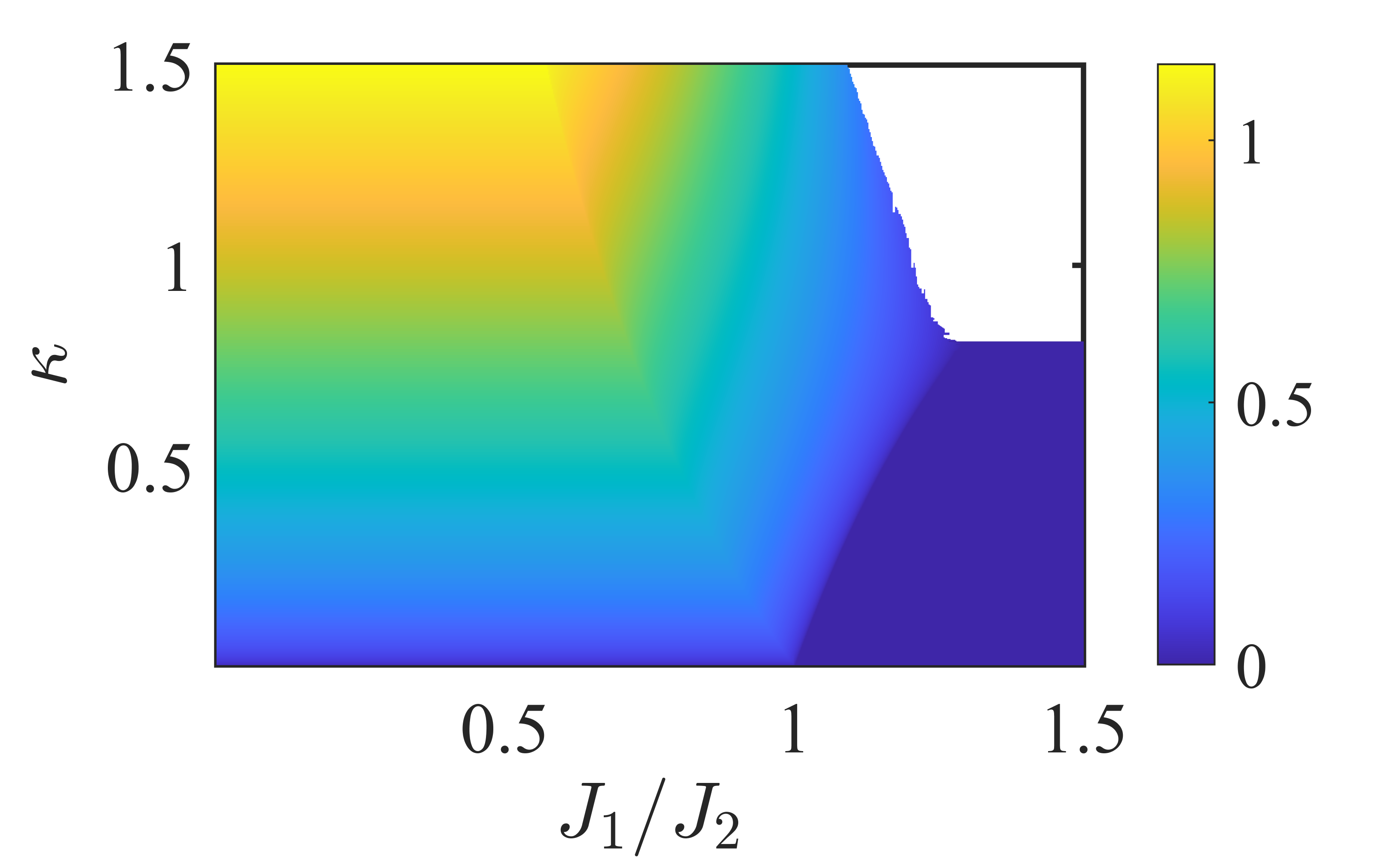}
            \caption{$A_2$}
        \end{subfigure}
        \begin{subfigure}[h]{0.3\textwidth}
            \includegraphics[width=\textwidth]{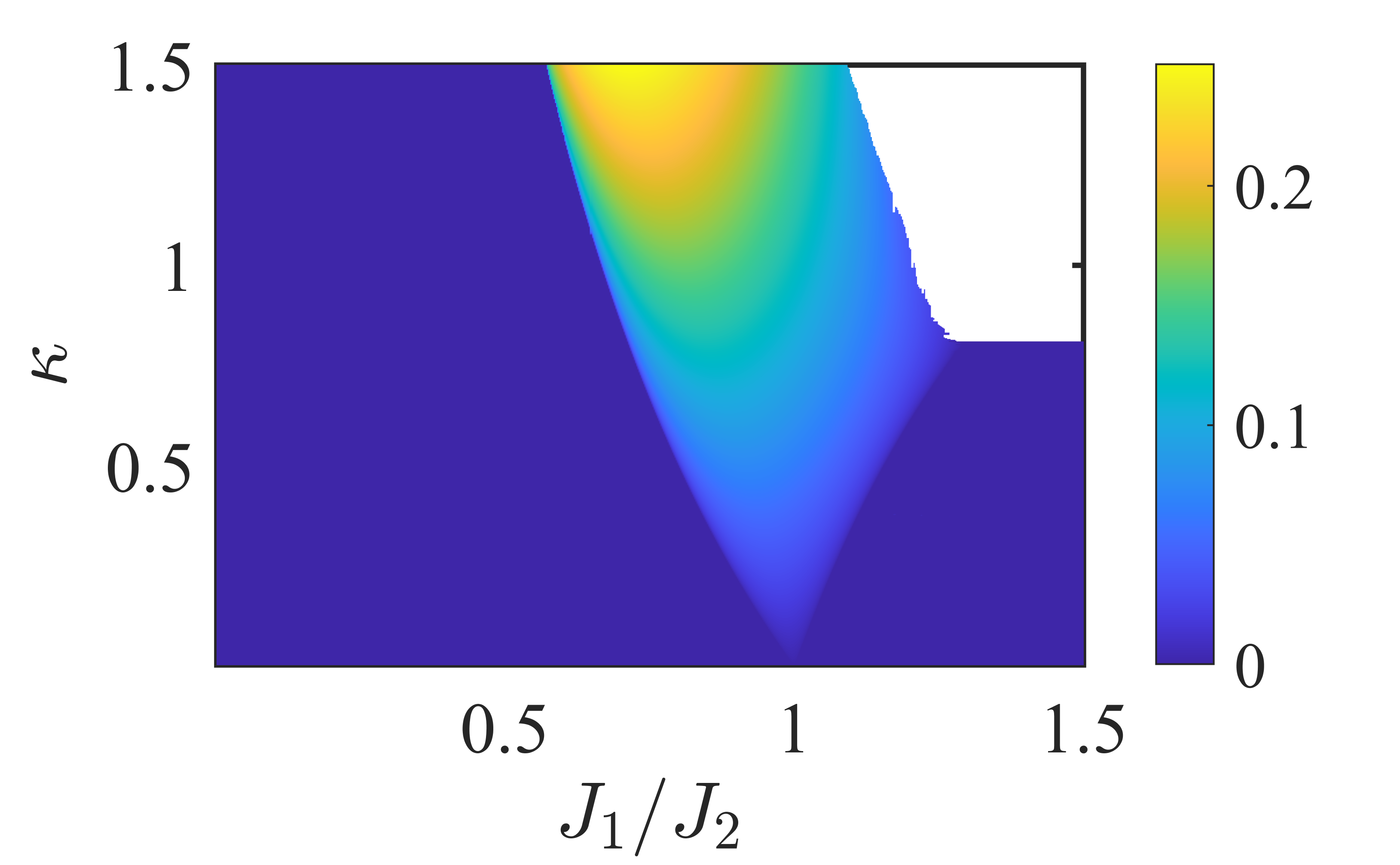}
            \caption{$B_1$}
        \end{subfigure}
        \caption{The ansatz value (a) $A_1$, (b) $A_2$ and (c) $B_1$ of $(\pi,\pi)$-flux.
            $A_1$ is finite at $J_1$ is large and $A_2$ is large at $J_1$ is large.
            $B_1$ is finite only when $A_1$ and $A_2$ are both finite.
            $B_2$ is always zero in this condition.
        }
        \label{pi-pi ansatz}
    \end{figure}
    With these ansatz values,
    the spinon dispersion and the structure factor can be obtained,
    which are shown in Fig.~\ref{pi-pi structure}.
    \begin{figure}[H]
        \centering
        \begin{subfigure}[h]{0.3\textwidth}
            \includegraphics[width=\textwidth]{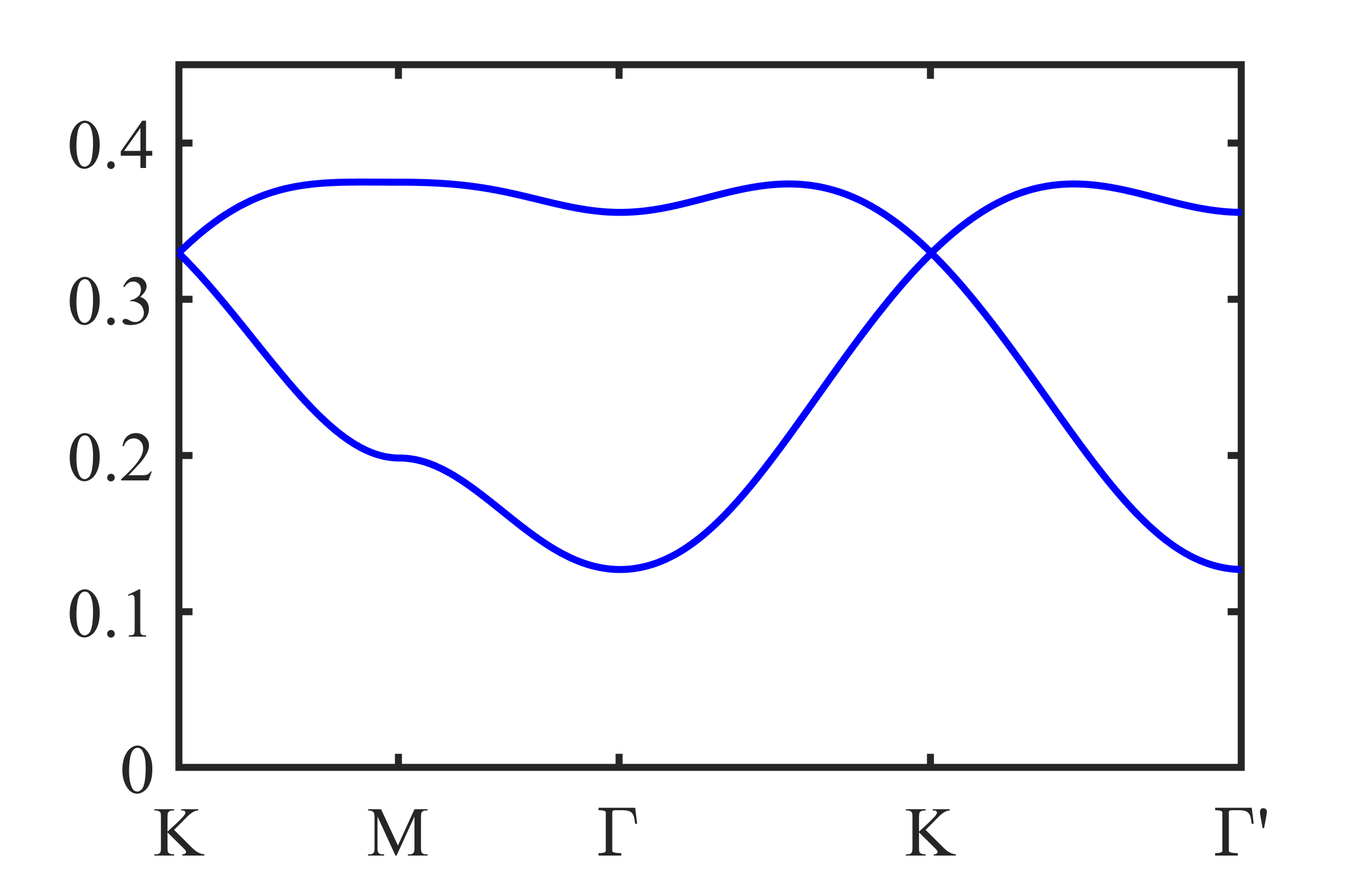}
            %\label{fig:a}
            \caption{Spinon dispersion}
        \end{subfigure}
        \begin{subfigure}[h]{0.3\textwidth}
            \includegraphics[width=\textwidth]{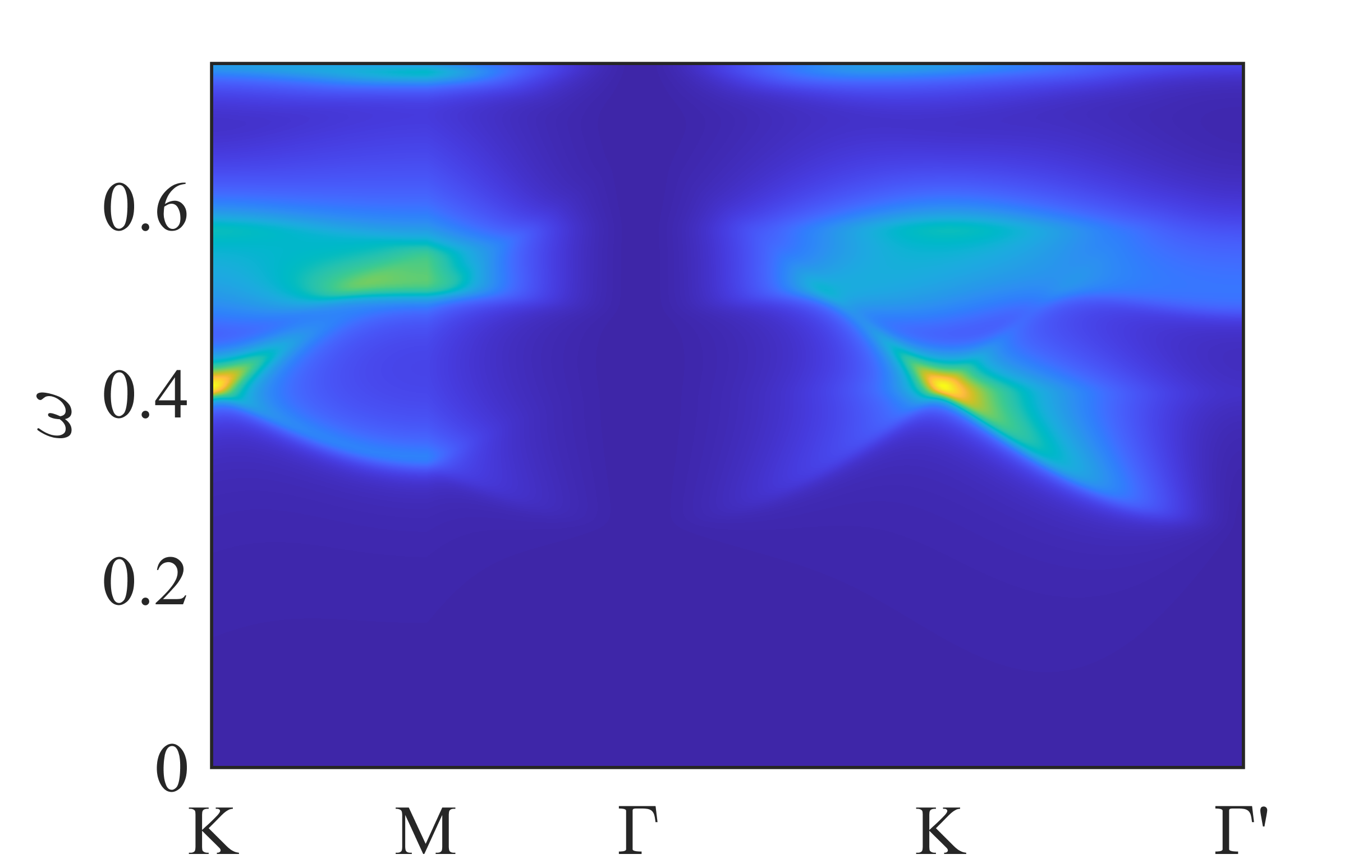}
            \caption{dynamic structure factor}
        \end{subfigure}
        \begin{subfigure}[h]{0.3\textwidth}
            \includegraphics[width=\textwidth]{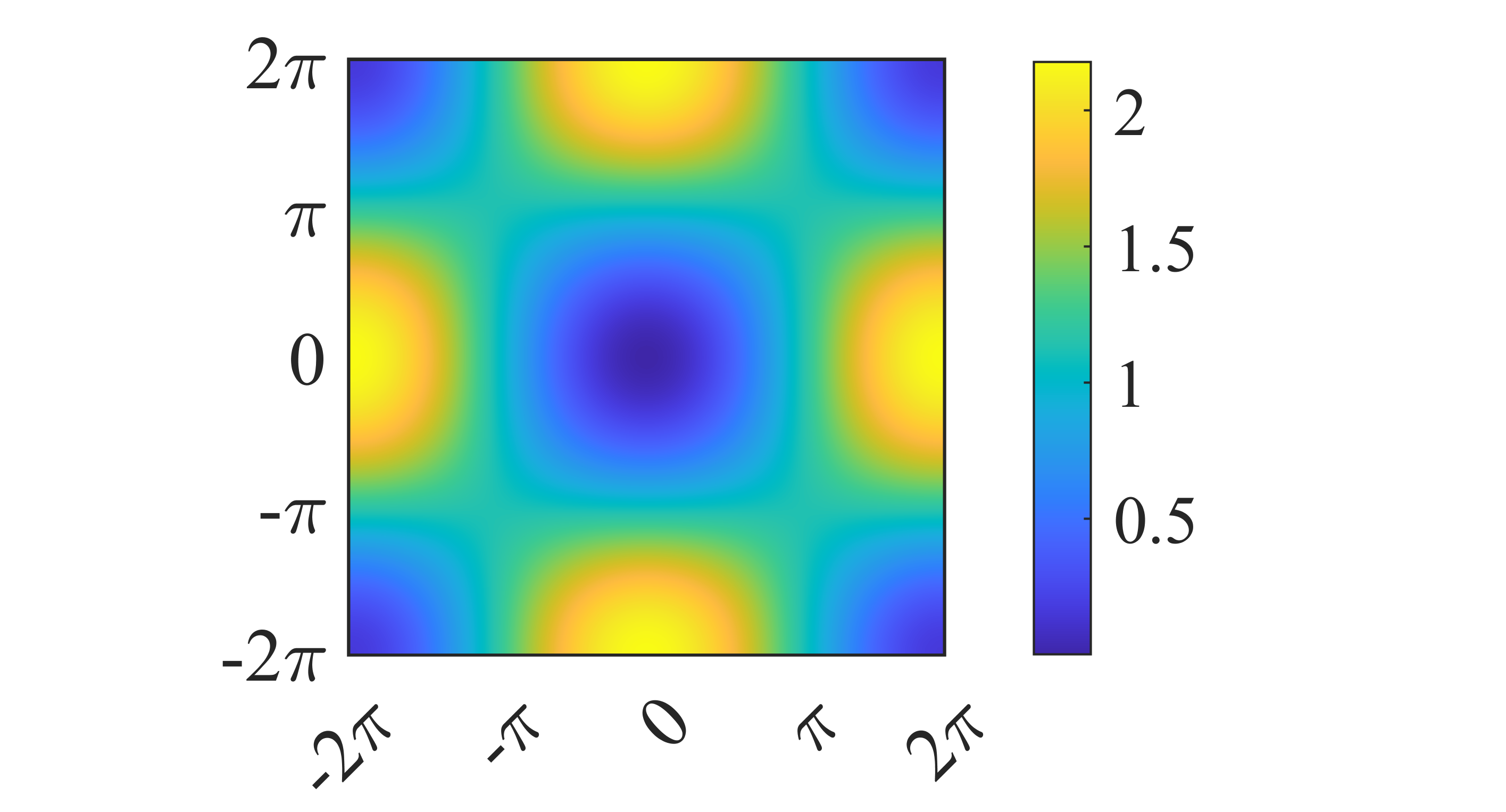}
            \caption{static structure factor}
        \end{subfigure}
        \caption{(a) is the spinon dispersion,
            (b) is the dynamic structure factor, and (c) is the static structure factor of $(\pi,\pi)$-flux state ($J_1/J_2=0.7$).
            The plot path of (a) and (b) is shown in Fig.~\ref{lattice-BZ} (b).
        }
        \label{pi-pi structure}
    \end{figure}
    As shown in Fig.~\ref{pi-pi structure} (a),
    the minima of the spinon dispersion is located at $\mathbf{Q}=(0,0)$,
    and magnetic order will form when $\kappa>\kappa_c$,
    and the details are discussed in Appendix \ref{App:4}.
    % \begin{eqnarray}
    %     \boldsymbol{A}_{\mathbf{k}}&=&
    %     \left(\begin{array}{cccc}
    %     0 & 0 &0& 0 \\
    %     0 & 0 &0& 0 \\
    %     0 & 0 &0& 0 \\
    %     0 & 0 &0& 0 \\
    %     \end{array}\right),\\
    %     \boldsymbol{B}_{\mathbf{k}}&=&
    %     \left(\begin{array}{cccccccc}
    %     0 & 0 &0& 0 \\
    %     0 & 0 &0& 0 \\
    %     0 & 0 &0& 0 \\
    %     0 & 0 &0& 0
    %     \end{array}\right).
    % \end{eqnarray}

    % \begin{eqnarray}
    %     D_{\mathbf{k}}&=&
    %     \left(\begin{array}{cccccccc}
    %         -\mu & 0 &\frac{1}{2}J_2 B_2 e^{-i k_2}& 0& 0& \frac{1}{2}J_1A_1(-1+e^{-i k_1})& 0& \frac{1}{2}J_1A_1(1-e^{-ik_2}) \\
    %         0 & -\mu &0& \frac{1}{2}J_2 B_2& -\frac{1}{2}J_1A_1& 0& -\frac{1}{2}J_1A_1& 0 \\
    %         \frac{1}{2}J_2 B_2 & 0 &-\mu& 0& 0& \frac{1}{2}J_1A_1& 0& \frac{1}{2}J_1A_1 \\
    %         0 & \frac{1}{2}J_2 B_2 &0& -\mu& -\frac{1}{2}J_1A_1& 0& -\frac{1}{2}J_1A_1& 0 \\
    %         0 &  -\frac{1}{2}J_1A_1 &0&  -\frac{1}{2}J_1A_1& -\mu& 0& \frac{1}{2}J_2 B_2& 0 \\
    %         \frac{1}{2}J_1A_1 & 0 & \frac{1}{2}J_1A_1& 0& 0& -\mu& 0& \frac{1}{2}J_2 B_2 \\
    %         0 &  -\frac{1}{2}J_1A_1 & 0&  -\frac{1}{2}J_1A_1& \frac{1}{2}J_2 B_2& 0& -\mu& 0\\
    %         \frac{1}{2}J_1A_1 & 0 & \frac{1}{2}J_1A_1& 0& 0& \frac{1}{2}J_2 B_2& 0& -\mu
    %         \end{array}\right),
    %         \label{Dq}
    % \end{eqnarray}

    \subsection{Plaquette-singlet states}
    \label{PS-state}
    To study the PS phase, we have tried different ansatz configurations which break glide symmetries.
    After solving the self-consistent equations,
    we find two plaquette-singlet state solutions depicted in Fig.~\ref{PS}.
    These PS mean-field ansatz have nonzero $A,B$ amplitudes only in the empty squares or $J_2$ squares.
    % which are in the the empty square and \(J_2\)-square respectively.
    % The configurations of these two states are depicted in Fig.~\ref{PS},
    % and are total decoupled.

    First we consider the plaquette-singlet state in the empty square.
    The mean-field Hamiltonian can be written as
    \begin{eqnarray}
        H_\mathrm{MF}=-\Ncell J_1\sum_s A_1\hat{A}_{s,s+1}+h.c.+4 \Ncell J_1\sum_s |A_1|^2-4 \Ncell \mu (\hat{n}-\kappa),
    \end{eqnarray}
    %where \(\Ncell\) is the number of unit cell and 
    where the summation is only for the sites in one unit cell,
    because the mean-field ansatz is decoupled into disconnected empty squares.
    The \(s=0,1,2,3\) in this summation is the atom index in the unit cell shown in Fig.~\ref{sslatice}.
    After Bogoliubov transformation,
    the mean-field Hamiltonian is diagonalized as
    \begin{eqnarray}
        H_{\mathrm{MF}}=\Ncell \sum_{s}\omega_{s}(\gamma_{s\uparrow}^\dagger\gamma_{s\uparrow}+\gamma_{s\downarrow}^\dagger\gamma_{s\downarrow}+1) +\Ncell \left[\mu+\mu \kappa+4J_1\left|A_1\right|^2\right],
    \end{eqnarray}
    where the spinon dispersions $\omega_{s}$ are
    \begin{eqnarray}
        \omega_{0}&=&\omega_{1}=|\mu|,\\
        \omega_{2}&=&\omega_{3}=\sqrt{\mu^2-|J_1A_1|^2}.
    \end{eqnarray}
    The self-consistent equations are
    \begin{eqnarray}
        8 J_1 A_1&=&-\sum_s \frac{\partial \omega_{s}}{\partial A_1},\\
        1+\kappa&=&-\sum_s\frac{\partial \omega_{s}}{\partial \mu}.
    \end{eqnarray}

    Then we consider the plaquette-singlet state in the \(J_2\)-square.
    The mean-field Hamiltonian can be written as
    \begin{eqnarray}
        H_\mathrm{MF}=-\Ncell J_1\sum_{\langle s s^\prime \rangle} A_1\hat{A}_{s,s^\prime}+\Ncell J_2 B_2 \hat{B}_2+h.c.+4\Ncell J_1\sum_s |A_1|^2-\Ncell J_2|B_2|^2-4\Ncell \mu (\hat{n}-\kappa),
    \end{eqnarray}
    where $s$ is the index of the sites in the \(J_2\)-square and \(s^\prime\) is the nearest neighbor site of the \(s\) site in the \(J_2\)-square.
    After Bogoliubov transformation,
    the mean-field Hamiltonian is diagonalized as
    \begin{eqnarray}
        H_{\mathrm{MF}}=\Ncell \sum_{s}\omega_{s}(\gamma_{s\uparrow}^\dagger\gamma_{s\uparrow}+\gamma_{s\downarrow}^\dagger\gamma_{s\downarrow}+1) +\Ncell \left[\mu+\mu \kappa+4J_1\left|A_1\right|^2-J_2\left|B_2\right|^2\right],
    \end{eqnarray}
    where the spinon dispersions are
    \begin{eqnarray}
        \omega_{0}&=&|\mu|,\\
        \omega_{1}&=&|\frac{1}{2}J_2B_2-\mu|,\\
        \omega_{2}&=&\sqrt{\left(\frac{1}{4}J_2B_2+\mu\right)^2-|J_1A_1|^2}+\frac{J_2B_2}{4},\\
        \omega_{3}&=&\sqrt{\left(\frac{1}{4}J_2B_2+\mu\right)^2-|J_1A_1|^2}-\frac{J_2B_2}{4},
    \end{eqnarray}
    and the self-consistent equations are
    \begin{eqnarray}
        8 J_1 A_1&=&-\sum_s \frac{\partial \omega_{s}}{\partial A_1},\\
        2 J_2 B_2&=&\sum_s \frac{\partial \omega_{s}}{\partial B_2},\\
        1+\kappa&=&-\sum_s\frac{\partial \omega_{s}}{\partial \mu}.
    \end{eqnarray}
\end{widetext}
\section{Mean-field phase diagram}\label{MF phase diagram}

\begin{figure}[ht]
    \centering
    \includegraphics[width=0.48\textwidth]{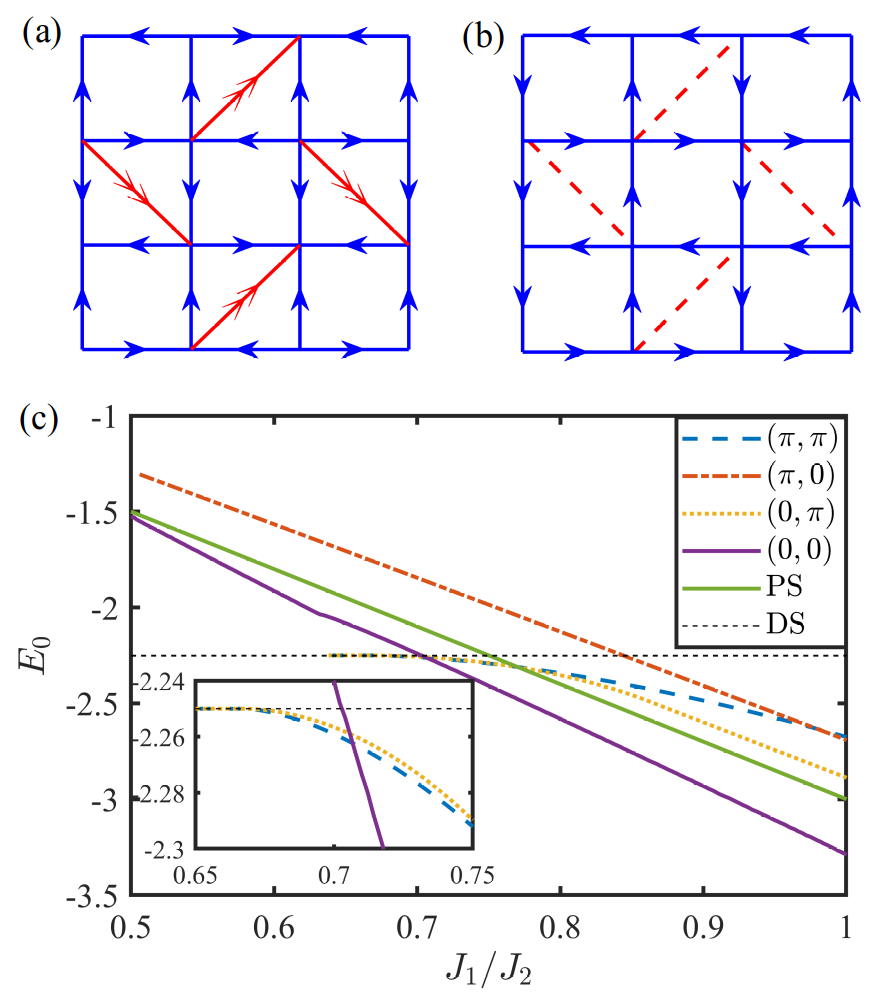}
    %\label{fig:a}
    \caption{(a) and (b) are the ansatz $A_{ij}$ of the zero-flux and $\pi$-flux states respectively.
        An arrow from site $i$ to $j$ means $A_{ij}>0$ and
        $A_{ij}$ in dashed lines are zero.
        The different arrows represent different amplitudes of $A_{ij}$.
        (c) is the ground state energies of the 4 gauge inequivalent symmetric spin liquid ansatz and two valence bound solid (DS and PS) states
        for $\kappa=1$.
        The inset in (c) is the details in $0.65<J_1/J_2<0.75$.
        It shows that only the DS state and $(\pi,\pi)$-flux and $(0,0)$-flux SL states appear as mean-field ground states for $\kappa=1$.
    }
    \label{ansatz and energy}
\end{figure}
In the last section we show the details and results of the four symmetric spin liquid states.
The $(\pi,\pi)$ and $(0,0)$-flux states are called $\pi$-flux and zero-flux hereafter,
Comparing the mean-field ground state energies of the 4 gauge inequivalent symmetric ansatz with the change of $J_1/J_2$ and $\kappa$,
% [see Fig.~\ref{ansatz and energy} (c)],
we get a mean-field phase diagram of the Heisenberg model in Shastry-Sutherland lattice,
which is shown in Fig.~\ref{phase diagram} (c).
We find only two ansatz with zero or $\pi$-flux in each plaquettes as mean-field ground states under physical condition ($\kappa=1$).
The ansatz configuration of these two state are shown in Fig.~\ref{ansatz and energy} (a) and (b).
% The $\pi$-flux state is dominant when $J_2$ is large and the zero-flux state is dominant when $J_1$ is large. 

% revised
The zero-flux state is the mean-field ground state for large $J_1$ ($J_1/J_2 > 0.71$ at $\kappa=1$).
The emergent gauge field for the zero-flux state is a staggered $U(1)$ gauge field,
and the spinons are either confined (possibly forming valence bond solid) when $\kappa$ is small\cite{confinement}
or condensed when $\kappa$ is large.
For physical $\kappa=1$ the spinons of zero-flux state condense and form the N\'eel AFM order.
The zero-flux Schwinger boson state and related phases has been discussed before in the context of square lattice antiferromagnets\cite{PhysRevLett.66.1773}.

The $\pi$-flux state is the mean-field ground state for small $J_1$ ($J_1/J_2<0.71$ at $\kappa=1$).
For intermediate $J_1$ ($0.66 < J_1/J_2 < 0.71$ at $\kappa=1$) this state has $Z_2$ gauge field.
For physical $\kappa=1$ the spinons of this state are gapped and form a gapped $Z_2$ spin liquid.
For very high $\kappa$ ($\kappa \gtrsim 2.5$ for $J_1/J_2\sim 0.7$) the bosons will condense and likely form
a 4-sublattice antiferromagnetic order similar to the $\pi$-flux Schwinger boson state of square lattice antiferromagnets\cite{piFluxOrder}.

% and the details is shown in the Supplemental Material.
In the lowest $J_1$ region ($J_1/J_2<0.66$ at $\kappa=1$)
the $\pi$-flux state reduces to the confined dimer-singlet state with only next-nearest-neighbor boson pairing $A_2\neq 0$.

Therefore, there are three distinct phases with the change of $J_1/J_2$ for physical $\kappa=1$ under mean-field approximation:
dimer-singlet phase for $J_1/J_2<0.66$, $\pi$-flux $Z_2$ spin liquid state for $0.66<J_1/J_2<0.71$ and N\'eel phase for $J_1/J_2>0.71$.
We note that the ground state energy of the $(0,\pi)$-flux state is slightly higher than the energy of $\pi$-flux ($(\pi,\pi)$-flux)  as shown in inset of Fig.~\ref{ansatz and energy} (c).
Therefore,
the $(0,\pi)$-flux state is not shown in the phase diagram in Fig.~\ref{phase diagram}.
However,
the energy difference of $(\pi,\pi)$ and $(0,\pi)$-flux state is very small.
Therefore,
this state may also exist beyond the mean-field level.

The dynamic and static structure factors of the $\pi$-flux spin liquid state are also calculated by the Schwinger boson mean field theory
and shown in Fig.~\ref{pi-pi structure} (b) and (c) respectively,
which may be used as numerical and experimental signatures of this spin liquid state.
In particular the dominant short-range spin correlations in this $\pi$-flux SL state
is related to a 4-sublattice AFM order (see Appendix \ref{App:4}).

To further investigate the PS phase,
we also study mean-field ansatz with plaquette-singlet order.
% which breaks the glide symmetry.
% The mean-field ansatz
% are finite only in the plaquette bonds of Shastry-Sutherland lattice.
There are two kinds of plaquette-singlet states for the two kinds of plaquette in Shastry-Sutherland lattice.
Only the mean field energy of plaquette-singlet state in the ``empty'' square is plotted in Fig.~\ref{ansatz and energy} (c),
because it has lower ground state energy (See Fig.~(\ref{energy plot}) in Appendix~\ref{App:1D}).
We find that the ground state energy of this state is always higher comparing with the
minimum energy of $\pi$ and zero-flux SL states with the change of parameter $J_1/J_2$, which is shown in Fig.~\ref{ansatz and energy} (c).
However, the energy differences are small near the intermediate region of parameter $J_1/J_2$ with $\pi$-flux SL ground state.
Therefore, the PS state may emerge after considering the gauge fluctuations
and projecting the mean-field wave function by Gutzwiller projection,
which is beyond the scope of the current work.
Because of the absence of the PS state in the mean-field level,
the possible DQCP\cite{DQCPandEmergentO4inPSandNeel,PS&DQCPbyIDMRG2019} cannot be studied by the Schwinger boson mean field theory.

\section{Comparison to self-consistent spin wave theory and ED}\label{compare}
\begin{figure}[ht]
    \centering{\includegraphics[width=0.45\textwidth]{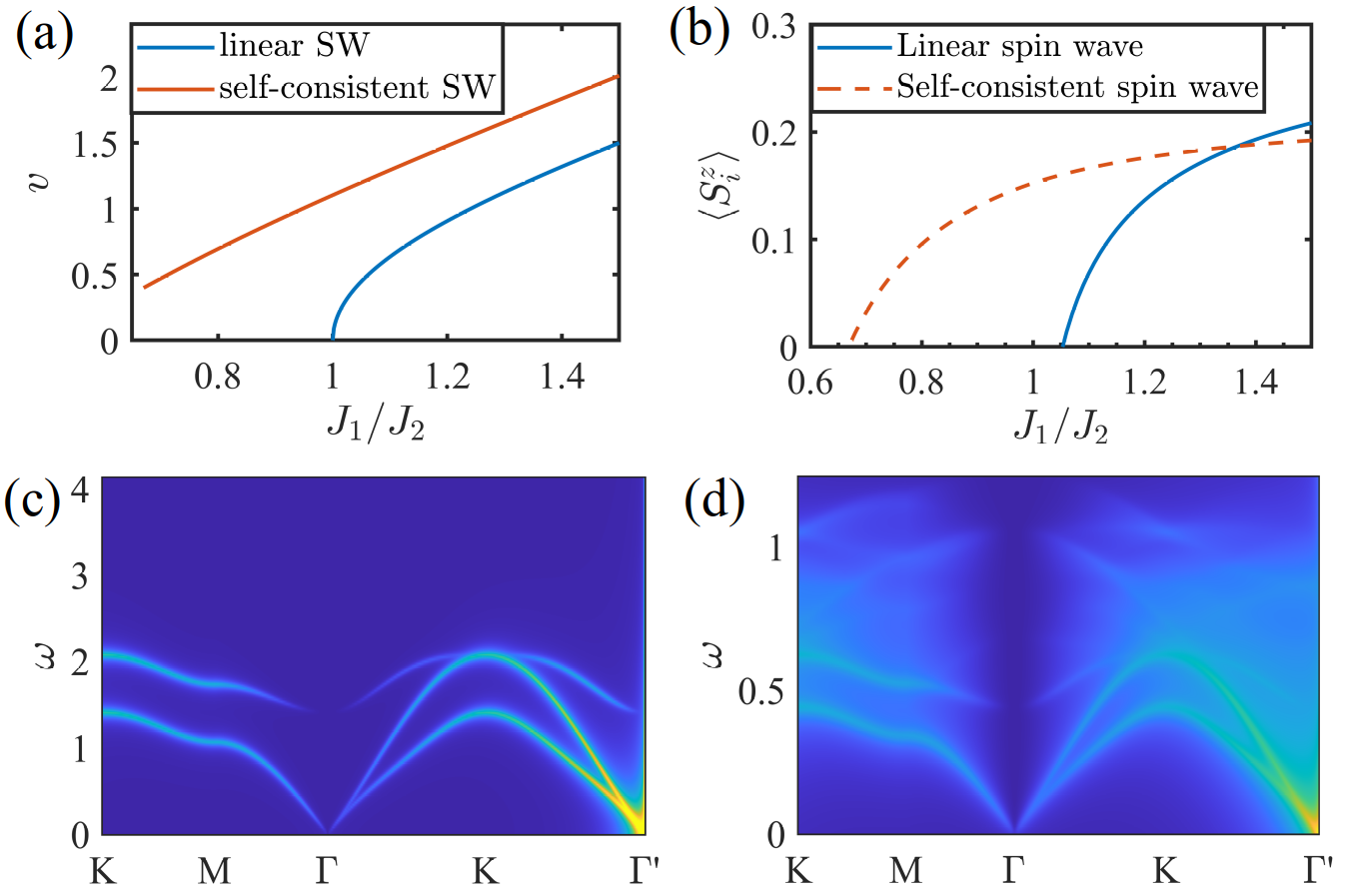}}
    %\label{fig:a}
    \caption{(a) is the magnon velocity near the Goldstone mode calculated by linear spin wave and self-consistent spin wave theory (see Appendix~\ref{App:5}).
        The magnon velocity of linear spin wave vanishes at $J_1/J_2=1$, where the magnon dispersion become $\propto k^2$ and linear spin wave theory breaks down.
        (b) is the N\'eel order parameter computed by self-consistent spin wave theory and linear spin wave theory,
        which vanishes at $J_1/J_2\approx 0.65$ and $J_1/J_2\approx 1.05$ respectively.
        (c) and (d) are the dynamic structure factor calculated by self-consistent spin wave theory and Schwinger boson mean-field theory in N\'eel phase.
        The parameters are $J_1=0.9$, $J_2=1$.}
    \label{Neel}
\end{figure}
The N\'eel phase can also be studied by the spin wave theory.
However, as shown in Fig.~\ref{Neel} (a) and (b), the linear spin wave theory breaks down at $J_1/J_2=1$
because the spin wave dispersion become $\epsilon_k\propto k^2$ under linear spin wave at $J_1/J_2=1$,
and the magnetic order parameter for spin-1/2 model vanishes at $J_1/J_2=1.05$,
which is much larger than the N\'eel phase boundary by other theories and numerics.
The details of the spin wave dispersion can be referred to Appendix~\ref{App:5}.
Near the N\'eel phase boundary,
the magnon interactions play important roles and need to be considered.
However, the conventional nonlinear spin wave theory by $1/S$ expansion also breaks down near $J_1/J_2=1$ because
the interaction correction depends on the linear spin wave Hamiltonian.
Therefore, we use the self-consistent spin wave theory\cite{SCSW1,SCSW2} to incorporate the effects of magnon interactions.
% at where the interaction of magnon plays important roles.
The idea of the self-consistent spin wave theory is to decouple the quartic terms of Holstein-Primakoff bosons
into all possible quadratic terms, and compute these corrections to linear spin wave Hamiltonian {self-consistently}.
% which has been used in Refs.~\cite{SCSW1,SCSW2}.
%The magnon dispersion and wave function are obtained by diagonalizing the total effective Hamiltonian.
%The mean-field pairs need to be solved self-consistently, 
% and therefore this method is named self-consistent spin wave theory.
The details of the self-consistent spin wave theory are given in the Appendix~\ref{App:5}.
The magnetic order parameter vanishes at $J_1/J_2\approx 0.65$ by self-consistent spin wave theory as shown in Fig.~\ref{Neel} (b),
which yields more accurate N\'eel phase boundary.

The dynamic structure factors in the N\'eel phase are calculated to compare the self-consistent spin wave theory and the Schwinger boson mean-field theory.
Fig.~\ref{Neel} (c) and (d)
are the dynamic structure factors at $J_1/J_2$=0.9
calculated by these two theories.
%by Schwinger boson and self-consistent spin wave theory respectively.
In the N\'eel phase,
the Schwinger boson condenses and yields sharp magnon peaks in the dynamic structure factor,
while the non-condensing spinons contribute to the continuum.
Although the magnon interaction is considered in the self-consistent spin wave theory,
the magnon damping channel is not included.
Therefore, the magnon peaks are always sharp and there is no continuum in the dynamic structure factor calculated by self-consistent spin wave theory.
The dynamic structure factor calculated by self-consistent spin wave theory is
more consistent with the schwinger boson mean field theory than the linear spin wave theory.
We note that the energy scales of dynamic structure factors are different in the Schwinger boson mean-field
theory and spin wave theory,
and they should become identical in the large $S$ limit.

\begin{figure}[ht]
    \includegraphics[width=0.47\textwidth]{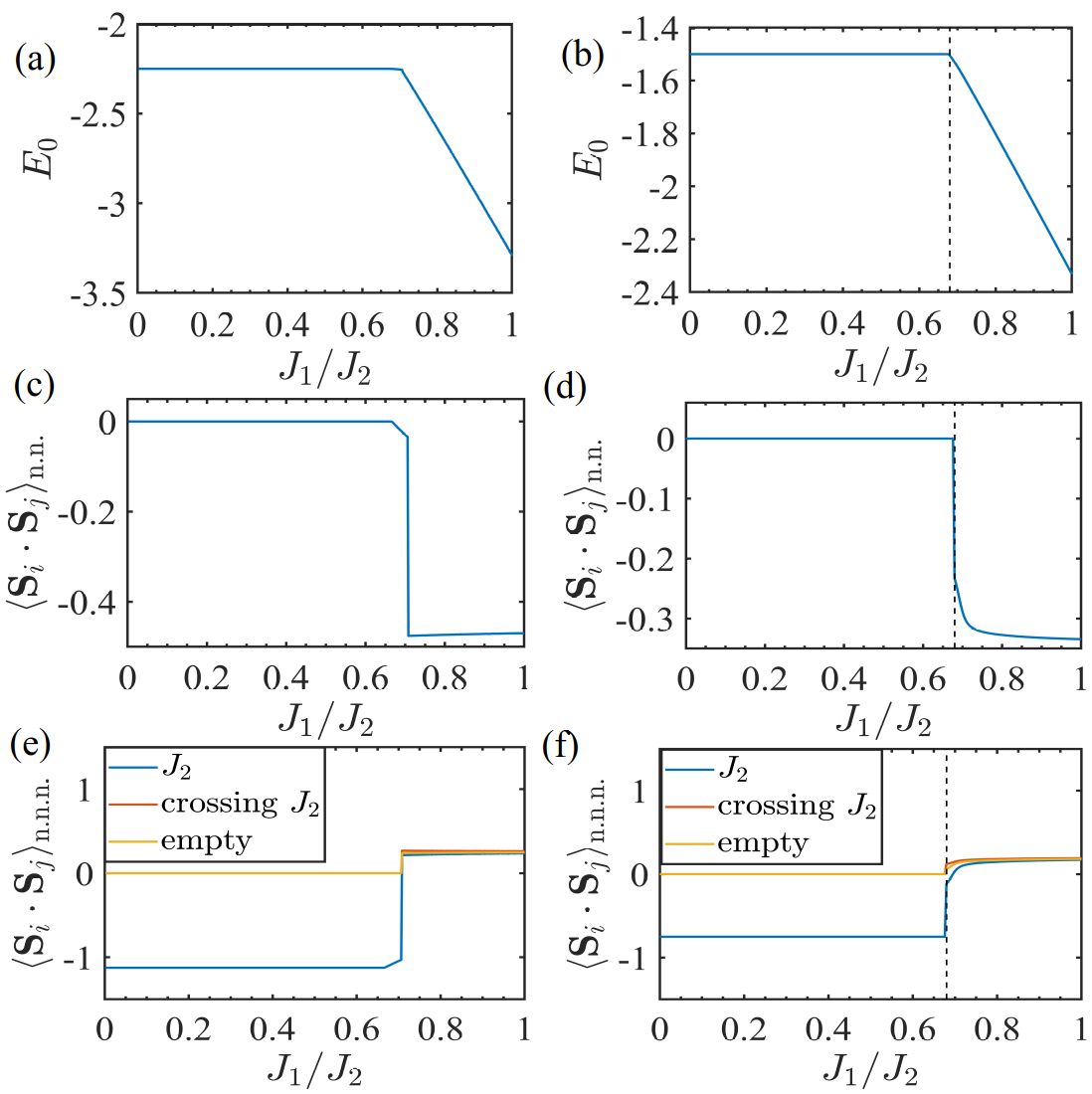}
    \caption{(a), (c), (e) and (b), (d), (f) are observables calculated by Schwinger boson mean-field theory and 32-site exact diagonalization respectively.
        (a) and (b) are the ground state energy per unit cell.
        (c) and (d) are the nearest-neighbor (n.n.) spin correlation.
        (e) and (f) are the next-nearest-neighbor (n.n.n.) spin correlation.
        The Shastry-Sutherland lattice has three inequivalent n.n.n. bonds,
        which are $J_2$ bond, bond crossing $J_2$ bond, and empty bond.
        The dashed lines in (b), (d) and (f) are $J_1/J_2=0.68$.
        The discontinuity of spin correlation functions and slope of ground state energies at $J_1/J_2=0.71$ in(a)(c)(e) indicates
        that the transition from $\pi$-flux spin liquid to N\'eel order is first order under mean-field approximation,
        while the transition from DS to $\pi$-flux spin liquid at $J_1/J_2=0.66$ seems continuous.
        Note that in (e) $\langle \boldsymbol{S}_i\cdot\boldsymbol{S}_j\rangle_{\text{n.n.n}}$ is less than $-3/4$ (the minimal possible value for spin-$1/2$) in some parameter range, this is due to the spin size(boson number) fluctuation in the mean-field approximation\cite{PSG-boson}.
    }
    \label{ground state properties}
\end{figure}

% The spin wave theory can only be used to find the boundary of magnetic ordered phase.
% The another side of spin liquid phase boundary should be verified by other methods.
% Exact diagonalization method is appropriate to check the boundary because
% the dimer-singlet phase has distinguishing ground state property.
We also compare the ground state properties computed by SBMFT and exact diagonalization [see Appendix~\ref{App:7}].
The results are shown in Fig.~\ref{ground state properties}.
% (a)-(c) and (d)-(e) are spin correlations calculated by Schwinger boson mean-field theory and 32 sites exact diagonalization respectively.
The ground state from 32-site exact diagonalization is
the exact dimer-single state for $J_1/J_2 <0.68$,
which is consistent with previous exact diagonalization studies\cite{EDforPhaseBoundaryOfDSandNeel,doi:10.7566/JPSCP.38.011166}
and very close to the SBMFT phase boundary $J_1/J_2\sim0.66$ for DS phase.
The near-neighbor and next-nearest-neighbor spin correlations
also show similar behavior in SBMFT and exact diagonalization.
Note that the $\langle \boldsymbol{S}_i\cdot\boldsymbol{S}_j\rangle_{\text{n.n.n}}$
in Fig.~\ref{ground state properties} (e) is \(-1.25\), which is less than \(-3/4\) (the minimal possible value for spin-1/2)
in the DS phase. This is due to the spin size(boson number) fluctuation in the mean-field approximation\cite{PSG-boson}.

% The measurements calculated by these two methods has similar behaviors,
% and these spin correlations have great jump at $J_1/J_2=0.68$,
%which indicates the boundary of the dimer-singlet phase,
% and is consistent with the result of Schwinger boson mean-field spin theory.

\section{Discussion and conclusion}\label{Conclusion}
We studied the Shastry-Sutherland model by the Schwinger boson mean field theory.
Using the projective symmetry group method, we find two kinds of possible symmetric ansatz.
Comparing the energy of the two ansatz with the change of $J_1/J_2$ and $\kappa$,
we get the Schwinger boson mean-field phase diagram of Shastry-Sutherland model.
We find a $\pi$-flux gapped $Z_2$ spin liquid state for
the intermediate parameter $0.66<J_1/J_2<0.71$ between the dimer-singlet phase and the N\'eel AFM phase.
This $\pi$-flux spin liquid state is continuously connected to the DS phase, but
has a first-order transition to the N\'eel AFM phase upon increasing $J_1/J_2$,
which can be seen from the discontinuity of spin correlation functions and slope of ground state energies
in the Schwinger boson mean-field results in Fig.~\ref{ground state properties}.
The continuous transition between the spin liquid and DS phases is an example of the confinement transition of Ising gauge field\cite{PhysRevB.65.024504},
which can be described by the condensation of gauge flux excitations (``visons'') and should be dual to an 3D Ising transition.
The short-range spin correlations in the $\pi$-flux state is closely related to a 4-sublattice AFM order instead of N\'eel AFM order.
We expect that ring-exchange coupling with opposite sign to that derived from the Hubbard model
would further stabilize this $\pi$-flux spin liquid state\cite{PSG-boson}.

To investigate the possibility of plaquette-singlet state\cite{PhaseTransitionDS-PS-Neel1,PhaseTransitionDS-PS-Neel2,SCBOTheoryHelixAndPS,PS&DQCPbyIDMRG2019,PSbyTensorNetwork,PSIndeucedByDifferentPlaqutte,PSIndeucedByDifferentPlaqutte2},
we studied PS ordered ansatz which break the glide symmetry.
% and do not belong to the PSG solutions.
We found that the ground state energy of these PS ordered ansatz are higher compared with
the symmetric spin liquid ansatz under mean-field approximation.
Therefore, the PS phase does not exist in our mean-field phase diagram.
However the energy difference between the PS phase and spin liquids are small in the spin liquid phase.
So it is possible that the PS phase may emerge after considering gauge fluctuations and
%wave-function projection.
Gutzwiller projection of mean-field wave functions,
which we leave for future studies.

To further investigate the N\'eel AFM phase,
we used a self-consistent spin wave theory because the linear spin wave theory breaks down for $J_1/J_2<1$.
% while the self-consistent spin wave theory breaks down for $J_1/J_2\lesssim 0.65$.
The self-consistent spin wave theory renormalizes the magnon dispersion by the magnon interactions,
and further stabilizes the magnetic order down to $J_1/J_2 \sim 0.65$.
The dynamic structure factor calculated by this theory is more consistent with
the results of Schwinger boson mean-field theory except for an overall energy scale.
% Therefore, the self-consistent spin wave theory yields more accurate
% than the linear spin wave theory and can be applied to a wider range of parameters.
% We also calculated the order parameter by self-consistent spin wave theory,
% and it vanishes at $J_1/J_2\approx 0.65$,
% The N\'eel phase boundary produced by the self-consistent spin wave theory, $J_1/J_2 \sim 0.65$, 
% is roughly consistent with the Schwinger boson mean-field result.

We have also performed exact diagonalization of the Shastry-Sutherland model with $32$ sites.
The results indicate that the phase boundary of dimer-singlet phase is $J_1/J_2 \sim 0.68$,
which is roughly consistent with the Schwinger boson mean-field result and
previous exact diagonalization studies of larger system sizes\cite{doi:10.7566/JPSCP.38.011166}.
The behavior of spin correlation functions from the exact diagonalization results is
similar to those of the Schwinger boson mean-field theory,
which provides partial support of our mean-field picture of the Shastry-Sutherland model.

Some previous numerical studies\cite{SLbyDMRG, SLbyED, PhysRevB.105.L041115} suggest that
the spin liquid phase in Shastry-Sutherland model is gapless
and possibly described by fermionic spinons with Dirac-cone dispersions similar to the DQCP\cite{PS&DQCPbyIDMRG2019}.
This gapless spin liquid phase cannot be captured by our Schwinger boson mean-field theory.
% However strong interactions between fermionic spinons induced by gapless $U(1)$ gauge field might produce Cooper pairing of spinons,
% which may open spin gap and reduce the $U(1)$ gauge field to $Z_2$.
The possible transition from gapless $U(1)$ spin liquids to gapped $Z_2$ spin liquids may be studied in controlled large-$N$ approximation\cite{PhysRevB.98.035137}.
The resulting gapped $Z_2$ fermionic spinon spin liquid is likely a dual description of the Schwinger boson symmetric $Z_2$ spin liquids considered here\cite{Z2}, and may be an interesting direction for future theoretical and numerical studies.
However it should be noted that the ``$\pi$-flux'' for Abrikosov fermion hoppings in this $U(1)$ spin liquid\cite{PS&DQCPbyIDMRG2019} is not directly related to the ``$\pi$-flux'' of boson pairing terms in our Schwinger boson formalism.
In our honest opinion, our Schwinger boson formalism as well as other slave particle formalism are just ``phenomenological'' low energy effective theories for quantum spin liquids,
that may or may not be realized in a particular model, and should be justified by further numerical and experimental studies.

\section{ACKNOWLEDGEMENTS}
KL thanks Fang-Yu Xiong, Xue-Mei Wang and Jie-Ran Xue for helpful discussions.
FW acknowledges support from National Natural Science Foundation of China (No. 12274004), and National Natural Science Foundation of China (No. 11888101).
\appendix
\section{Solutions of the algebraic PSG}\label{App:1}
In the following we will solve the algebraic PSGs by using these algebraic constraints on the generators of the Shastry-Sutherland lattice space group.
\begin{widetext}
    We only consider the condition where the invariant gauge group (IGG) is $Z_2$.
    For a space ground element $g$,
    the Schwinger boson operators are transformed as
    \begin{eqnarray}
        \hat{b}_{\boldsymbol{r},s}\to
        \exp[\mathbbm{i} \phi_g(g\boldsymbol{r})] \hat{b}_{g\boldsymbol{r},s}.
    \end{eqnarray}
    For a defining relation $g_1 g_2\cdots g_k=\mathbbm{1}$,
    we have
    \begin{eqnarray}
        \phi_{g_1g_2\cdot g_k}(g_1g_2\cdot g_k\boldsymbol{r})
        +\phi_{g_2\cdot g_k}(g_2\cdot g_k\boldsymbol{r})
        +\dots
        +\phi_{g_k}(g_k\boldsymbol{r})
        =(\text{an IGG element})=p\pi, p\in\mathbbm{Z}_2.
    \end{eqnarray}
    Note that all these equations about $\phi$s are implicitly modulo $2\pi$.

    We consider a gauge transformation $\hat{b}_{\boldsymbol{r},s}=\exp[\mathbbm{i}\phi(\boldsymbol{r})]\hat{b}'_{\boldsymbol{r},s}$,
    then the generators of PSG transformed as
    \begin{eqnarray}
        \phi'_{g}(g\boldsymbol{r}) = \phi_g(g\boldsymbol{r})+\phi(g\boldsymbol{r})-\phi(\boldsymbol{r}).
    \end{eqnarray}
    With this relation,
    by choosing $\phi(X,Y,s)$ for all $X\neq 0$, we can make $\phi'_{\TX}(X,Y,s)=0$ for all $X,Y,s$;\\
    then by choosing $\phi(X=0,Y,s)$, we can make $\phi'_{\TY}(X=0,Y,s)=0$ for all $Y,s$.
    we are still left three gauge freedoms.
    The first one is the global constant phase,
    \begin{eqnarray}
        \phi(X,Y,s)_1=\phi_s,
    \end{eqnarray}
    this does not change $\phi_{\TX}$ and $\phi_{\TY}$,
    but will change $\phi_{\Cf}$ and $\phi_{\Mm}$ as
    \begin{eqnarray}
        \phi'_{\Cf}(X,Y,s)&=&\phi_{\Cf}(X,Y,s)+\phi_s-\phi_{s-1},\\
        \phi'_{\Mm}(X,Y,s)&=&\phi_{\Mm}(X,Y,s)+\phi_s-\phi_{-s}.
    \end{eqnarray}
    The second gauge freedom is
    \begin{eqnarray}
        \phi(X,Y,s)_2=\pi\cdot X,
    \end{eqnarray}
    which also does not change $\phi_{\TX}$ and $\phi_{\TY}$ modulo IGG,
    but will change $\phi_{\Cf}$ and $\phi_{\Mm}$ as
    \begin{eqnarray}
        \phi'_{\Cf}(X,Y,s)&=&\phi_{\Cf}(X,Y,s)+\pi\cdot (X-Y),\\
        \phi'_{\Mm}(X,Y,s)&=&\phi_{\Mm}(X,Y,s)+\pi\cdot (X-Y-x_{-s}).
    \end{eqnarray}
    The third gauge freedom is
    \begin{eqnarray}
        \phi(X,Y,s)_3=\pi\cdot (X+Y),
    \end{eqnarray}
    which does not change $\phi_{\TX}$ and $\phi_{\TY}$ and $\phi_{\Cf}$ modulo IGG,
    but will change $\phi_{\Mm}$ as
    \begin{eqnarray}
        \phi'_{\Mm}(X,Y,s)&=&\phi_{\Mm}(X,Y,s)-\pi\cdot (x_{-s}+y_{-s})\nonumber\\
        &=&\phi_{\Mm}(X,Y,s)+\pi\cdot s.
    \end{eqnarray}

    We then consider the relation of the generators of the space group.
    From the relation of Eq.~(\ref{re1}),
    the algebraic constraint is ($\phi_{\TX}$ omitted hereafter),
    \begin{eqnarray}
        \phi_{\TY}(X+1,Y,s)-\phi_{\TY}(X,Y,s)=p_1\pi,
    \end{eqnarray}
    then the solution is
    \begin{equation}
        \phi_{\TY}(X,Y,s)=p_1\pi\cdot X.
        \label{equ:PSG1}
    \end{equation}
    The algebraic constraint from relation of Eq.~(\ref{re2}) is
    \begin{eqnarray}
        -\phi_{\TY}(X,Y+1,s)+\phi_{\Cf}(X+1,Y+1,s)-\phi_{\Cf}(X,Y,s)=p_2\pi,
    \end{eqnarray}
    then we have
    \begin{equation}
        \phi_{\Cf}(X+1,Y+1,s)-\phi_{\Cf}(X,Y,s)=p_2\pi+p_1\pi\cdot X
        \label{equ:PSG2}
    \end{equation}
    From the relation of Eq.~(\ref{re3}),
    we have
    \begin{eqnarray}
        \phi_{\Cf}(X+1,Y,s)-\phi_{\TY}(Y,-X,s-1)-\phi_{\Cf}(X,Y,s)=p_3\pi,
    \end{eqnarray}
    which yields
    \begin{equation}
        \phi_{\Cf}(X+1,Y,s)-\phi_{\Cf}(X,Y,s)=p_3\pi+p_1\pi\cdot Y.
        \label{equ:PSG3}
    \end{equation}
    Combining with Eq.~(\ref{equ:PSG2}), the solution is
    \begin{eqnarray}
        \phi_{\Cf}(X,Y,s)=\phi_{\Cf}(X,Y,s)+p_2\pi\cdot Y+p_3\pi\cdot (X-Y)+p_1\pi\cdot [XY-\frac{1}{2}Y(Y+1)]
    \end{eqnarray}
    Note that we can use the gauge freedom $\phi_2$ to set $p_3=0$.

    From the relation of Eq.~(\ref{re4}),
    the algebraic constraint is
    \begin{eqnarray}
        \phi_{\Cf}(X,Y,s)+\phi_{\Cf}(Y,-X,s+3)+\phi_{\Cf}(-X,-Y,s+2)+\phi_{\Cf}(-Y,X,s+1)=p_4\pi,
    \end{eqnarray}
    from which we have
    \begin{equation}
        \sum_{s} \phi_{\Cf}(0,0,s)+p_1\pi\cdot (X^2+Y^2)=p_4\pi.
        \label{equ:PSG4}
    \end{equation}
    Therefore we must have $p_1=0$, then $\phi_{\TY}(X,Y,s)=0$.
    For simplicity, we define $\phi_{\Cf,s}=\phi_{\Cf}(0,0,s)$,
    then we have the solution
    \begin{eqnarray}
        \phi_{\Cf}(X,Y,s)=p_2\pi\cdot Y+\phi_{\Cf,s},\label{solutionC4}
    \end{eqnarray}
    with $\sum_{s}\phi_{\Cf,s}=p_4\pi$.

    The relation of Eq.~(\ref{re5}) yields
    \begin{eqnarray}
        \phi_{\Mm}(X+1,Y,s)-\phi_{\TY}(-Y+x_{-s},-X+y_{-s},-s)-\phi_{\Mm}(X,Y,s)=p_5\pi,
    \end{eqnarray}
    then we have
    \begin{equation}
        \phi_{\Mm}(X+1,Y,s)-\phi_{\Mm}(X,Y,s)=p_5\pi.
        \label{equ:PSG5}
    \end{equation}
    Then we consider the relation of Eq.~(\ref{re6}),
    which yields
    \begin{eqnarray}
        -\phi_{\TY}(X,Y+1,s)+\phi_{\Mm}(X,Y+1,s)-\phi_{\Mm}(X,Y,s)=p_6\pi,
    \end{eqnarray}
    then we have
    \begin{equation}
        \phi_{\Mm}(X,Y+1,s)-\phi_{\Mm}(X,Y,s)=p_6\pi.
        \label{equ:PSG6}
    \end{equation}
    Combined with Eq.~(\ref{equ:PSG5}),
    we get the solution
    \begin{eqnarray}
        \phi_{\Mm}(X,Y,s)=\phi_{\Mm}(0,0,s)+p_5\pi\cdot X+p_6\pi \cdot Y.\label{solutionSigma}
    \end{eqnarray}

    Finally, we consider the relations of Eq.~(\ref{re7}) and Eq.~(\ref{re8}).
    The algebraic constraint of Eq.~(\ref{re7}) is
    \begin{equation}
        \phi_{\Mm}(X,Y,s)-\phi_{\Mm}(-Y-y_s,-X-x_s,-s)=p_7\pi.
        \label{equ:PSG7}
    \end{equation}
    Substitute Eq.~(\ref{solutionSigma}) to this equation,
    we get
    \begin{eqnarray}
        \phi_{\Mm}(0,0,s)+\phi_{\Mm}(0,0,-s)+(p_5-p_6)\pi\cdot (X-Y)
        -p_5\pi y_s-p_y\pi x_s =p_7\pi
    \end{eqnarray}
    then
    we must have $p_5=p_6$.
    For simplicity, we define $\phi_{\Mm,s}=\phi_{\Mm}(0,0,s)$,
    then $\phi_{\Mm}(X,Y,s)=p_5\pi\cdot (X+Y)+\phi_{\Mm,s}$,
    and
    $\phi_{\Mm,s}+\phi_{\Mm,-s}=p_7\pi+p_5\pi\cdot s$.
    The relation Eq.~(\ref{re8}) yields
    \begin{equation}
        \phi_{\Cf}(X+1,Y,s)+\phi_{\Mm}(Y,-X-1,s-1)+\phi_{\Cf}(X+x_s,-Y-y_s,1-s)-\phi_{\Mm}(X,Y,s)=p_8\pi.
        \label{equ:PSG8}
    \end{equation}
    Substitute Eq.~(\ref{solutionC4}) and (\ref{solutionSigma}),
    we have
    \begin{eqnarray}
        \phi_{\Cf,s}+\phi_{\Cf,1-s}
        +p_2\pi \cdot (-y_s)+
        \phi_{\Mm,s-1}-\phi_{\Mm,s}-p_5\pi=p_8\pi,
    \end{eqnarray}
    which yields the following 4 equations by setting the value of $s$,
    \begin{eqnarray}
        \phi_{\Cf,0}+\phi_{\Cf,1}+\phi_{\Mm,3}-\phi_{\Mm,0} & = & p_8\pi +p_5\pi \label{equ:PSG8-1}\\
        \phi_{\Cf,1}+\phi_{\Cf,0}+\phi_{\Mm,0}-\phi_{\Mm,1} & = & p_8\pi +p_5\pi\label{equ:PSG8-2}\\
        \phi_{\Cf,2}+\phi_{\Cf,3}+\phi_{\Mm,1}-\phi_{\Mm,2} & = & p_8\pi +p_5\pi +p_2\pi\label{equ:PSG8-3}\\
        \phi_{\Cf,3}+\phi_{\Cf,2}+\phi_{\Mm,2}-\phi_{\Mm,3} & = & p_8\pi +p_5\pi +p_2\pi\label{equ:PSG8-4}
    \end{eqnarray}
    With Eq.~$(\ref{equ:PSG8-1})$-Eq.~$(\ref{equ:PSG8-2})$,
    we get
    $(\phi_{\Mm,3}+\phi_{\Mm,1})-2\phi_{\Mm,0}=p_5\pi=0$,
    then we have $p_5=0$.
    Then we can set $p_8=0$ by using gauge freedom $\phi_3$.
    With the gauge freedom $\phi_1$,
    we can also set $\phi_{\Cf,1}=\phi_{\Cf,2}=\phi_{\Cf,3}=0$,
    then we have $\phi_{\Cf,0}=p_4\pi$.
    To conclude,
    the solutions to  $\phi_{\Mm,s}$ are (modulo IGG),
    \begin{eqnarray}
        \phi_{\Mm,0}&=&\frac{p_7\pi}{2}+p_4\pi, \\
        \phi_{\Mm,3}&=&\frac{p_7\pi}{2}, \\
        \phi_{\Mm,1}&=&\frac{p_7\pi}{2}, \\
        \phi_{\Mm,2}&=&\frac{p_7\pi}{2}+p_2\pi.
    \end{eqnarray}

    Finally,
    we get the final solution to algebraic PSG:
    \begin{eqnarray}
        \phi_{\TX}(X,Y,s) & = & 0 \label{equ:PSG-sol-1},\\
        \phi_{\TY}(X,Y,s) & = & 0 \label{equ:PSG-sol-2},\\
        \phi_{\Cf}(X,Y,s) & = & p_2\pi\cdot Y+p_4\pi\cdot \delta_{s,0} \label{equ:PSG-sol-3},\\
        \phi_{\Mm}(X,Y,s) & = & \frac{p_7\pi}{2}+p_2\pi\cdot x_s y_s +p_4\pi \cdot \delta_{s,0}, \label{equ:PSG-sol-4}
    \end{eqnarray}
    with three remaining free $Z_2$ integer parameters $p_2,p_4,p_7$.
    Therefore, there are at most 8 kinds of PSGs.

    Then we need to consider the constraints on PSG by ansatz.
    The nn bond poses no constraint, because there is no nontrivial space group element that maps one nn bond to itself or its reverse.
    For the nnn bond, if $A_{\mathrm{n.n.n.}}\neq 0$,
    consider $(0,0,0)-(0,-1,2)$, which is invariant under $\Mm$,
    then $\phi_{\Mm}(0,0,0)+\phi_{\Mm}(0,-1,2)=p_7\pi+p_2\pi+p_4\pi=0$,
    namely $p_2+p_4+p_7=0$;
    if $B_{\mathrm{n.n.n.}}\neq 0$,
    consider $(0,0,0)-(0,-1,2)$, which is invariant under $\Mm$,
    then $\phi_{\Mm}(0,0,0)-\phi_{\Mm}(0,-1,2)=-p_2\pi-p_4\pi=0$,
    namely $p_2+p_4=0$, this is incompatible with $A_{\mathrm{n.n.n.}}\neq 0$;
    consider $(-1,0,1)-(0,0,3)$, which is reverted by $\Mm$,
    then $\phi_{\Mm}(-1,0,1)-\phi_{\Mm}(0,0,3)=0$,
    then if $B_{\mathrm{n.n.n.}}\neq 0$, it must be real.
    If we only consider the condition where at least one of $A_{\mathrm{n.n.n.}}$ and $B_{\mathrm{n.n.n.}}$ is not zero,
    there are at most 6 kinds of PSGs with these constraints.
    If we assume the nearest-neighbor ansatz $A_1$ is nonzero,
    these 6 states can be classified by two gauge invariant phase $\Phi_1$ and $\Phi_2$,
    which are defined on empty square plaquettes and $J_2$ square plaquettes respectively,
    \begin{eqnarray}
        A_{i j}\left(-A_{j k}^*\right) A_{k l}\left(-A_{l i}^*\right)=\left|A_1\right|^4 e^{i \Phi}.
    \end{eqnarray}
    We find these ansatz with same gauge flux $(\Phi_1,\Phi_2)$ are gauge equivalent,
    therefore, we have 4 gauge inequivalent ansatz.
    We will show the ansatz and the properties of these 6 kinds of PSGs in the following.
\end{widetext}

\subsection{$A_{\rm n.n.n.}\neq 0$ and $B_{\rm n.n.n.}= 0$}
\label{App:1A}

In this condition, $p_2+p_4=1$ and $p_7=1$, and there are two possible ansatz.

\begin{figure}[H]
    \centering
    \begin{subfigure}[h]{0.23\textwidth}
        \includegraphics[width=\textwidth]{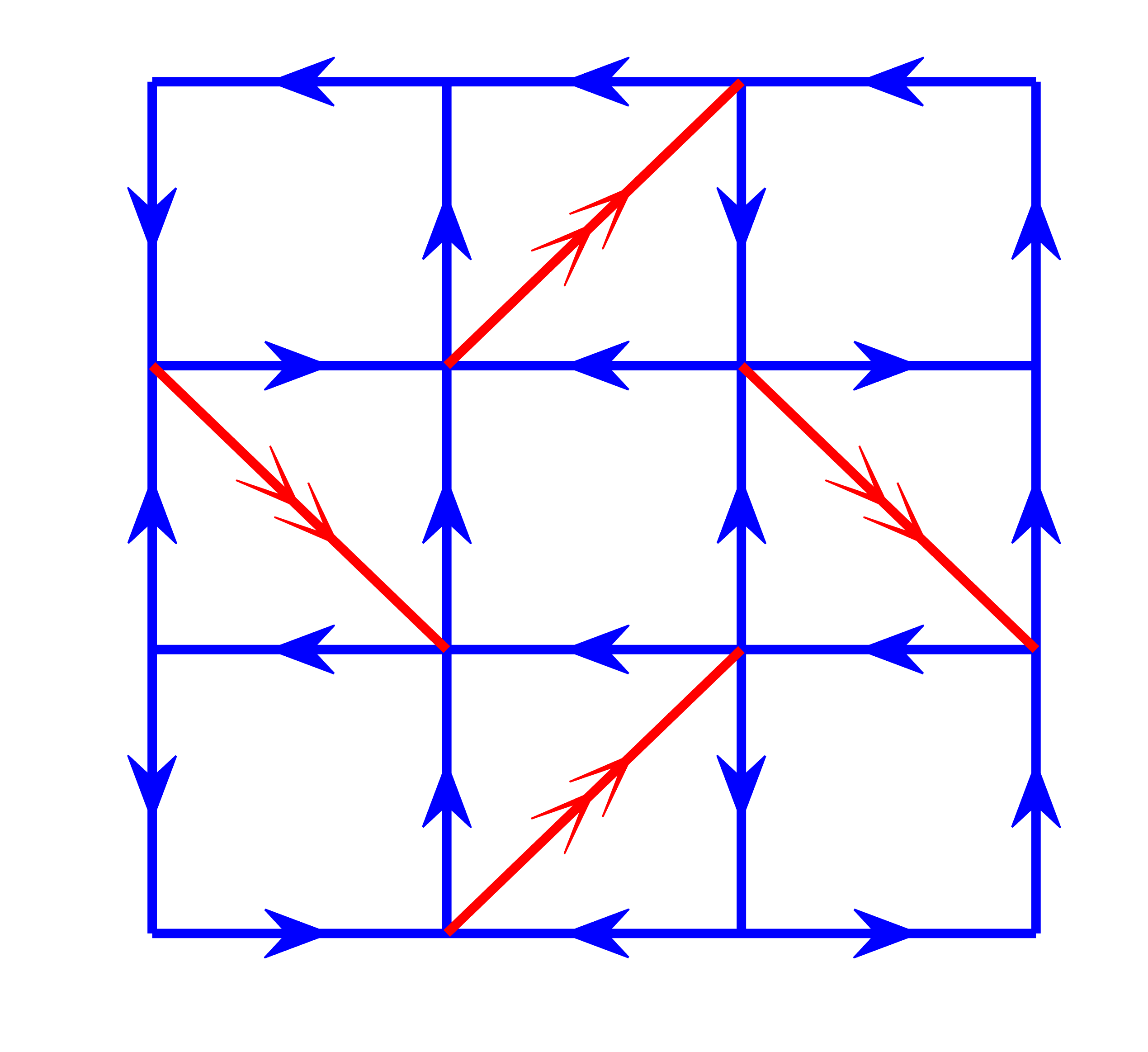}
        %\label{fig:a}
        \caption{$A$}
    \end{subfigure}
    \begin{subfigure}[h]{0.23\textwidth}
        \includegraphics[width=\textwidth]{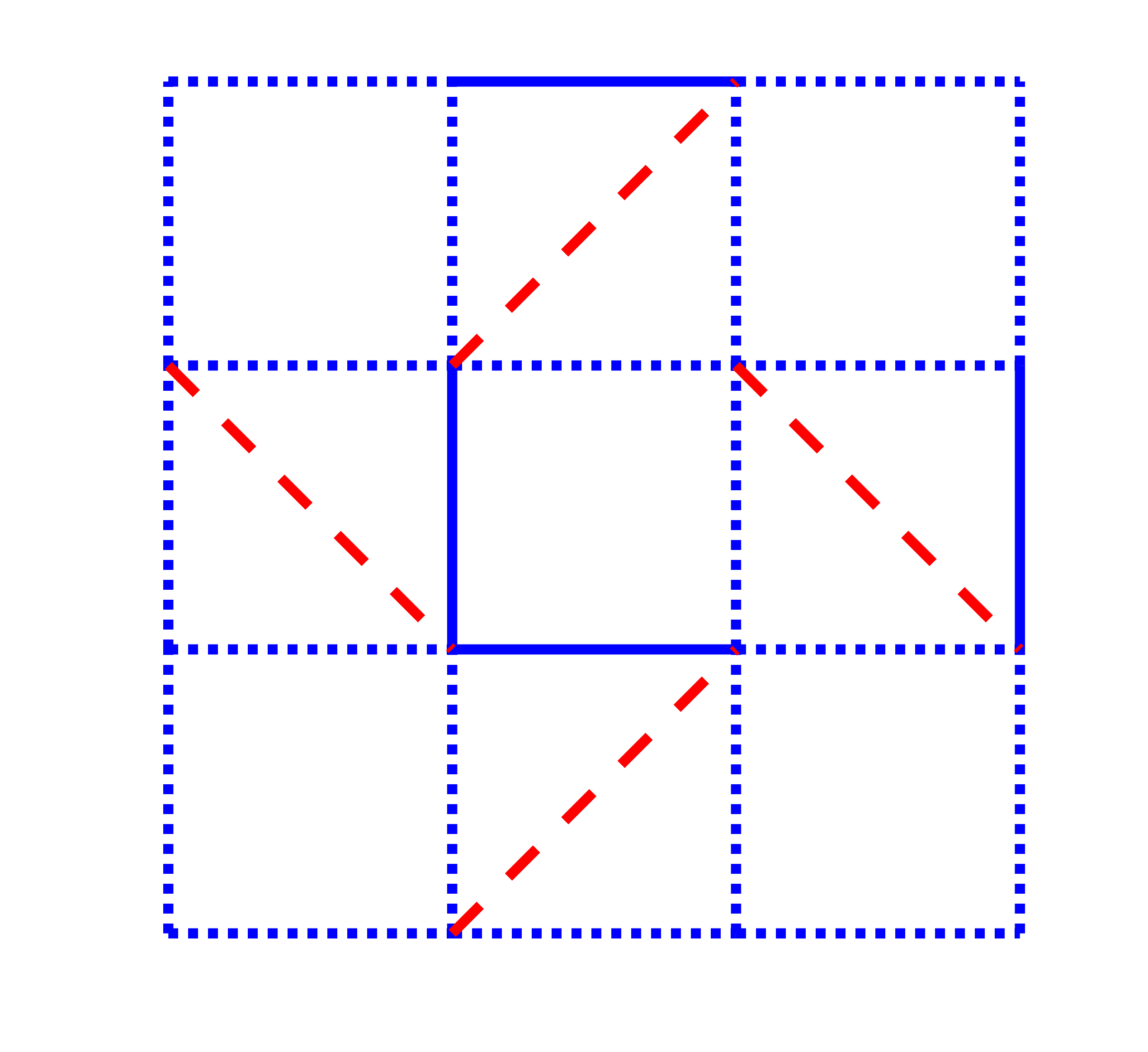}
        \caption{$B$}
    \end{subfigure}
    \caption{The ansatz in dashed lines are 0. The ansatz $B$ are all real and in solid and dotted lines have opposite sign.
        It is 0-flux in empty square and $\pi$-flux in $J_2$ square.
    }
    \label{0-Pi}
\end{figure}

For the $(p_2,p_4,p_7)=(1,0,1)$ condition, we have PSG solution
\begin{eqnarray}
    \phi_{T_x}(X,Y,s)&=&0,\\
    \phi_{T_y}(X,Y,s)&=&0,\\
    \phi_{C_4}(X,Y,s)&=&\pi\cdot Y,\\
    \phi_{\sigma}(X,Y,s)&=&\frac{\pi}{2}+x_sy_s\pi,
\end{eqnarray}
from which we get the $(0,\pi)$-flux ansatz in this condition,
which is shown in Fig.~\ref{0-Pi}.
Therefore, after considering the fluctuation of gauge field and Gutzwiller projection,
the $(0,\pi)$-flux may also exist in the phase diagram.

\begin{figure}[H]
    \centering
    \begin{subfigure}[h]{0.23\textwidth}
        \includegraphics[width=\textwidth]{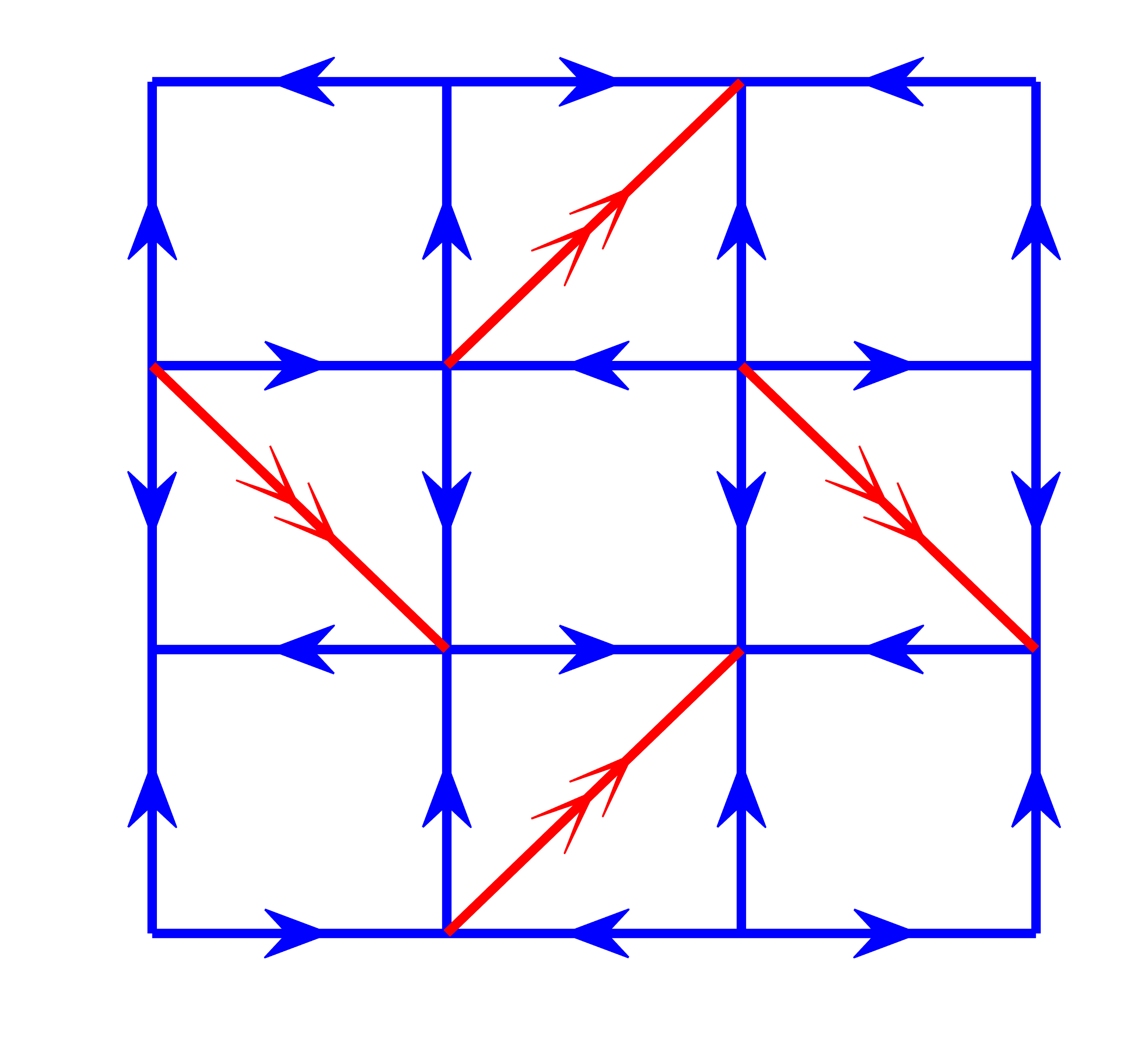}
        %\label{fig:a}
        \caption{$A$}
    \end{subfigure}
    \begin{subfigure}[h]{0.23\textwidth}
        \includegraphics[width=\textwidth]{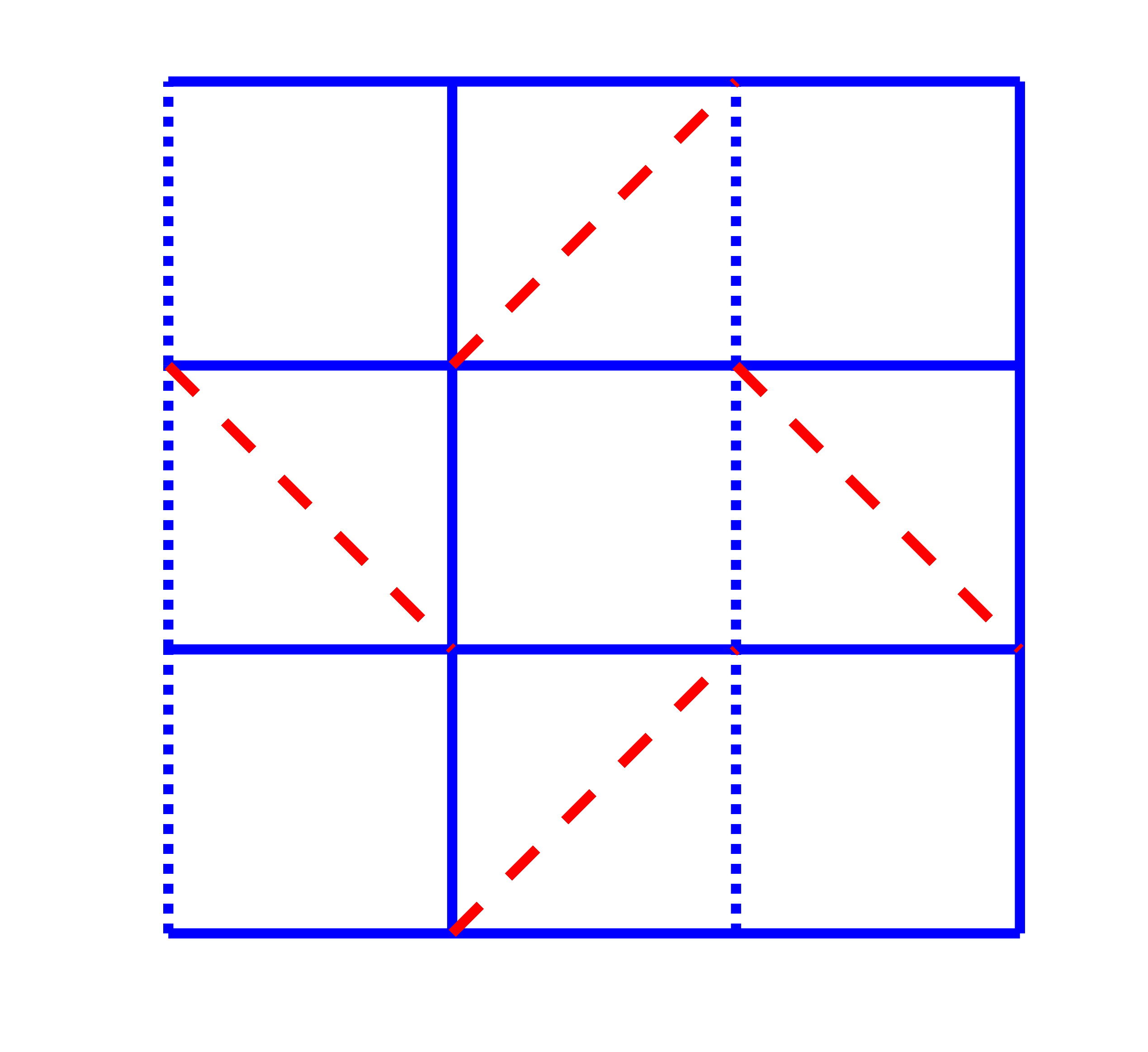}
        \caption{$B$}
    \end{subfigure}
    \caption{The ansatz on dashed lines are 0. The ansatz $B$ are all real and on solid and dotted lines have opposite sign.
        It is $\pi$-flux in empty square and $\pi$-flux in $J_2$ square.
    }
    \label{pi-pi}
\end{figure}

For the $(p_2,p_4,p_7)=(0,1,1)$ condition, we have
\begin{eqnarray}
    \phi_{T_x}(X,Y,s)&=&0,\\
    \phi_{T_y}(X,Y,s)&=&0,\\
    \phi_{C_4}(X,Y,s)&=&\delta_{s,0}\pi,\\
    \phi_{\sigma}(X,Y,s)&=&\frac{\pi}{2}+\delta_{s,0}\pi.
\end{eqnarray}
The ansatz are $(\pi,\pi)$-flux, which is shown in Fig.~\ref{pi-pi}

\subsection{$A_{\rm n.n.n.}= 0$ and $B_{\rm n.n.n.}\neq 0$}
\label{App:1B}

In this condition, $p_2+p_4=0$ and there are no constraint on $p_7$.
Therefore, there are four types of possible ansatz.

\begin{figure}[h]
    \centering
    \begin{subfigure}[h]{0.215\textwidth}
        \includegraphics[width=\textwidth]{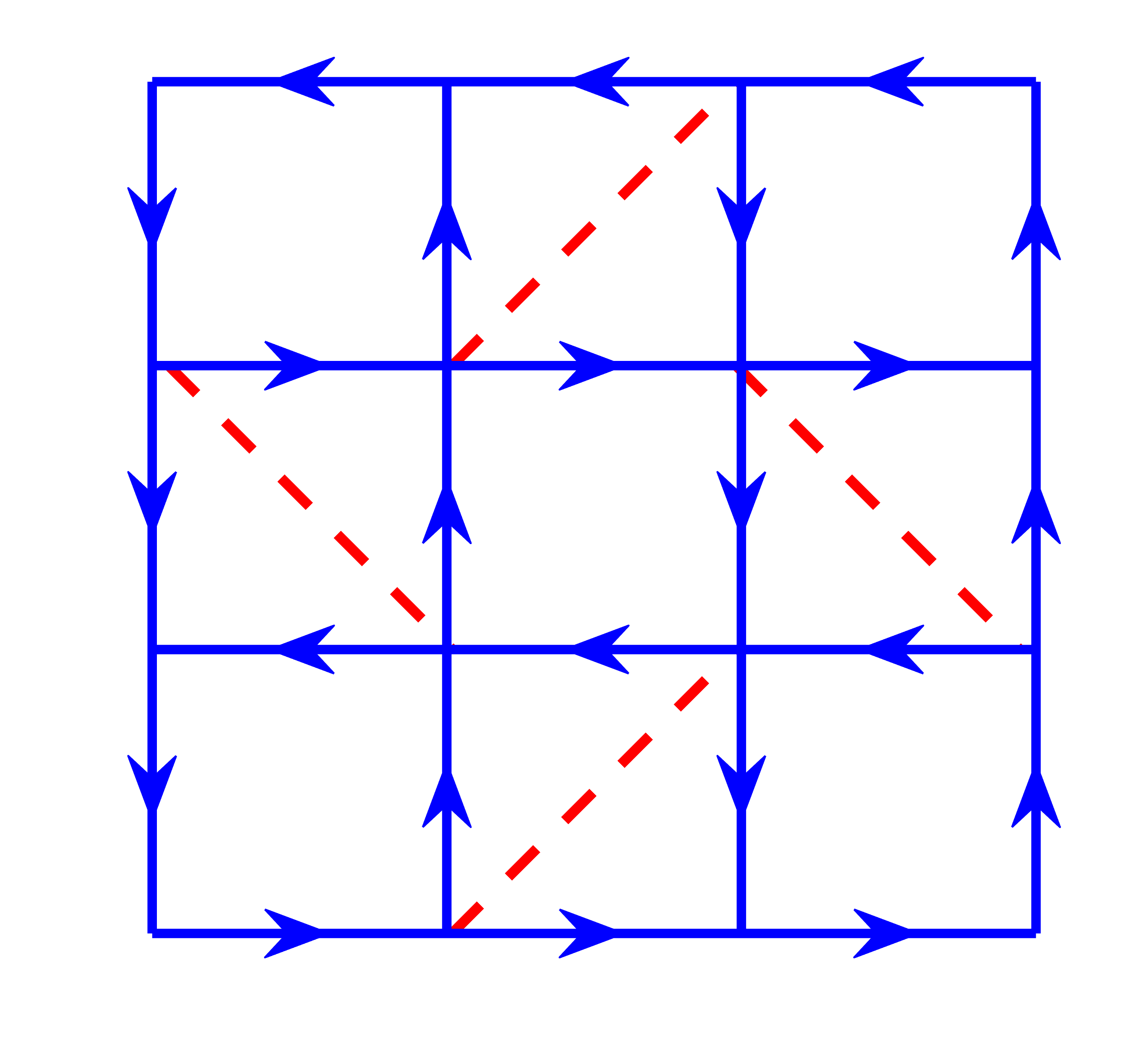}
        %\label{fig:a}
        \caption{$A$ of $(p_2,p_4,p_7)=(0,0,0)$}
    \end{subfigure}
    \begin{subfigure}[h]{0.25\textwidth}
        \includegraphics[width=\textwidth]{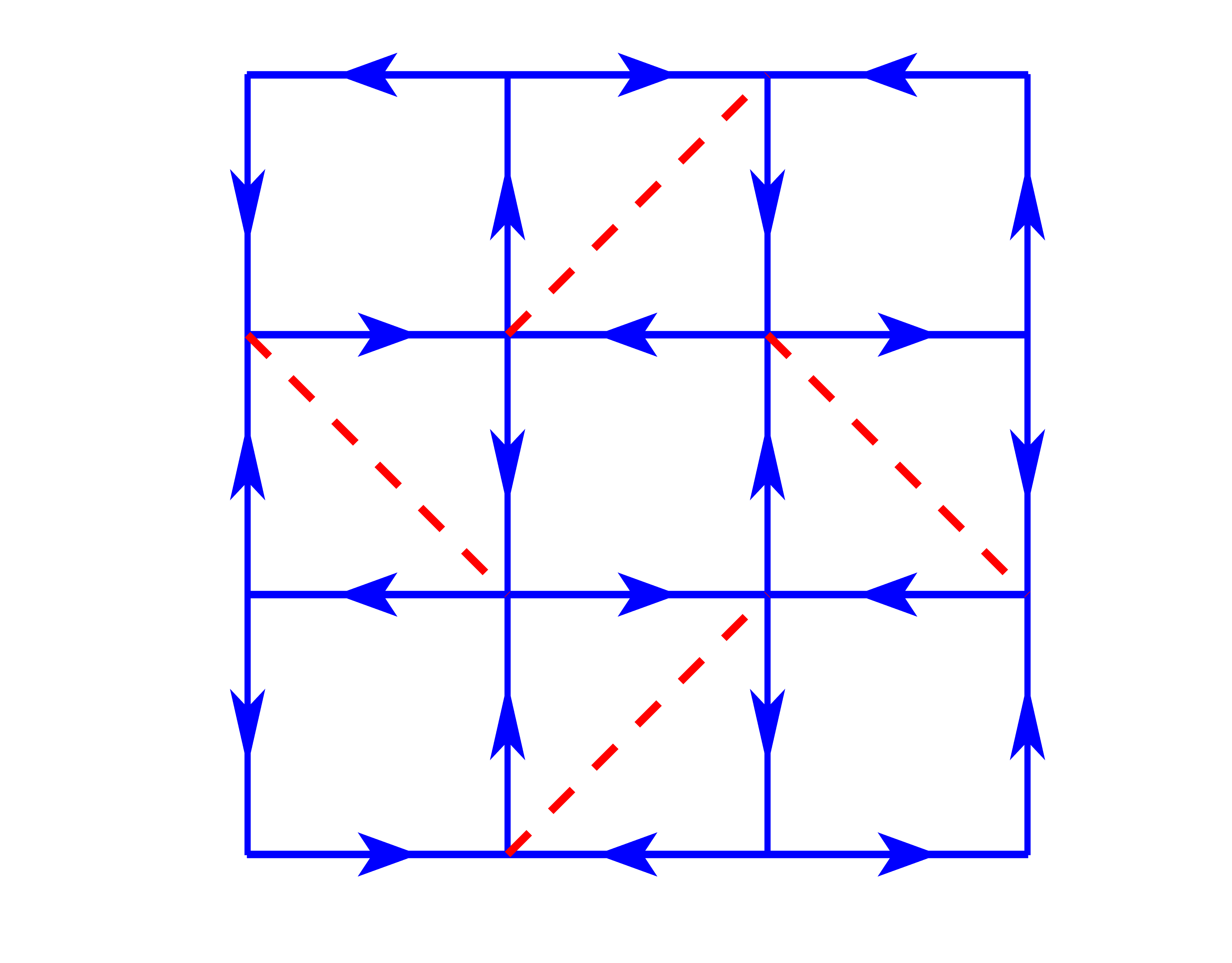}
        \caption{$A$ of $(p_2,p_4,p_7)=(0,0,1)$}
    \end{subfigure}
    \caption{The ansatz on dashed lines are 0.
        The two ansatz are gauge equivalent.
        It is $0$-flux in empty square and $0$-flux in $J_2$ square.
        $B_{\rm n.n}=0$ and $B_{\rm n.n.n}$ are real and uniform.
    }
    \label{00}
\end{figure}

First we consider $p_2=p_4=0$ condition.
For $(p_2,p_4,p_7)=(0,0,0)$ condition, the PSG solution is
\begin{eqnarray}
    \phi_{C_4}(X,Y,s)&=&0,\\
    \phi_{\sigma}(X,Y,s)&=&0.
\end{eqnarray}
while $(p_2,p_4,p_7)=(0,0,1)$ is
\begin{eqnarray}
    \phi_{C_4}(X,Y,s)&=&0,\\
    \phi_{\sigma}(X,Y,s)&=&\frac{\pi}{2}.
\end{eqnarray}
The ansatz of these two conditions are both $(0,0)$-flux and gauge equivalent,
which are shown in Fig.~\ref{00}.

\begin{figure}[h]
    \centering
    \begin{subfigure}[h]{0.215\textwidth}
        \includegraphics[width=\textwidth]{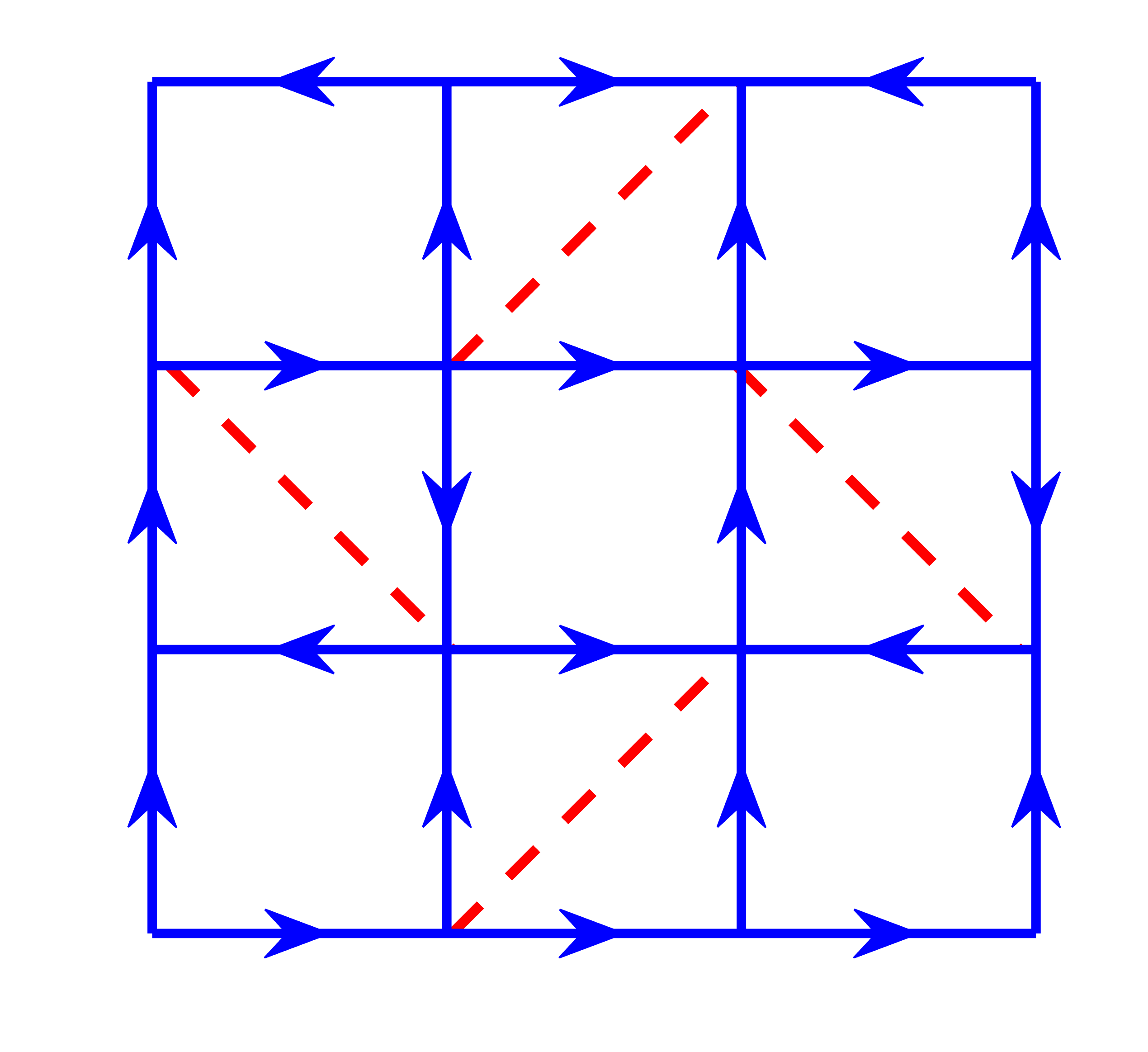}
        %\label{fig:a}
        \caption{$A$ of $(p_2,p_4,p_7)=(1,1,0)$}
    \end{subfigure}
    \begin{subfigure}[h]{0.25\textwidth}
        \includegraphics[width=\textwidth]{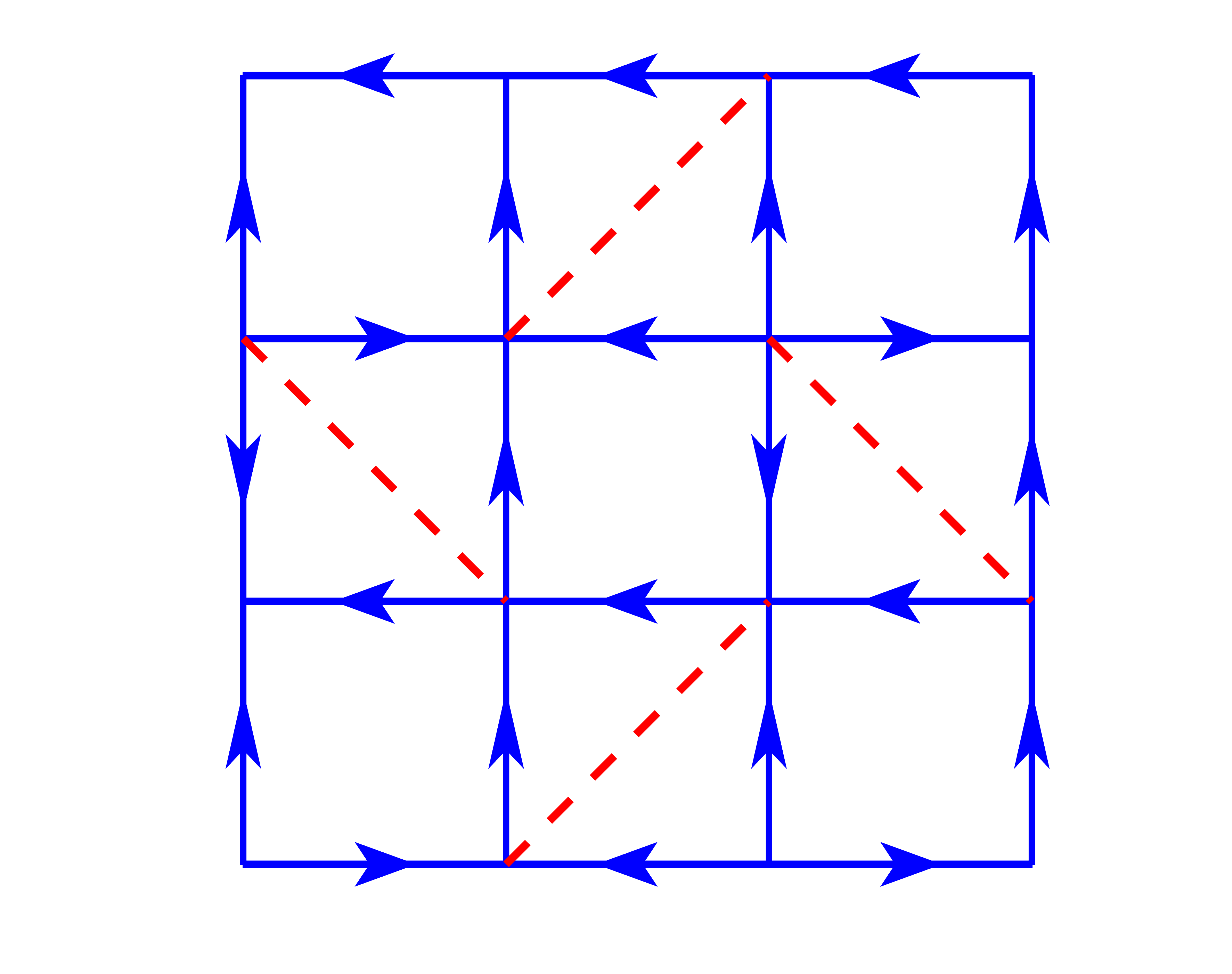}
        \caption{$A$ of $(p_2,p_4,p_7)=(1,1,1)$}
    \end{subfigure}
    \caption{The ansatz on dashed lines are 0.
        The two ansatz are gauge equivalent.
        It is $\pi$-flux in empty square and $0$-flux in $J_2$ square.
        $B_{\rm n.n}=0$ and $B_{\rm n.n.n}$ are real and uniform.
    }
    \label{pi-0}
\end{figure}

Then we consider $p_2=p_4=1$ condition.
For $(p_2,p_4,p_7)=(1,1,0)$ condition, the PSG solution is
\begin{eqnarray}
    \phi_{C_4}(X,Y,s)&=&Y\pi+\delta_{s,0},\\
    \phi_{\sigma}(X,Y,s)&=&x_sy_s\pi+\delta_{s,0}\pi.
\end{eqnarray}
while $(p_2,p_4,p_7)=(1,1,1)$ is
\begin{eqnarray}
    \phi_{C_4}(X,Y,s)&=&Y\pi+\delta_{s,0},\\
    \phi_{\sigma}(X,Y,s)&=&\frac{\pi}{2}+x_sy_s\pi+\delta_{s,0}\pi.
\end{eqnarray}
The ansatz are both $(\pi,0)$-flux and also gauge equivalent,
which are shown in Fig.~\ref{pi-0}.

\subsection{Plaquette-singlet state}
\label{App:1D}

\begin{figure}[]
    \centering
    \begin{subfigure}[h]{0.23\textwidth}
        \includegraphics[width=\textwidth]{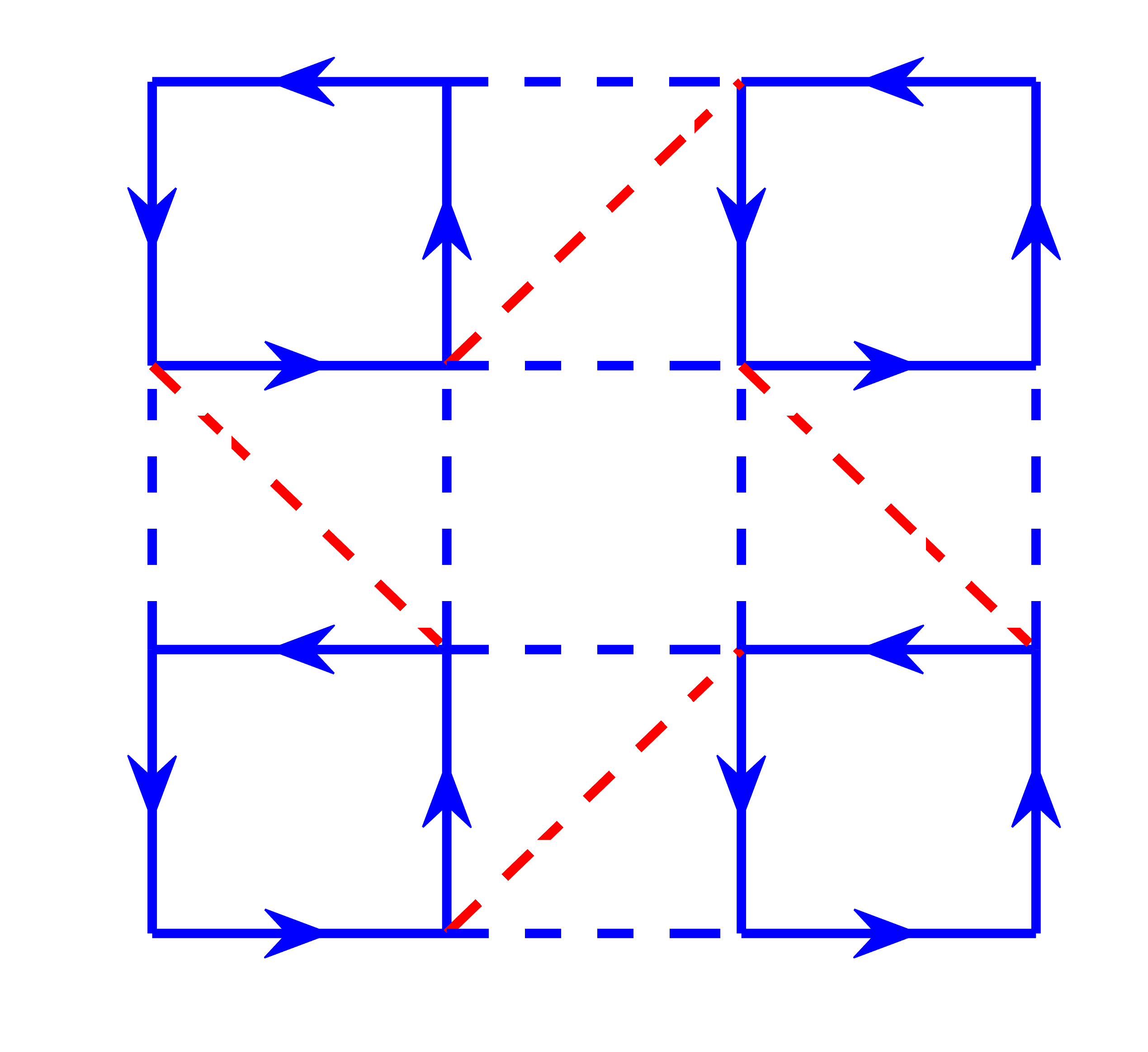}
        %\label{fig:a}
        \caption{$A$ of PS in empty-square}
    \end{subfigure}
    \begin{subfigure}[h]{0.23\textwidth}
        \includegraphics[width=\textwidth]{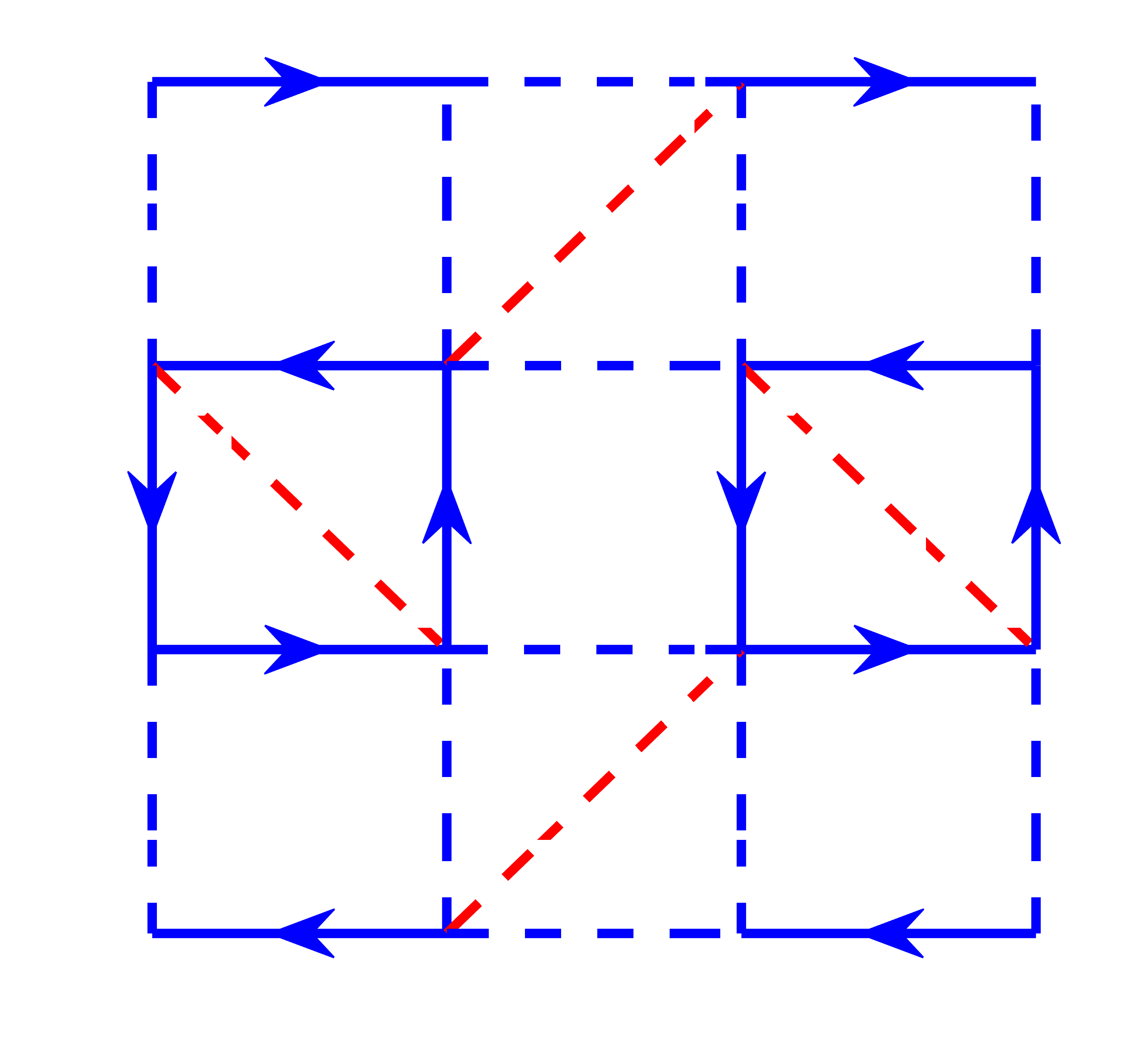}
        \caption{$A$ of PS in \(J_2\)-square}
    \end{subfigure}
    \caption{The ansatz $A$ of plaquette-singlet state in (a) empty-square and (b) $J_2$-square.
        $B$ of the PS state in empty-square is always zero.
        For the $B$ of the PS state in \(J_2\)-square,
        only the $B_2$ in the \(J_2\) bonds of the \(J_2\)-square is nonzero.
        Ansatz \(A\) on dashed bonds are vanishing in the self-consistent solution.
    }
    \label{PS}
\end{figure}

\begin{figure}[H]
    \centering{\includegraphics[scale=0.5]{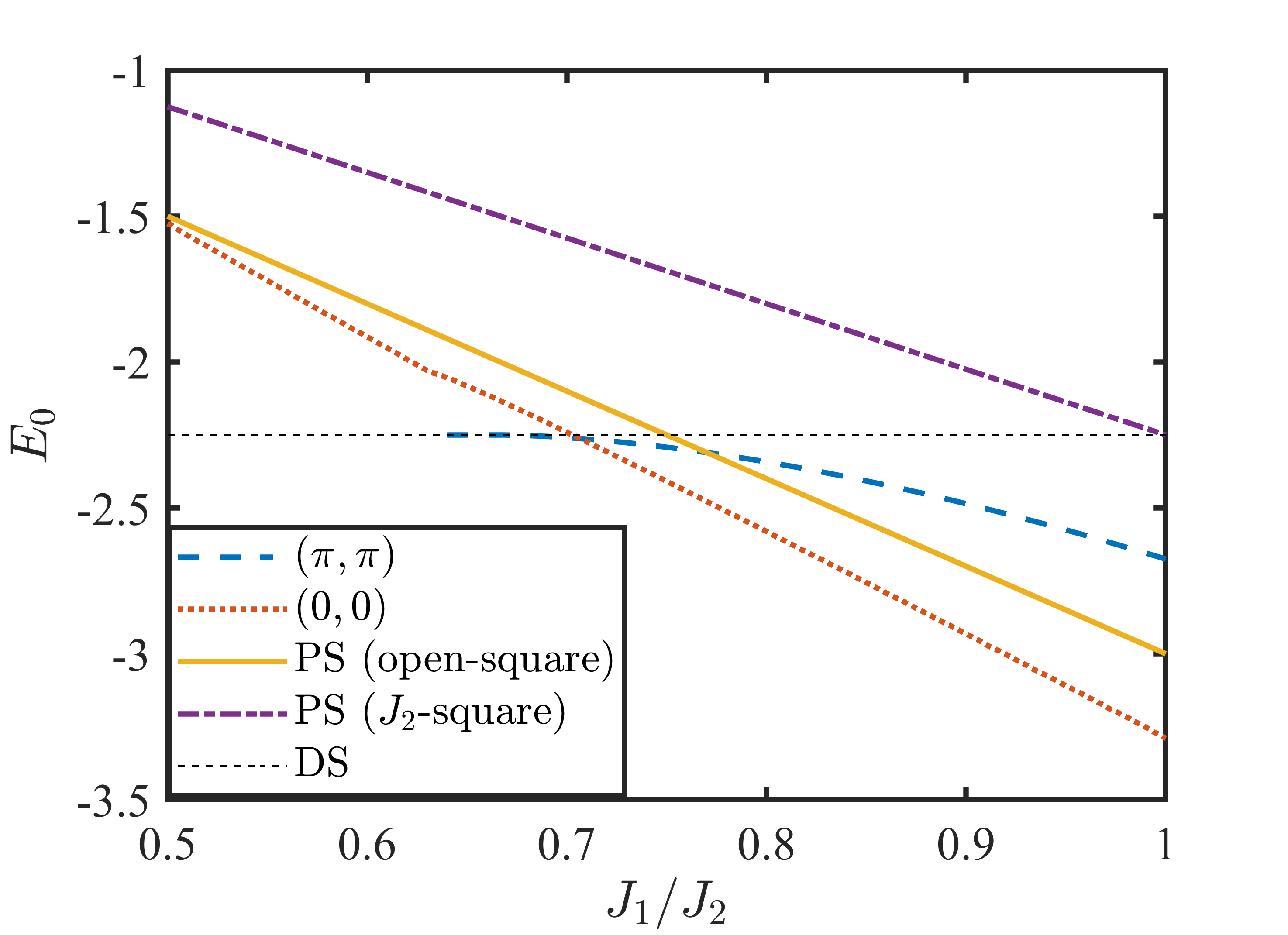}}
    \caption{The mean-field ground state energies in the physical condition \(\kappa=1\).
        The ground state energy of plaquette-singlet (PS) in the \(J_2\)-square is much larger than that in the empty-square condition
        because of the existence of \(B_2\).
    }
    \label{energy plot}
\end{figure}

Now we consider the plaquette-singlet states,
which break the glide symmetry and are out of the algebraic PSG solutions.
After self-consistent calculation,
we find only the zero-flux plaquette-singlet ansatz can be solved self-consistently,
% added
and the self-consistent solutions are such that only the ansatz($A$ and $B$) within the selected plaquettes are nonzero.
Because there are two inequivalent plaquette (empty-square and \(J_2\)-square) in the Shastry-Sutherland lattice,
only two inequivalent plaquette-singlet ansatz exist,
which are shown in Fig.~\ref{PS}.
Calculating the self-consistent equations in Section \ref{PS-state},
we get the ground state energy with the change of \(J_1/J_2\).
The energies of these two PS states and (0,0) and (\(\pi,\pi\)) states in physical condition \(\kappa=1\) are shown in Fig.~\ref{energy plot}.
\begin{widetext}
    \section{Details of Schwinger boson mean-field Hamiltonian}
    \label{App:3}
    In the previous section,
    we showed the mean-field ansatz configurations of the 6 PSGs,
    which can classified by 4 gauge inequivalent solutions.
    We also showed the configurations of 2 PS states.
    In this section,
    we show the Hamiltonian details of the 4 symmetric spin liquid states and 2 PS states.

    \subsection{$(0,0)$-flux state}\label{App:3 A}
    After Fourier transformation,
    the mean-field Hamiltonian of the symmetric spin liquids are formally written as Eq.~(\ref{eq:Hk}).
    The mean-field configurations of $(0,\pi)$-flux state are shown in Fig.~\ref{00}.
    Because the two configurations in Fig.~\ref{00} are gauge equivalent and have same physical properties,
    we just need to consider one of the configurations.
    If we consider the configuration of Fig.~\ref{00}~(a),
    the $4\times 4$ matrices $\boldsymbol{A}_{\mathbf{k}}$ and $\boldsymbol{B}_{\mathbf{k}}$
    of the $(0,0)$-flux state are
    \begin{eqnarray}
        \boldsymbol{A}_{\mathbf{k}}&=&
        \left(\begin{array}{cccc}
                0                              & \frac{1}{2}J_1A_1(-1+e^{-ik_1}) & 0                               & \frac{1}{2}J_1A_1(1-e^{-ik_2}) \\
                \frac{1}{2}J_1A_1(1-e^{ik_1})  & 0                               & \frac{1}{2}J_1A_1(-1+e^{-ik_2}) & 0                              \\
                0                              & \frac{1}{2}J_1A_1(1-e^{ik_2})   & 0                               & \frac{1}{2}J_1A_1(-1+e^{ik_1}) \\
                \frac{1}{2}J_1A_1(-1+e^{ik_2}) & 0                               & \frac{1}{2}J_1A_1(1-e^{-ik_1})  & 0                              \\
            \end{array}\right),\\
        \boldsymbol{B}_{\mathbf{k}}&=&
        \left(\begin{array}{cccccccc}
                0                           & 0                            & \frac{1}{2}J_2 B_2 e^{-ik_2} & 0                           \\
                0                           & 0                            & 0                            & \frac{1}{2}J_2 B_2 e^{ik_1} \\
                \frac{1}{2}J_2 B_2 e^{ik_2} & 0                            & 0                            & 0                           \\
                0                           & \frac{1}{2}J_2 B_2 e^{-ik_1} & 0                            & 0
            \end{array}\right).
    \end{eqnarray}
    \subsection{$(0,\pi)$-flux state}\label{App:3 B}
    The mean-field configurations of $(0,\pi)$-flux state are shown in Fig.~\ref{0-Pi}.
    The $\boldsymbol{A}_{\mathbf{k}}$ and $\boldsymbol{B}_{\mathbf{k}}$
    of the $(0,0)$-flux state are
    \begin{eqnarray}
        \boldsymbol{A}_{\mathbf{k}}&=&
        \left(\begin{array}{cccc}
                0                              & \frac{1}{2}J_1A_1(-1-e^{-ik_1}) & -\frac{1}{2}J_2A_2e^{-ik_2}     & \frac{1}{2}J_1A_1(1+e^{-ik_2}) \\
                \frac{1}{2}J_1A_1(1+e^{ik_1})  & 0                               & \frac{1}{2}J_1A_1(-1+e^{-ik_2}) & -\frac{1}{2}J_2A_2e^{ik_1}     \\
                \frac{1}{2}J_2A_2e^{ik_2}      & \frac{1}{2}J_1A_1(1-e^{ik_2})   & 0                               & \frac{1}{2}J_1A_1(-1+e^{ik_1}) \\
                \frac{1}{2}J_1A_1(-1-e^{ik_2}) & \frac{1}{2}J_2A_2e^{-ik_1}      & \frac{1}{2}J_1A_1(1-e^{-ik_1})  & 0                              \\
            \end{array}\right),\\
        \boldsymbol{B}_{\mathbf{k}}&=&
        \left(\begin{array}{cccccccc}
                0                           & \frac{1}{2}B_1(-1-e^{-ik_1}) & 0                            & \frac{1}{2}B_1(-1-e^{-ik_2}) \\
                \frac{1}{2}B_1(-1-e^{ik_1}) & 0                            & \frac{1}{2}B_1(-1+e^{-ik_2}) & 0                            \\
                0                           & \frac{1}{2}B_1(-1+e^{ik_2})  & 0                            & \frac{1}{2}B_1(-1+e^{ik_1})  \\
                \frac{1}{2}B_1(-1-e^{ik_2}) & 0                            & \frac{1}{2}B_1(-1+e^{-ik_1}) & 0
            \end{array}\right).
    \end{eqnarray}
    \subsection{$(\pi,0)$-flux state}\label{App:3 C}
    The mean-field configurations of $(\pi,0)$-flux state are shown in Fig.~\ref{pi-0}.
    Because the two configurations in Fig.~\ref{pi-0} are gauge equivalent and have same physical properties,
    we just need to consider one of the configurations.
    If we consider the configuration of Fig.~\ref{pi-0}~(a),
    the $\boldsymbol{A}_{\mathbf{k}}$ and $\boldsymbol{B}_{\mathbf{k}}$ of the $(\pi,0)$-flux state are
    \begin{eqnarray}
        \boldsymbol{A}_{\mathbf{k}}&=&
        \left(\begin{array}{cccc}
                0                             & \frac{1}{2}J_1A_1(-1-e^{-ik_1}) & 0                               & \frac{1}{2}J_1A_1(-1-e^{-ik_2}) \\
                \frac{1}{2}J_1A_1(1+e^{ik_1}) & 0                               & \frac{1}{2}J_1A_1(-1+e^{-ik_2}) & 0                               \\
                0                             & \frac{1}{2}J_1A_1(1-e^{ik_2})   & 0                               & \frac{1}{2}J_1A_1(-1+e^{ik_1})  \\
                \frac{1}{2}J_1A_1(1+e^{ik_2}) & 0                               & \frac{1}{2}J_1A_1(1-e^{-ik_1})  & 0                               \\
            \end{array}\right),\label{eq:pi-pi-Ak}\\
        \boldsymbol{B}_{\mathbf{k}}&=&
        \left(\begin{array}{cccccccc}
                0                           & 0                            & \frac{1}{2}J_2 B_2 e^{-ik_2} & 0                           \\
                0                           & 0                            & 0                            & \frac{1}{2}J_2 B_2 e^{ik_1} \\
                \frac{1}{2}J_2 B_2 e^{ik_2} & 0                            & 0                            & 0                           \\
                0                           & \frac{1}{2}J_2 B_2 e^{-ik_1} & 0                            & 0
            \end{array}\right)\label{eq:pi-pi-Bk}.
    \end{eqnarray}
    \subsection{$(\pi,\pi)$-flux state}\label{App:3 D}
    The mean-field configurations of $(\pi,\pi)$-flux state are shown in Fig.~\ref{pi-pi}.
    The $\boldsymbol{A}_{\mathbf{k}}$ and $\boldsymbol{B}_{\mathbf{k}}$
    of the $(\pi,\pi)$-flux state are
    \begin{eqnarray}
        \boldsymbol{A}_{\mathbf{k}}&=&
        \left(\begin{array}{cccc}
                0                             & \frac{1}{2}J_1A_1(-1-e^{-ik_1}) & -\frac{1}{2}J_2A_2e^{-ik_2}     & \frac{1}{2}J_1A_1(-1-e^{-ik_2}) \\
                \frac{1}{2}J_1A_1(1+e^{ik_1}) & 0                               & \frac{1}{2}J_1A_1(-1-e^{-ik_2}) & -\frac{1}{2}J_2A_2e^{ik_1}      \\
                \frac{1}{2}J_2A_2e^{ik_2}     & \frac{1}{2}J_1A_1(1+e^{ik_2})   & 0                               & \frac{1}{2}J_1A_1(-1-e^{ik_1})  \\
                \frac{1}{2}J_1A_1(1+e^{ik_2}) & \frac{1}{2}J_2A_2e^{-ik_1}      & \frac{1}{2}J_1A_1(1+e^{-ik_1})  & 0                               \\
            \end{array}\right),\\
        \boldsymbol{B}_{\mathbf{k}}&=&
        \left(\begin{array}{cccccccc}
                0                           & \frac{1}{2}B_1(1+e^{-ik_1}) & 0                           & \frac{1}{2}B_1(-1-e^{-ik_2}) \\
                \frac{1}{2}B_1(1+e^{ik_1})  & 0                           & \frac{1}{2}B_1(1+e^{-ik_2}) & 0                            \\
                0                           & \frac{1}{2}B_1(1+e^{ik_2})  & 0                           & \frac{1}{2}B_1(1+e^{ik_1})   \\
                \frac{1}{2}B_1(-1-e^{ik_2}) & 0                           & \frac{1}{2}B_1(1+e^{-ik_1}) & 0
            \end{array}\right).
    \end{eqnarray}
    \subsection{Plaquette-singlet states}
    The mean-field configurations of the plaquette-singlet states are shown in Fig.~\ref{PS}.
    It shows that the mean-field ansatz are decoupled into disconnected empty or \(J_2\) squares.
    In the Nambu spinor $\Psi_{\mathbf{k}}=(b_{0\mathbf{k}\uparrow},b_{1\mathbf{k}\uparrow},b_{2\mathbf{k}\uparrow},b_{3\mathbf{k}\uparrow},b_{0\mathbf{-k}\downarrow}^\dagger,b_{1\mathbf{-k}\downarrow}^\dagger,b_{2\mathbf{-k}\downarrow}^\dagger,b_{3\mathbf{-k}\downarrow}^\dagger)^T$,
    the Hamiltonian matrices of the PS states of open square and \(J_2\) square are
    \begin{eqnarray}
        H_{\text{open-square}}&=&\left(\begin{array}{cccccccc} -\mu  & 0 & 0 & 0 & 0 & -\frac{1}{2}J_1A_1 & 0 & -\frac{1}{2}J_1A_1\\ 0 & -\mu  & 0 & 0 & \frac{1}{2}J_1A_1 & 0 & -\frac{1}{2}J_1A_1 & 0\\ 0 & 0 & -\mu  & 0 & 0 & \frac{1}{2}J_1A_1 & 0 & \frac{1}{2}J_1A_1\\ 0 & 0 & 0 & -\mu  & \frac{1}{2}J_1A_1 & 0 & -\frac{1}{2}J_1A_1 & 0\\ 0 & \frac{1}{2}J_1A_1 & 0 & \frac{1}{2}J_1A_1 & -\mu  & 0 & 0 & 0\\ -\frac{1}{2}J_1A_1 & 0 & \frac{1}{2}J_1A_1 & 0 & 0 & -\mu  & 0 & 0\\ 0 & -\frac{1}{2}J_1A_1 & 0 & -\frac{1}{2}J_1A_1 & 0 & 0 & -\mu  & 0\\ -\frac{1}{2}J_1A_1 & 0 & \frac{1}{2}J_1A_1 & 0 & 0 & 0 & 0 & -\mu  \end{array}\right),\\
        H_{\text{$J_2$-square}}&=&\left(\begin{array}{cccccccc} -\mu  & 0 & 0 & 0 & 0 & -\frac{1}{2}J_1A_1 & 0 & -\frac{1}{2}J_1A_1\\ 0 & -\mu  & 0 & \frac{1}{2}J_2B_2 & \frac{1}{2}J_1A_1 & 0 & -\frac{1}{2}J_1A_1 & 0\\ 0 & 0 & -\mu  & 0 & 0 & \frac{1}{2}J_1A_1 & 0 & \frac{1}{2}J_1A_1\\ 0 & \frac{1}{2}J_2B_2 & 0 & -\mu  & \frac{1}{2}J_1A_1 & 0 & -\frac{1}{2}J_1A_1 & 0\\ 0 & \frac{1}{2}J_1A_1 & 0 & \frac{1}{2}J_1A_1 & -\mu  & 0 & 0 & 0\\ -\frac{1}{2}J_1A_1 & 0 & \frac{1}{2}J_1A_1 & 0 & 0 & -\mu  & 0 & \frac{1}{2}J_2B_2\\ 0 & -\frac{1}{2}J_1A_1 & 0 & -\frac{1}{2}J_1A_1 & 0 & 0 & -\mu  & 0\\ -\frac{1}{2}J_1A_1 & 0 & \frac{1}{2}J_1A_1 & 0 & 0 & \frac{1}{2}J_2B_2 & 0 & -\mu  \end{array}\right).
    \end{eqnarray}
    Note that we choose the the $J_2$-square as the unit cell of \(H_{\text{$J_2$-square}}\) to get decoupled Hamiltonian.
\end{widetext}

\section{Magnetic order from the zero-flux state}
\label{App:2}

In the Schwinger boson formalism,
the magnetic order is formed by the condensation of the Schwinger bosons.
When the density of the boson is large than the critical $\kappa_c$,
the dispersion of the spinon will become zero at several $\mathbf{Q}$ points,
and  the spinon will condense at these $\mathbf{Q}$ points and develop magnetic orders.

For the Shastry-Sutherland model,
the N\'eel phase is formed by the condensation of the zero-flux ((0,0)-flux) state.
As shown in Fig.~\ref{00 ansatz} (a),
the dispersion of spinon become zero at $\mathbf{Q}=(\pi,\pi)$.
Note that $\mathbf{Q}=-\mathbf{Q}$.
If we choose the empty square as the unit cell,
after Fourier transformation $b_{\mathbf{r}}=\frac{1}{\sqrt{\Ncell }} \sum_{\mathbf{r}} e^{-i \mathbf{k} \cdot \mathbf{r}} b_{\mathbf{k}}$,
the mean-field Hamiltonian becomes
\begin{eqnarray}
    H_{\mathrm{MF}}&=&\sum_{\mathbf{k}} \Psi_{\mathbf{k}}^{\dagger} D_{\mathbf{k}} \Psi_{\mathbf{k}}\nonumber\\ &&+\Ncell \left[\mu+\mu \kappa+8J_1\left|A_1\right|^2-2J_2\left|B_2\right|^2\right].
\end{eqnarray}
where we have used the Nambu spinor $\Psi_{\mathbf{k}}=(b_{0\mathbf{k}\uparrow},b_{1\mathbf{k}\uparrow},b_{2\mathbf{k}\uparrow},b_{3\mathbf{k}\uparrow},b_{0\mathbf{-k}\downarrow}^\dagger,b_{1\mathbf{-k}\downarrow}^\dagger,b_{2\mathbf{-k}\downarrow}^\dagger,b_{3\mathbf{-k}\downarrow}^\dagger)^T$.
The subscript $s=0,1,2,3$ in $b_{s \bm{k}\sigma}$ is the atom index in the unit cell,
which is shown in Fig.~\ref{sslatice}.
The $8\times 8$ matrix $D_{\mathbf{k}}$ at $\mathbf{Q}$ is
\begin{widetext}
    \begin{eqnarray}
        D_{\mathbf{Q}}&=&
        \left(\begin{array}{cccccccc}
                -\mu    & 0       & J_2 B_2 & 0       & 0       & J_1A_1  & 0       & J_1A_1  \\
                0       & -\mu    & 0       & J_2 B_2 & -J_1A_1 & 0       & -J_1A_1 & 0       \\
                J_2 B_2 & 0       & -\mu    & 0       & 0       & J_1A_1  & 0       & J_1A_1  \\
                0       & J_2 B_2 & 0       & -\mu    & -J_1A_1 & 0       & -J_1A_1 & 0       \\
                0       & -J_1A_1 & 0       & -J_1A_1 & -\mu    & 0       & J_2 B_2 & 0       \\
                J_1A_1  & 0       & J_1A_1  & 0       & 0       & -\mu    & 0       & J_2 B_2 \\
                0       & -J_1A_1 & 0       & -J_1A_1 & J_2 B_2 & 0       & -\mu    & 0       \\
                J_1A_1  & 0       & J_1A_1  & 0       & 0       & J_2 B_2 & 0       & -\mu
            \end{array}\right).
        \label{Dq}
    \end{eqnarray}
\end{widetext}

In the critical density $\kappa_c$,
it satisfies
\begin{eqnarray}
    -\mu+J_2B_2-2A_1=0.
\end{eqnarray}
At this point,
$D_{\mathbf{Q}}$ has two eigenvectors with zero eigenvalue,
\begin{eqnarray}
    \Psi_1&=&\frac{1}{\sqrt{2}}\left(\begin{array}{cccccccc}
            0 & 1 & 0 & 1 & 1 & 0 & 1 & 0
        \end{array}\right)^T,\nonumber \\
    \Psi_2&=&\frac{1}{\sqrt{2}}\left(\begin{array}{cccccccc}
            -1 & 0 & -1 & 0 & 0 & 1 & 0 & 1
        \end{array}\right)^T.
\end{eqnarray}
Therefore,
the condensation at $\mathbf{Q}$ is $\langle \Psi_{\mathbf{Q}} \rangle=c_1\Psi_1+c_2\Psi_2$,
where $c_{1,2}$ are two complex numbers.
We define $x_s\equiv (\langle b_{(X,Y,s)\uparrow} \rangle,\langle b_{(X,Y,s)\downarrow} \rangle)^T$,
then condensation on site $(X,Y,s)$ is given by
\begin{eqnarray}
    x_0&=&x_2= \frac{(-1)^{X+Y}}{\sqrt{2}}\left(\begin{array}{c}
            -c_2 \\ c_1^*
        \end{array}\right),\nonumber \\
    x_1&=&x_3 = \frac{(-1)^{X+Y}}{\sqrt{2}}\left(\begin{array}{c}
            c_1 \\ c_2^*
        \end{array}\right),
\end{eqnarray}
and the ordered magnetic momentum is then $\mathbf{S}(X,Y,s)=\frac{1}{2}x_s^\dagger \bm{\sigma} x_s$.
Thus we have
\begin{eqnarray}
    \mathbf{S}(X,Y,0)&=&\mathbf{S}(X,Y,2)=\mathbf{n},\\
    \mathbf{S}(X,Y,1)&=&\mathbf{S}(X,Y,3)=-\mathbf{n},
\end{eqnarray}
where $\mathbf{n}$ is the $SO(3)$ vector corresponded with the $SU(2)$ vector $(-c_2,c_1^*)^T$.
This gives the N\'eel magnetic order momentum,
\begin{eqnarray}
    \mathbf{S}(\mathbf{r})&=&(-1)^{\mathbf{r}}\mathbf{n}.
\end{eqnarray}
\section{Magnetic order from the $\pi$-flux state}
\label{App:4}

In the large $\kappa$ limit,
the spinon in the $\pi$-flux spin liquid phase will also condense and form magnetic order.
In this section, we discuss the magnetic order from the $\pi$-flux state.

The mean-field Hamiltonian of $(\pi,\pi)$-flux state is
\begin{eqnarray}
    H_{\mathrm{MF}}&=&\sum_{\mathbf{k}} \Psi_{\mathbf{k}}^{\dagger} D_{\mathbf{k}} \Psi_{\mathbf{k}}+\Ncell \left[\mu+\mu \kappa\nonumber\right.\\ &&\left.+8J_1(\left|A_1\right|^2-\left|B_1\right|^2)+2J_2\left|A_2\right|^2\right],
\end{eqnarray}
where the expression $D_\mathbf{k}$ is shown in Eq.~(\ref{eq:Hk}),
 and the $\boldsymbol{A}_{\mathbf{k}}$ and $\boldsymbol{B}_{\mathbf{k}}$ are shown in Eq.~(\ref{eq:pi-pi-Ak}) and Eq.~(\ref{eq:pi-pi-Bk}) respectively.

For simplicity, we only discuss the large $J_1$ condition, where only $A_1$ is finite.
In the mean-field level, the Shastry-Sutherland lattice is equivalent to the square lattice in this condition where $A_2$ and $B_2$ vanishes.
The magnetic order of this condition is
\begin{eqnarray}
    \mathbf{S}(X,Y,0)&=&\mathbf{n}_1+\mathbf{n}_2+\mathbf{n}_3,\\
    \mathbf{S}(X,Y,1)&=&-\mathbf{n}_1+\mathbf{n}_2-\mathbf{n}_3,\\
    \mathbf{S}(X,Y,2)&=&-\mathbf{n}_1-\mathbf{n}_2+\mathbf{n}_3,\\
    \mathbf{S}(X,Y,3)&=&\mathbf{n}_1-\mathbf{n}_2-\mathbf{n}_3,
\end{eqnarray}
which is discussed in Ref.~\cite{piFluxOrder},
and $n_1^2+n_2^2+n_3^2=n^2$.
The details of the calculation can be referred to Appendix C in Ref.~\cite{piFluxOrder}.
\section{Self-consistent spin wave theory}
\label{App:5}

In this section, we introduce the self-consistent spin wave theory in details.
Near the phase boundary of the N\'eel phase,
the quantum fluctuation is significant and the interaction of magnons takes important role.
For the Shastry-Sutherland model,
the linear spin wave Hamiltonian is not positive or semi-positive definite in $J_1/J_2<1$,
and the linear spin wave theory breaks down in this region.
The nonlinear spin wave theory also breaks down because the
correction of magnon interaction depends on the magnon wave function of linear spin wave theory.
Therefore, we need to use a self-consistent spin wave theory
to study N\'eel phase of the Shastry-Sutherland model in $J_1/J_2<1$.

Spin wave theory is to describe the ordered phase in terms of small fluctuations of the
spins about their expectation values, which can be regarded as the classical ground state of
the Hamiltonian.
For the Shastry-Sutherland model,
the magnetic ordered state is a N\'eel state,
and the classical ground state is the N\'eel antiferromagnetic state.
With the classical ground state, the spin Hamiltonian can be expressed by the boson operators by the
Holstein-Primakoff transformation\cite{HP}.
Expanding the Hamiltonian by $1/S$ as we regarding $S$ as a large number,
the Hamiltonian is transformed to
\begin{eqnarray}
    H=H_0+H_2+H_4+H_6+\dots,
\end{eqnarray}
where linear term $H_1$ vanishes if the classical ground state is proper.
Keeping up to quadratic terms $H_2$, one obtains the noninteracting spin wave Hamiltonian.
The higher order terms $H_4+H_6+\dots$ introduce the interactions of magnons.
For the Shastry-Sutherland model,
if we choose the ``$J_2$-square'' as the unit cell,
the linear spin wave Hamiltonian $H_2$ in the momentum space can be written as
\begin{widetext}

    \begin{eqnarray}
        H_2&=&\sum_{\bm{k}}\left(\begin{array}{llllllll}
                a_{\bm{k}}^\dagger & b_{\bm{k}}^\dagger & c_{\bm{k}}^\dagger & d_{\bm{k}}^\dagger & a_{-\bm{k}} & b_{-\bm{k}} & c_{-\bm{k}} & d_{-\bm{k}}
            \end{array}\right)
        \mathcal{H}_{\bm{k}}\left(\begin{array}{c}
                a_{\bm{k}}          \\
                b_{\bm{k}}          \\
                c_{\bm{k}}          \\
                d_{\bm{k}}          \\
                a_{-\bm{k}}^\dagger \\
                b_{-\bm{k}}^\dagger \\
                c_{-\bm{k}}^\dagger \\
                d_{-\bm{k}}^\dagger
            \end{array}\right),
    \end{eqnarray}
    here $a,b,c,d$ are the Holstein-Primakoff bosons for the unit cell has 4 sites.
    The $\mathcal{H}_{\bm{k}}$ satisfies
    \begin{eqnarray}
        \mathcal{H}_{\bm{k}}&=&\left(\begin{array}{cc}
                \boldsymbol{B}_{\bm{k}}           & \boldsymbol{A}_{\bm{k}}      \\
                \boldsymbol{A}_{\bm{k}}^{\dagger} & \boldsymbol{B}_{-\bm{k}}^{T}
            \end{array}\right),
    \end{eqnarray}
    \begin{eqnarray}
        \boldsymbol{A}_{\bm{k}}&=&
        \left(\begin{array}{cccccccc}
                0                                    & \frac{1}{2}J_1(1+\mathrm{e}^{ik_1})  & 0                                   & \frac{1}{2}J_1(1+\mathrm{e}^{ik_2})  \\
                \frac{1}{2}J_1(1+\mathrm{e}^{-ik_1}) & 0                                    & \frac{1}{2}J_1(1+\mathrm{e}^{ik_2}) & 0                                    \\
                0                                    & \frac{1}{2}J_1(1+\mathrm{e}^{-ik_2}) & 0                                   & \frac{1}{2}J_1(1+\mathrm{e}^{-ik_1}) \\
                \frac{1}{2}J_1(1+\mathrm{e}^{-ik_2}) & 0                                    & \frac{1}{2}J_1(1+\mathrm{e}^{ik_1}) & 0
            \end{array}\right),\\
        \boldsymbol{B}_{\bm{k}}&=&
        \left(\begin{array}{cccc}
                2J_1-\frac{1}{2}J_2 & 0                                     & 0                   & 0                                    \\
                0                   & 2J_1-\frac{1}{2}J_2                   & 0                   & \frac{1}{2}J_2\mathrm{e}^{i(k_1-k2)} \\
                0                   & 0                                     & 2J_1-\frac{1}{2}J_2 & 0                                    \\
                0                   & \frac{1}{2}J_2\mathrm{e}^{i(-k_1+k2)} & 0                   & 2J_1-\frac{1}{2}J_2                  \\
            \end{array}\right).
    \end{eqnarray}
\end{widetext}
% \begin{eqnarray}
%     \mathcal{H}_{\bm{k}}&=&
%     \left(\begin{array}{cccccccc}
%         2J_1-\frac{1}{2}J_2 & 0 &0& 0& 0& 0& 0& 0 \\
%         0 & 2J_1-\frac{1}{2}J_2 &0& \frac{1}{2}J_2\mathrm{e}^{i(k_1-k2)}& 0& 0& 0& 0 \\
%         0 & 0 &2J_1-\frac{1}{2}J_2& 0& 0& 0& 0& 0 \\
%         0 & \frac{1}{2}J_2\mathrm{e}^{i(-k_1+k2)} &0& 2J_1-\frac{1}{2}J_2& 0& 0& 0& 0 \\
%         0 & \frac{1}{2}J_1(1+\mathrm{e}^{ik_1}) &0& 0& 2J_1-\frac{1}{2}J_2& 0& 0& 0 \\
%         \frac{1}{2}J_1(1+\mathrm{e}^{-ik_1}) & 0 &\frac{1}{2}J_1(1+\mathrm{e}^{ik_2})& 0& 0& 2J_1-\frac{1}{2}J_2& 0& \frac{1}{2}J_2\mathrm{e}^{i(k_1-k2)} \\
%         0 & \frac{1}{2}J_1(1+\mathrm{e}^{-ik_2}) &0& \frac{1}{2}J_1(1+\mathrm{e}^{-ik_1})& 0& 0& 2J_1-\frac{1}{2}J_2& 0 \\
%         \frac{1}{2}J_1(1+\mathrm{e}^{-ik_2}) & 0 &\frac{1}{2}J_1(1+\mathrm{e}^{ik_1})& 0& 0& \frac{1}{2}J_2\mathrm{e}^{i(-k_1+k2)}& 0 & 2J_1-\frac{1}{2}J_2
%         \end{array}\right).
% \end{eqnarray}
Using Bogoliubov transformation to diagonalize the $\mathcal{H}_{\bm{k}}$,
we will get the linear spin wave dispersion and wave functions,
which breaks down at $J_1/J_2<1$.
The self-consistent spin wave theory is to use the boson-pair expectation values
to decouple the quartic terms.
We use a specific term for example:
\begin{eqnarray}
    \nonumber
    &&\sum_{\bm{k},\bm{p},\bm{q}}V(\bm{k},\bm{p},\bm{q})a_{\bm{k}}^\dagger b_{\bm{p}}^\dagger a_{\bm{q}} b_{\bm{k}+\bm{p}-\bm{q}}\\\nonumber
    &\approx&\sum_{\bm{k}}\left[\sum_{\bm{p},\bm{q}}V(\bm{k},\bm{p},\bm{k})\langle b_{\bm{p}}^\dagger b_{\bm{p}}\rangle\right]a_{\bm{k}}^\dagger a_{\bm{k}}+\dots\\
    &=&\sum_{\bm{k}}f(\bm{k})a_{\bm{k}}^\dagger a_{\bm{k}}+\dots.
\end{eqnarray}
Here $a$ and $b$ are the boson operators of different flavors,
and we set $\sum_{\bm{k}}f(\bm{k})\equiv \sum_{\bm{p},\bm{q}}V(\bm{k},\bm{p},\bm{k})\langle b_{\bm{p}}^\dagger b_{\bm{p}}\rangle$.
The symbol $(\dots)$ represents the other decoupling types.
Therefore,
the Hamiltonian is transformed into
\begin{eqnarray}
    H&=&H_2+H_4+\dots\nonumber\\
    &=&H_2+\widetilde{H} _{2}+H_4-\widetilde{H} _{2}+\dots \nonumber\\
    &=&H_2^\prime+H_4^\prime,
\end{eqnarray}
and $H_4^\prime=0$ in the mean-field level.
We call this theory a self-consistent spin wave theory because the
decoupled quartic term $\widetilde{H}_2$ need to be calcultaed self-consistently.
The effects of magnon interaction is considered in this theory.
When $H_2$ is not positive or semi-positive definite,
this method may also work because we only need to keep $H_2^\prime$ to be positive and semi-positive definite.
For the Shastry-Sutherland model,
the self-consistent spin wave theory works in the $J_1/J_2>0.65$,
where the magnetic order parameter vanishes,
which indicates the phase boundary of the N\'eel state.
\begin{widetext}
    \section{Continuum limit of linear spin wave on Shastry-Sutherland model}
    \label{App:6}

    In this section,
    we follow the prescription in Refs.~\cite{Sachdev2012}
    to derive the continuum field theory of the linear spin wave on Shastry-Sutherland model.
    The magnon velocity can be obtained from this continuum field theory.

    We assume the N\'eel order is along the \(z\) direction
    \begin{equation}
        \langle \bm{S}_i \rangle=(-1)^sS\hat{\bm{z}},
    \end{equation}
    here \(s=0,1,2,3\), which is the sublattice label shown in Fig.~\ref{sslatice}.
    The lattice vectors are \(2\hat{\bm{x}}\) and \(2\hat{\bm{y}}\).
    The site position in the Shastry-Sutherland lattice is expressed by \((\bm{r},s)\),
    where \(\bm{r}\) is the unit cell position and \(s\) is the sublattice label.
    After Holstein-Primakoff transformation,
    the linear spin wave Hamiltonian is
    \begin{eqnarray}
        H_{LSW}=S\sum_{\bm{r}}&&\left\{(4J_1-J_2)\sum_s b_{s\bm{r}}^\dagger b_{s\bm{r}}+J_2\left(b_{1\bm{r}}^\dagger b_{3\bm{r}+\hat{\bm{x}}+\hat{\bm{y}}}+b_{0\bm{r}}^\dagger b_{2\bm{r}+\hat{\bm{x}}-\hat{\bm{y}}}+\mathrm{h.c.}\right)\right.\nonumber\\
        &&\left. -J_1\left[b_{0\bm{r}}\left(b_{1\bm{r}+\hat{\bm{x}}}+b_{1\bm{r}-\hat{\bm{x}}}+b_{3\bm{r}+\hat{\bm{y}}}+b_{3\bm{r}-\hat{\bm{y}}}\right)+b_{2\bm{r}}\left(b_{1\bm{r}+\hat{\bm{y}}}+b_{1\bm{r}-\hat{\bm{y}}}+b_{3\bm{r}+\hat{\bm{x}}}+b_{3\bm{r}-\hat{\bm{x}}}\right)+\mathrm{h.c.}\right]\right\}.\label{LSWH}
    \end{eqnarray}
    In the imaginary time path integral formalism,
    the bosonic \(b_{s\bm{r}}\) operators become complex fields.
    For later convenience,
    the operators are transformed to complex fields as
    \begin{eqnarray}
        b_{0\bm{r}}&\rightarrow&\phi_0(\bm{r}),\\
        b_{1\bm{r}}&\rightarrow&\phi_1^*(\bm{r}),\\
        b_{2\bm{r}}&\rightarrow&\phi_2(\bm{r}),\\
        b_{3\bm{r}}&\rightarrow&\phi_3^*(\bm{r}).
    \end{eqnarray}
    The gradient expansion is used to get the continuum field theory from the linear spin wave Hamiltonian in Eq.~(\ref{LSWH}),
    \begin{eqnarray}
        \phi^*_s(\bm{r})\phi_{s^\prime}(\bm{r}+\bm{a})=
        \phi_s^*(\bm{r})\left[\phi_{s^\prime}(\bm{r})+\bm{a}\cdot\nabla_{\bm{r}}\phi_{s^\prime}(\bm{r})+\frac{1}{2}(\bm{a}\cdot\nabla_{\bm{r}})^2\phi_{s^\prime}(\bm{r})+\dots\right].
    \end{eqnarray}
    Substituting the gradient expansion to the linear spin wave Hamiltonian and keep up to the 2nd order,
    we obtains the Hamiltonian density,
    \begin{eqnarray}
        \mathcal{H}=\mathcal{H}_0+\mathcal{H}_1+\mathcal{H}_2,
    \end{eqnarray}
    here \(\mathcal{H}_0\) has no gradient, \(\mathcal{H}_0\) has the 1nd order gradient terms and \(\mathcal{H}_2\) has the 2st order gradient terms.
    The expression of \(\mathcal{H}_0\) is
    \begin{eqnarray}
        \mathcal{H}_0=\left(\begin{array}{llll}
                \phi_0^*(\bm{r}) & \phi_1^*(\bm{r}) & \phi_2^*(\bm{r}) & \phi_3^*(\bm{r})
            \end{array}\right)    \left(\begin{array}{cccc}
                4J_1-J_2 & -2J_1    & J_2      & -2J_1    \\
                -2J_1    & 4J_1-J_2 & -2J_1    & J_2      \\
                J_2      & -2J_1    & 4J_1-J_2 & -2J_1    \\
                -2J_1    & J_2      & -2J_1    & 4J_1-J_2
            \end{array}\right)\left(\begin{array}{c}
                \phi_0(\bm{r}) \\
                \phi_1(\bm{r}) \\
                \phi_2(\bm{r}) \\
                \phi_3(\bm{r})
            \end{array}\right),
    \end{eqnarray}
    and the eigenvalues and eigenvectors of \(\mathcal{H}_0\) are
    \begin{eqnarray}
        0,\Psi_0=\left(\begin{array}{c}
            1/2 \\
            1/2 \\
            1/2 \\
            1/2
        \end{array}\right);\ 8J_1S,\Psi_1=\left(\begin{array}{c}
            1/2  \\
            -1/2 \\
            1/2  \\
            -1/2
        \end{array}\right);\ (4J1-2J_2)S,\Psi_2=\left(\begin{array}{c}
            1/2  \\
            1/2  \\
            -1/2 \\
            -1/2
        \end{array}\right);\ (4J1-2J_2)S,\Psi_3=\left(\begin{array}{c}
            1/2  \\
            -1/2 \\
            -1/2 \\
            1/2
        \end{array}\right);
    \end{eqnarray}
    The eigenvector \(\Psi_0\) with zero eigenvalue is the Goldstone mode.
    Then the \(\mathcal{H}_0\) is diagonalized as
    \begin{eqnarray}
        \mathcal{H}_0=8SJ_1\Psi_1^*\Psi_1+S(4J_1-2J_2)(\Psi_2^*\Psi_2+\Psi_3^*\Psi_3).
    \end{eqnarray}
    For later convenience,
    we set \(\bm{\phi}\equiv\left(\begin{array}{llll}
            \phi_0(\bm{r}) & \phi_1(\bm{r}) & \phi_2(\bm{r}) & \phi_3(\bm{r})
        \end{array}\right)^T\)
    and \(\bm{\Psi}\equiv\left(\begin{array}{llll}
            \Psi_0(\bm{r}) & \Psi_1(\bm{r}) & \Psi_2(\bm{r}) & \Psi_3(\bm{r})
        \end{array}\right)^T\).
    The expression of \(\mathcal{H}_1\) is
    \begin{eqnarray}
        \mathcal{H}_1&=&SJ_2[\phi_1(\partial_x+\partial_y)\phi_3^*+\phi_0^*(\partial_x-\partial_y)\phi_2-\phi_3(\partial_x+\partial_y)\phi_1^*-\phi_2^*(\partial_x-\partial_y)\phi_0]\nonumber\\
        &=&SJ_2\left[\bm{\phi}^*\left(\begin{array}{cccc}
                    0  & 0  & 1 & 0 \\
                    0  & 0  & 0 & 1 \\
                    -1 & 0  & 0 & 0 \\
                    0  & -1 & 0 & 0
                \end{array}\right)\partial_x\bm{\phi}+\bm{\phi}^*\left(\begin{array}{cccc}
                    0 & 0  & -1 & 0 \\
                    0 & 0  & 0  & 1 \\
                    1 & 0  & 0  & 0 \\
                    0 & -1 & 0  & 0
                \end{array}\right)\partial_y\bm{\phi}\right]\nonumber\\
        &=&SJ_2\left[\bm{\Psi}^*\left(\begin{array}{cccc}
                    0 & 0 & -1 & 0  \\
                    0 & 0 & 0  & -1 \\
                    1 & 0 & 0  & 0  \\
                    0 & 1 & 0  & 0
                \end{array}\right)\partial_x\bm{\Psi}+\bm{\Psi}^*\left(\begin{array}{cccc}
                    0 & 0 & 0  & -1 \\
                    0 & 0 & -1 & 0  \\
                    0 & 1 & 0  & 0  \\
                    1 & 0 & 0  & 0
                \end{array}\right)\partial_y\bm{\Psi}\right].
    \end{eqnarray}

    Now we consider the 2nd order gradient terms \(\mathcal{H}_2\),
    whose expression is
    \begin{eqnarray}
        \mathcal{H}_2=&&\frac{1}{2}SJ_2[\phi_1(\partial_x+\partial_y)^2\phi_3^*+\phi_0(\partial_x-\partial_y)^2\phi_2^*+\phi_3(\partial_x+\partial_y)^2\phi_1^*+\phi_2(\partial_x-\partial_y)^2\phi_0^*]\nonumber\\
        &&-SJ_1[\phi_0(\partial_x^2\phi_1^*+\partial_y^2\phi_3^*)+\phi_2(\partial_y^2\phi_1^*+\partial_x^2\phi_3^*)+\mathrm{c.c.}].
    \end{eqnarray}
    In the continuum limit, the low energy effective theory is only contributed by the Goldstone mode \( \Psi_0 \).
    % which is the \(\Psi_0^*\Psi_0\) terms.
    Therefore,
    we need to integrate the high energy modes \(\Psi_1,\Psi_2\) and \(\Psi_3\) to get the effective field theory of \(\Psi_0\).
    For this purpose,
    we only keep the \(\Psi_0^* (\dots) \Psi_0\) terms in \(\mathcal{H}_2\)
    because the other terms in \(\mathcal{H}_2\) do not contribute the effective field theory of \(\Psi_0\) up to 2nd order gradient terms.
    Then we just need to replace \(\phi_s\) by \(\frac{1}{2}\Psi_0\),
    \begin{eqnarray}
        \mathcal{H}_2&\cong & \frac{1}{4}SJ_2[\Psi_0(\partial_x+\partial_y)^2\Psi_0^*+\Psi_0(\partial_x-\partial_y)^2\Psi_0^*]-\frac{1}{2}SJ_1[\Psi_0(\partial_x^2+\partial_y^2)\Psi_0^*+\Psi_0^*(\partial_x^2+\partial_y^2)\Psi_0]+\dots \nonumber\\
        &=&S(J_1-\frac{1}{2}J_2)\left[(\partial_x\Psi_0^*)(\partial_x\Psi_0)+(\partial_y\Psi_0^*)(\partial_y\Psi_0)\right].
    \end{eqnarray}

    The Lagrangian density of the linear spin wave theory is
    \begin{eqnarray}
        \mathcal{L}&=&\phi_0^* \partial_\tau\phi_0-\phi_1^* \partial_\tau\phi_1+\phi_2^* \partial_\tau\phi_2-\phi_3^* \partial_\tau\phi_3-\mathcal{H}\nonumber\\
        &=&\Psi_0^*\partial_\tau \Psi_1+\Psi_1^*\partial_\tau \Psi_0+\Psi_2^*\partial_\tau \Psi_3+\Psi_3^*\partial_\tau \Psi_2-\mathcal{H}.
    \end{eqnarray}
    By integrating the high energy modes by Gaussian part of their action in \(\mathcal{H}\),
    we get the effective action for \(\Psi_0\), whose Lagrangian density is
    \begin{eqnarray}
        \mathcal{L}_\mathrm{eff}=\frac{1}{8SJ_1}(\partial_\tau \Psi_0^*)(\partial_\tau \Psi_0)+S\left[J_1-\frac{J_2}{2}-\frac{J_2^2}{4J_1-2J_2}\right]\left[(\partial_x\Psi_0^*)(\partial_x\Psi_0)+(\partial_y\Psi_0^*)(\partial_y\Psi_0)\right],
    \end{eqnarray}
    which yields the dispersion relation
    \begin{eqnarray}
        \frac{1}{8SJ_1}\omega^2=S\left[J_1-\frac{J_2}{2}-\frac{J_2^2}{4J_1-2J_2}\right]\bm{k}^2.
    \end{eqnarray}
    Finally, we get the magnon velocity \(v_{\mathrm{sw}}\) of linear spin wave theory,
    \begin{eqnarray}
        v_{\mathrm{sw}}=4SJ_1\sqrt{\frac{J_1-J_2}{2J_1-J_2}},
    \end{eqnarray}
    which vanishes at \(J_1=J_2\), where the linear spin wave theory breaks down.
    For the magnon velocity of the self-consistent spin wave theory,
    it can only be calculated numerically.
    The magnon velocities and magnetic moment sizes calculated by self-consistent and linear spin wave theory are shown in Table~\ref{tab:magnon velocity}.
    \begin{table}[h]
        \begin{tabular}{|l|l|l|l|l|l|l|l|l|l|l|l|l|l|l|l|l|l|l|l|}
            \hline
            $J_1/J_2$         & 0.7   & 0.75  & 0.8   & 0.85  & 0.9   & 0.95  & 1     & 1.05  & 1.1   & 1.15  & 1.2   & 1.25  & 1.13  & 1.35  & 1.4   \\ \hline
            $v_\mathrm{SCSW}$ & 0.470 & 0.585 & 0.695 & 0.802 & 0.905 & 1.005 & 1.103 & 1.199 & 1.294 & 1.387 & 1.478 & 1.569 & 1.659 & 1.748 & 1.836 \\ \hline
            $v_\mathrm{LSW}$  & n.a.  & n.a.  & n.a.  & n.a.  & n.a.  & n.a.  & 0     & 0.447 & 0.635 & 0.781 & 0.907 & 1.021 & 1.126 & 1.225 & 1.320 \\ \hline
            $m_\mathrm{SCSW}$ & 0.032 & 0.069 & 0.096 & 0.116 & 0.130 & 0.143 & 0.153 & 0.160 & 0.166 & 0.171 & 0.176 & 0.180 & 0.183 & 0.186 & 0.188 \\ \hline
            $m_\mathrm{LSW}$  & n.a.  & n.a.  & n.a.  & n.a.  & n.a.  & n.a.  & n.a.  & n.a.  & 0.067 & 0.109 & 0.136 & 0.156 & 0.171 & 0.183 & 0.193 \\ \hline
        \end{tabular}
        \caption{Magnon velocities and magnetic moment sizes from self-consistent spin wave(SCSW) and linear spin wave(LSW).
            The unit of the velocities is
            %\(2\pi J_2/a\),
            \( J_2 a \),
            where \(a\) is the square lattice constant.
            The magnetic moment is calculated by \(1/2-\langle n_m\rangle \),
            here \( \langle n_m\rangle \) is the average magnon density.
            ``n.a.'' means not applicable.}\label{tab:magnon velocity}
    \end{table}

\end{widetext}
\section{Exact diagonalization of 32-site Shastry-Sutherland model}
\label{App:7}

\begin{figure}[h]
    \centering{\includegraphics[scale=0.4]{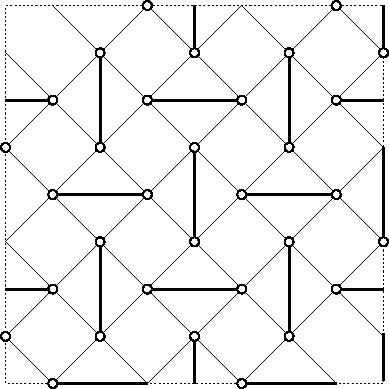}}
    \caption{The $32$-site Shastry-Sutherland lattice used in the ED study.
        Open circles indicate distinct sites under periodic boundary condition.
        $J_2$ and $J_1$ bonds are depicted as thick and thin solid lines respectively.
    }
    \label{fig:lattice-32}
\end{figure}

We study the ground state properties of the Shastry-Sutherland model [see Eq.~(\ref{Eq:SS Hamiltonian})] by exact diagonalization,
on finite-size($2\sqrt{2}\times 2\sqrt{2}\times 4$) lattice with $32$ sites (see Fig.~\ref{fig:lattice-32}).
With periodic boundary conditions this preserves the full lattice symmetries of the Shastry-Sutherland model.

The ground state energies and wave functions are obtained by the Lanczos method.
The following lattice symmetries are exploited to reduce the Hilbert space sizes,
(a) lattice translations (with respect to 4-site unit cell);
(b) four-fold rotation $C_4$ around the center of an empty square, in the translation trivial sector.
The ground states are found in the sector with trivial translations and $C_{4}$ eigenvalue $(+1)$.
The reduced Hilbert space sizes is $18788230$ for this $32$-site lattice.

We consider antiferromagntic $J_1$ and $J_2$ couplings and set $J_2=1$.
The ground state energy and some of the ground state spin-spin correlation functions are shown in
%% need fix
Fig.~\ref{ground state properties}.
We note that for $J_1/J_2 \leq 0.67578125$ the exact ground state is the dimer-singlet state,
and a level crossing happens for $J_1/J_2$ between $0.67578125$ and $0.677734375$.
This level crossing point is consistent with previous exact diagonalization studies
%% need fix
\cite{ED}
%\cite{Wang,Ling, CPL2022}

\begin{figure}[h]
    \centering
    \begin{subfigure}[h]{0.23\textwidth}
        \includegraphics[width=\textwidth]{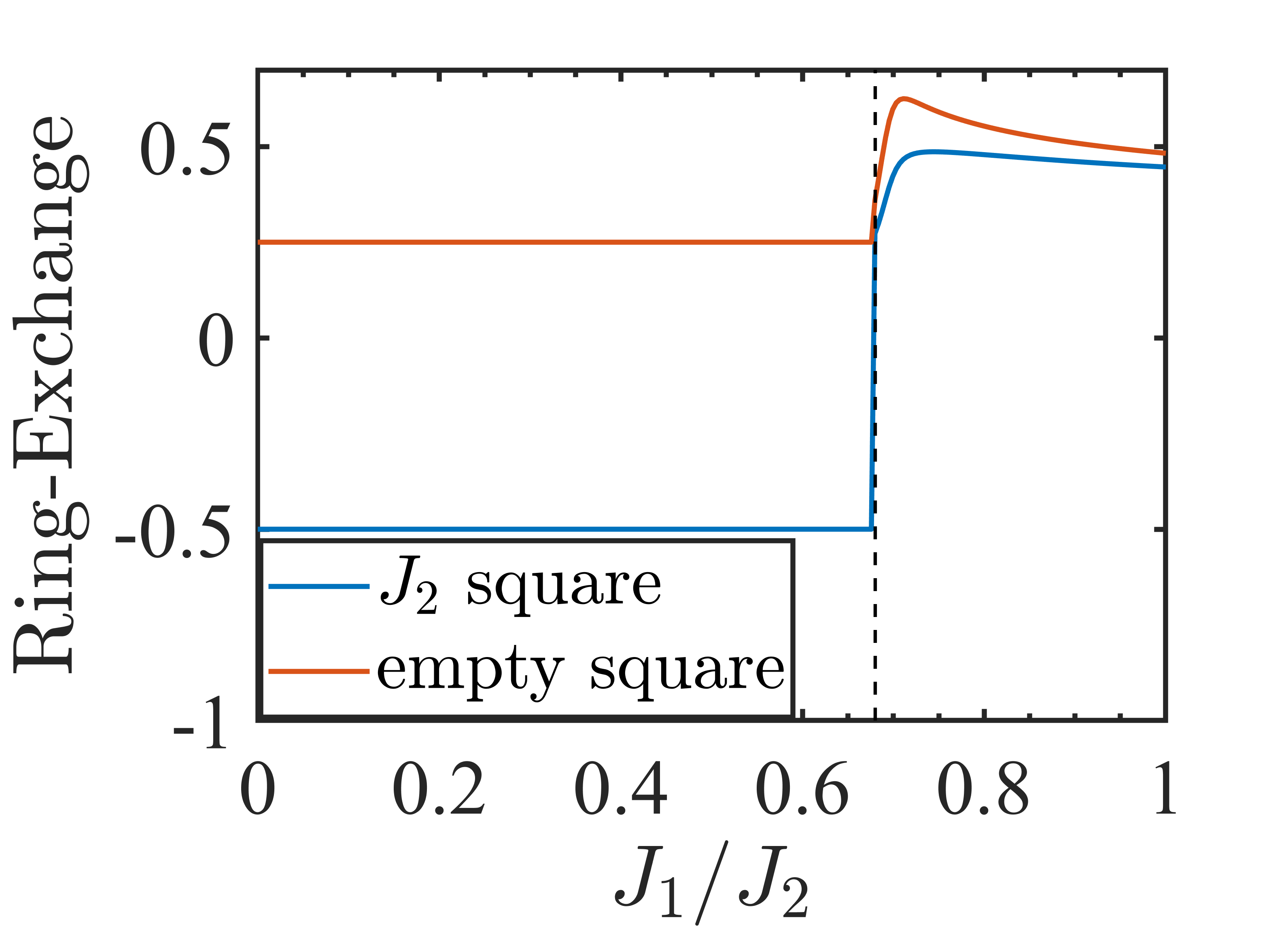}
        %\label{fig:a}
        \caption{Exact diagonalization}
    \end{subfigure}
    \begin{subfigure}[h]{0.23\textwidth}
        \includegraphics[width=\textwidth]{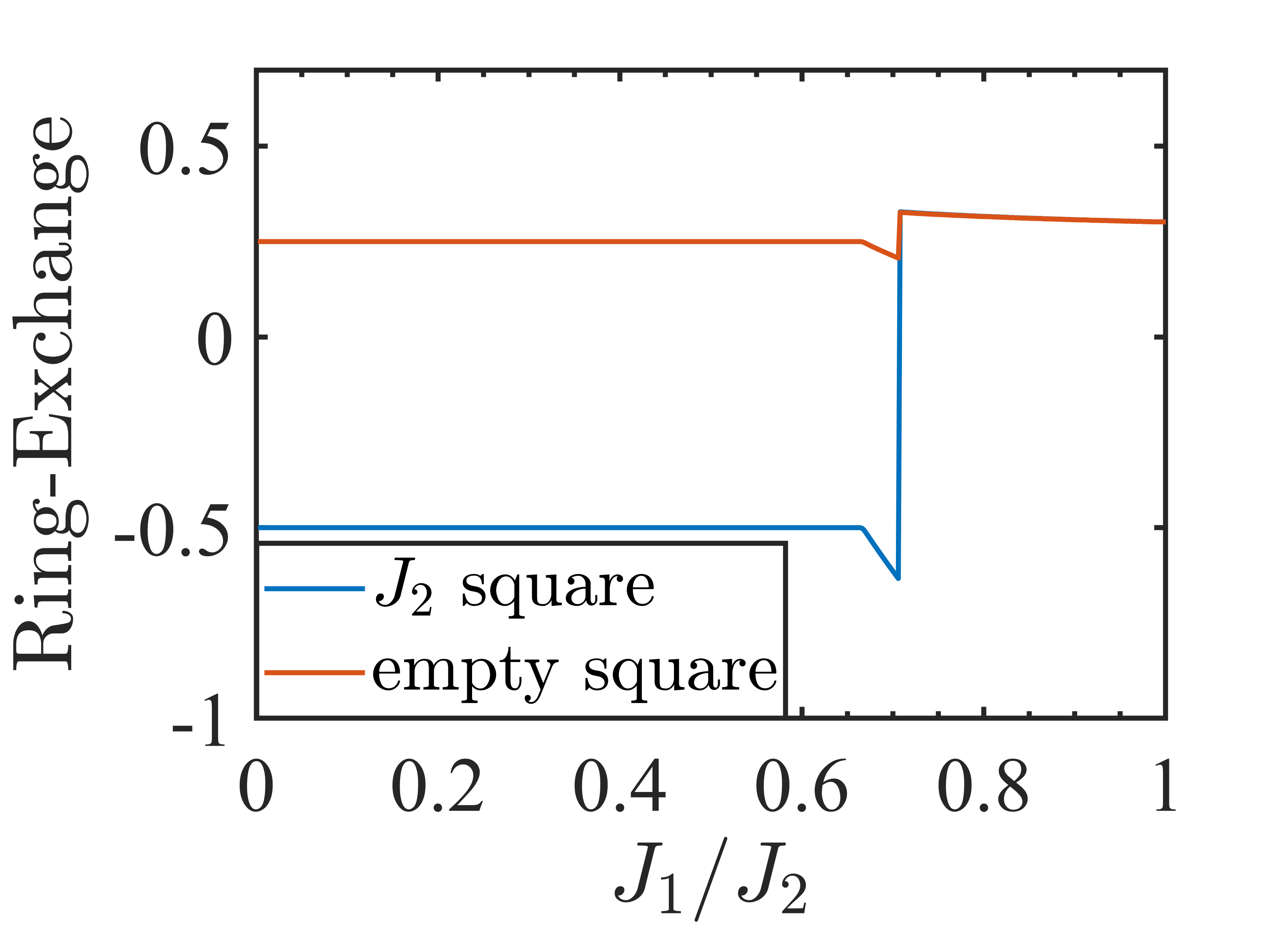}
        \caption{Schwinger boson mean field theory}
    \end{subfigure}
    \caption{
        Expectation values of 4-site ring exchange operators around ``$J_2$ square'' and ``empty square''
        calculated by (a) exact diagonalization and (b) Schwinger boson mean field theory.
    }
    \label{fig:ring-exchange}
\end{figure}

Here we also present the ground state expectation values of the 4-site ring exchange operators around ``$J_2$ square'' and ``empty square''
from the exact diagonalization, and compare them to the Schwinger boson mean-field theory results
(see Fig.~\ref{fig:ring-exchange}).
This operator cyclically permutes spins on the four sites involved,
namely it is $\sum_{ \{s_i\} } (|s_1,s_2,s_3,s_4\rangle\langle s_2,s_3,s_4,s_1|+\text{h.c.})$.

% \bibliography{}

\bibliography{cite}

\begin{thebibliography}{45}%
\makeatletter
\providecommand \@ifxundefined [1]{%
 \@ifx{#1\undefined}
}%
\providecommand \@ifnum [1]{%
 \ifnum #1\expandafter \@firstoftwo
 \else \expandafter \@secondoftwo
 \fi
}%
\providecommand \@ifx [1]{%
 \ifx #1\expandafter \@firstoftwo
 \else \expandafter \@secondoftwo
 \fi
}%
\providecommand \natexlab [1]{#1}%
\providecommand \enquote  [1]{``#1''}%
\providecommand \bibnamefont  [1]{#1}%
\providecommand \bibfnamefont [1]{#1}%
\providecommand \citenamefont [1]{#1}%
\providecommand \href@noop [0]{\@secondoftwo}%
\providecommand \href [0]{\begingroup \@sanitize@url \@href}%
\providecommand \@href[1]{\@@startlink{#1}\@@href}%
\providecommand \@@href[1]{\endgroup#1\@@endlink}%
\providecommand \@sanitize@url [0]{\catcode `\\12\catcode `\$12\catcode
  `\&12\catcode `\#12\catcode `\^12\catcode `\_12\catcode `\%12\relax}%
\providecommand \@@startlink[1]{}%
\providecommand \@@endlink[0]{}%
\providecommand \url  [0]{\begingroup\@sanitize@url \@url }%
\providecommand \@url [1]{\endgroup\@href {#1}{\urlprefix }}%
\providecommand \urlprefix  [0]{URL }%
\providecommand \Eprint [0]{\href }%
\providecommand \doibase [0]{https://doi.org/}%
\providecommand \selectlanguage [0]{\@gobble}%
\providecommand \bibinfo  [0]{\@secondoftwo}%
\providecommand \bibfield  [0]{\@secondoftwo}%
\providecommand \translation [1]{[#1]}%
\providecommand \BibitemOpen [0]{}%
\providecommand \bibitemStop [0]{}%
\providecommand \bibitemNoStop [0]{.\EOS\space}%
\providecommand \EOS [0]{\spacefactor3000\relax}%
\providecommand \BibitemShut  [1]{\csname bibitem#1\endcsname}%
\let\auto@bib@innerbib\@empty
%</preamble>
\bibitem [{\citenamefont {Cui}\ \emph {et~al.}(2023)\citenamefont {Cui},
  \citenamefont {Liu}, \citenamefont {Lin}, \citenamefont {Wu}, \citenamefont
  {Hong}, \citenamefont {Liu}, \citenamefont {Li}, \citenamefont {Hu},
  \citenamefont {Xi}, \citenamefont {Li}, \citenamefont {Yu}, \citenamefont
  {Sandvik},\ and\ \citenamefont {Yu}}]{DQCPwithChangeOffield}%
  \BibitemOpen
  \bibfield  {author} {\bibinfo {author} {\bibfnamefont {Y.}~\bibnamefont
  {Cui}}, \bibinfo {author} {\bibfnamefont {L.}~\bibnamefont {Liu}}, \bibinfo
  {author} {\bibfnamefont {H.}~\bibnamefont {Lin}}, \bibinfo {author}
  {\bibfnamefont {K.-H.}\ \bibnamefont {Wu}}, \bibinfo {author} {\bibfnamefont
  {W.}~\bibnamefont {Hong}}, \bibinfo {author} {\bibfnamefont {X.}~\bibnamefont
  {Liu}}, \bibinfo {author} {\bibfnamefont {C.}~\bibnamefont {Li}}, \bibinfo
  {author} {\bibfnamefont {Z.}~\bibnamefont {Hu}}, \bibinfo {author}
  {\bibfnamefont {N.}~\bibnamefont {Xi}}, \bibinfo {author} {\bibfnamefont
  {S.}~\bibnamefont {Li}}, \bibinfo {author} {\bibfnamefont {R.}~\bibnamefont
  {Yu}}, \bibinfo {author} {\bibfnamefont {A.~W.}\ \bibnamefont {Sandvik}},\
  and\ \bibinfo {author} {\bibfnamefont {W.}~\bibnamefont {Yu}},\ }\bibfield
  {title} {\bibinfo {title} {Proximate deconfined quantum critical point in
  ${\mathrm{srcu}}_{2}({\mathrm{bo}}_{3}{)}_{2}$},\ }\href
  {https://doi.org/10.1126/science.adc9487} {\bibfield  {journal} {\bibinfo
  {journal} {Science}\ }\textbf {\bibinfo {volume} {380}},\ \bibinfo {pages}
  {1179} (\bibinfo {year} {2023})},\ \Eprint
  {https://arxiv.org/abs/https://www.science.org/doi/pdf/10.1126/science.adc9487}
  {https://www.science.org/doi/pdf/10.1126/science.adc9487} \BibitemShut
  {NoStop}%
\bibitem [{\citenamefont {Zayed}\ \emph {et~al.}(2017)\citenamefont {Zayed},
  \citenamefont {R{\"u}egg}, \citenamefont {Larrea~J.}, \citenamefont
  {L{\"a}uchli}, \citenamefont {Panagopoulos}, \citenamefont {Saxena},
  \citenamefont {Ellerby}, \citenamefont {McMorrow}, \citenamefont
  {Str{\"a}ssle}, \citenamefont {Klotz}, \citenamefont {Hamel}, \citenamefont
  {Sadykov}, \citenamefont {Pomjakushin}, \citenamefont {Boehm}, \citenamefont
  {Jim{\'e}nez-Ruiz}, \citenamefont {Schneidewind}, \citenamefont
  {Pomjakushina}, \citenamefont {Stingaciu}, \citenamefont {Conder},\ and\
  \citenamefont {R{\o}nnow}}]{PS&DQCP-E2017}%
  \BibitemOpen
  \bibfield  {author} {\bibinfo {author} {\bibfnamefont {M.~E.}\ \bibnamefont
  {Zayed}}, \bibinfo {author} {\bibfnamefont {C.}~\bibnamefont {R{\"u}egg}},
  \bibinfo {author} {\bibfnamefont {J.}~\bibnamefont {Larrea~J.}}, \bibinfo
  {author} {\bibfnamefont {A.~M.}\ \bibnamefont {L{\"a}uchli}}, \bibinfo
  {author} {\bibfnamefont {C.}~\bibnamefont {Panagopoulos}}, \bibinfo {author}
  {\bibfnamefont {S.~S.}\ \bibnamefont {Saxena}}, \bibinfo {author}
  {\bibfnamefont {M.}~\bibnamefont {Ellerby}}, \bibinfo {author} {\bibfnamefont
  {D.~F.}\ \bibnamefont {McMorrow}}, \bibinfo {author} {\bibfnamefont
  {T.}~\bibnamefont {Str{\"a}ssle}}, \bibinfo {author} {\bibfnamefont
  {S.}~\bibnamefont {Klotz}}, \bibinfo {author} {\bibfnamefont
  {G.}~\bibnamefont {Hamel}}, \bibinfo {author} {\bibfnamefont {R.~A.}\
  \bibnamefont {Sadykov}}, \bibinfo {author} {\bibfnamefont {V.}~\bibnamefont
  {Pomjakushin}}, \bibinfo {author} {\bibfnamefont {M.}~\bibnamefont {Boehm}},
  \bibinfo {author} {\bibfnamefont {M.}~\bibnamefont {Jim{\'e}nez-Ruiz}},
  \bibinfo {author} {\bibfnamefont {A.}~\bibnamefont {Schneidewind}}, \bibinfo
  {author} {\bibfnamefont {E.}~\bibnamefont {Pomjakushina}}, \bibinfo {author}
  {\bibfnamefont {M.}~\bibnamefont {Stingaciu}}, \bibinfo {author}
  {\bibfnamefont {K.}~\bibnamefont {Conder}},\ and\ \bibinfo {author}
  {\bibfnamefont {H.~M.}\ \bibnamefont {R{\o}nnow}},\ }\bibfield  {title}
  {\bibinfo {title} {4-spin plaquette singlet state in the shastry--sutherland
  compound srcu2(bo3)2},\ }\href {https://doi.org/10.1038/nphys4190} {\bibfield
   {journal} {\bibinfo  {journal} {Nature Physics}\ }\textbf {\bibinfo {volume}
  {13}},\ \bibinfo {pages} {962} (\bibinfo {year} {2017})}\BibitemShut
  {NoStop}%
\bibitem [{\citenamefont {Guo}\ \emph {et~al.}(2020)\citenamefont {Guo},
  \citenamefont {Sun}, \citenamefont {Zhao}, \citenamefont {Wang},
  \citenamefont {Hong}, \citenamefont {Sidorov}, \citenamefont {Ma},
  \citenamefont {Wu}, \citenamefont {Li}, \citenamefont {Meng}, \citenamefont
  {Sandvik},\ and\ \citenamefont {Sun}}]{PSExperimentThermodynamics1}%
  \BibitemOpen
  \bibfield  {author} {\bibinfo {author} {\bibfnamefont {J.}~\bibnamefont
  {Guo}}, \bibinfo {author} {\bibfnamefont {G.}~\bibnamefont {Sun}}, \bibinfo
  {author} {\bibfnamefont {B.}~\bibnamefont {Zhao}}, \bibinfo {author}
  {\bibfnamefont {L.}~\bibnamefont {Wang}}, \bibinfo {author} {\bibfnamefont
  {W.}~\bibnamefont {Hong}}, \bibinfo {author} {\bibfnamefont {V.~A.}\
  \bibnamefont {Sidorov}}, \bibinfo {author} {\bibfnamefont {N.}~\bibnamefont
  {Ma}}, \bibinfo {author} {\bibfnamefont {Q.}~\bibnamefont {Wu}}, \bibinfo
  {author} {\bibfnamefont {S.}~\bibnamefont {Li}}, \bibinfo {author}
  {\bibfnamefont {Z.~Y.}\ \bibnamefont {Meng}}, \bibinfo {author}
  {\bibfnamefont {A.~W.}\ \bibnamefont {Sandvik}},\ and\ \bibinfo {author}
  {\bibfnamefont {L.}~\bibnamefont {Sun}},\ }\bibfield  {title} {\bibinfo
  {title} {Quantum phases of ${\mathrm{srcu}}_{2}({\mathrm{bo}}_{3}{)}_{2}$
  from high-pressure thermodynamics},\ }\href
  {https://doi.org/10.1103/PhysRevLett.124.206602} {\bibfield  {journal}
  {\bibinfo  {journal} {Phys. Rev. Lett.}\ }\textbf {\bibinfo {volume} {124}},\
  \bibinfo {pages} {206602} (\bibinfo {year} {2020})}\BibitemShut {NoStop}%
\bibitem [{\citenamefont {Jim{\'e}nez}\ \emph {et~al.}(2021)\citenamefont
  {Jim{\'e}nez}, \citenamefont {Crone}, \citenamefont {Fogh}, \citenamefont
  {Zayed}, \citenamefont {Lortz}, \citenamefont {Pomjakushina}, \citenamefont
  {Conder}, \citenamefont {L{\"a}uchli}, \citenamefont {Weber}, \citenamefont
  {Wessel}, \citenamefont {Honecker}, \citenamefont {Normand}, \citenamefont
  {R{\"u}egg}, \citenamefont {Corboz}, \citenamefont {R{\o}nnow},\ and\
  \citenamefont {Mila}}]{PSExperimentThermodynamics2}%
  \BibitemOpen
  \bibfield  {author} {\bibinfo {author} {\bibfnamefont {J.~L.}\ \bibnamefont
  {Jim{\'e}nez}}, \bibinfo {author} {\bibfnamefont {S.~P.~G.}\ \bibnamefont
  {Crone}}, \bibinfo {author} {\bibfnamefont {E.}~\bibnamefont {Fogh}},
  \bibinfo {author} {\bibfnamefont {M.~E.}\ \bibnamefont {Zayed}}, \bibinfo
  {author} {\bibfnamefont {R.}~\bibnamefont {Lortz}}, \bibinfo {author}
  {\bibfnamefont {E.}~\bibnamefont {Pomjakushina}}, \bibinfo {author}
  {\bibfnamefont {K.}~\bibnamefont {Conder}}, \bibinfo {author} {\bibfnamefont
  {A.~M.}\ \bibnamefont {L{\"a}uchli}}, \bibinfo {author} {\bibfnamefont
  {L.}~\bibnamefont {Weber}}, \bibinfo {author} {\bibfnamefont
  {S.}~\bibnamefont {Wessel}}, \bibinfo {author} {\bibfnamefont
  {A.}~\bibnamefont {Honecker}}, \bibinfo {author} {\bibfnamefont
  {B.}~\bibnamefont {Normand}}, \bibinfo {author} {\bibfnamefont
  {C.}~\bibnamefont {R{\"u}egg}}, \bibinfo {author} {\bibfnamefont
  {P.}~\bibnamefont {Corboz}}, \bibinfo {author} {\bibfnamefont {H.~M.}\
  \bibnamefont {R{\o}nnow}},\ and\ \bibinfo {author} {\bibfnamefont
  {F.}~\bibnamefont {Mila}},\ }\bibfield  {title} {\bibinfo {title} {A quantum
  magnetic analogue to the critical point of water},\ }\href
  {https://doi.org/10.1038/s41586-021-03411-8} {\bibfield  {journal} {\bibinfo
  {journal} {Nature}\ }\textbf {\bibinfo {volume} {592}},\ \bibinfo {pages}
  {370} (\bibinfo {year} {2021})}\BibitemShut {NoStop}%
\bibitem [{\citenamefont {Senthil}\ \emph {et~al.}(2004)\citenamefont
  {Senthil}, \citenamefont {Vishwanath}, \citenamefont {Balents}, \citenamefont
  {Sachdev},\ and\ \citenamefont {Fisher}}]{DQCP1}%
  \BibitemOpen
  \bibfield  {author} {\bibinfo {author} {\bibfnamefont {T.}~\bibnamefont
  {Senthil}}, \bibinfo {author} {\bibfnamefont {A.}~\bibnamefont {Vishwanath}},
  \bibinfo {author} {\bibfnamefont {L.}~\bibnamefont {Balents}}, \bibinfo
  {author} {\bibfnamefont {S.}~\bibnamefont {Sachdev}},\ and\ \bibinfo {author}
  {\bibfnamefont {M.~P.~A.}\ \bibnamefont {Fisher}},\ }\bibfield  {title}
  {\bibinfo {title} {Deconfined quantum critical points},\ }\href
  {https://doi.org/10.1126/science.1091806} {\bibfield  {journal} {\bibinfo
  {journal} {Science}\ }\textbf {\bibinfo {volume} {303}},\ \bibinfo {pages}
  {1490} (\bibinfo {year} {2004})},\ \Eprint
  {https://arxiv.org/abs/https://www.science.org/doi/pdf/10.1126/science.1091806}
  {https://www.science.org/doi/pdf/10.1126/science.1091806} \BibitemShut
  {NoStop}%
\bibitem [{\citenamefont {Sachdev}(2008)}]{DQCP2}%
  \BibitemOpen
  \bibfield  {author} {\bibinfo {author} {\bibfnamefont {S.}~\bibnamefont
  {Sachdev}},\ }\bibfield  {title} {\bibinfo {title} {Quantum magnetism and
  criticality},\ }\href {https://doi.org/10.1038/nphys894} {\bibfield
  {journal} {\bibinfo  {journal} {Nature Physics}\ }\textbf {\bibinfo {volume}
  {4}},\ \bibinfo {pages} {173} (\bibinfo {year} {2008})}\BibitemShut {NoStop}%
\bibitem [{\citenamefont {Sandvik}(2007)}]{DQCP3}%
  \BibitemOpen
  \bibfield  {author} {\bibinfo {author} {\bibfnamefont {A.~W.}\ \bibnamefont
  {Sandvik}},\ }\bibfield  {title} {\bibinfo {title} {Evidence for deconfined
  quantum criticality in a two-dimensional heisenberg model with four-spin
  interactions},\ }\href {https://doi.org/10.1103/PhysRevLett.98.227202}
  {\bibfield  {journal} {\bibinfo  {journal} {Phys. Rev. Lett.}\ }\textbf
  {\bibinfo {volume} {98}},\ \bibinfo {pages} {227202} (\bibinfo {year}
  {2007})}\BibitemShut {NoStop}%
\bibitem [{\citenamefont {Xi}\ \emph {et~al.}(2023)\citenamefont {Xi},
  \citenamefont {Chen}, \citenamefont {Xie},\ and\ \citenamefont {Yu}}]{DQCP4}%
  \BibitemOpen
  \bibfield  {author} {\bibinfo {author} {\bibfnamefont {N.}~\bibnamefont
  {Xi}}, \bibinfo {author} {\bibfnamefont {H.}~\bibnamefont {Chen}}, \bibinfo
  {author} {\bibfnamefont {Z.~Y.}\ \bibnamefont {Xie}},\ and\ \bibinfo {author}
  {\bibfnamefont {R.}~\bibnamefont {Yu}},\ }\bibfield  {title} {\bibinfo
  {title} {Plaquette valence bond solid to antiferromagnet transition and
  deconfined quantum critical point of the shastry-sutherland model},\ }\href
  {https://doi.org/10.1103/PhysRevB.107.L220408} {\bibfield  {journal}
  {\bibinfo  {journal} {Phys. Rev. B}\ }\textbf {\bibinfo {volume} {107}},\
  \bibinfo {pages} {L220408} (\bibinfo {year} {2023})}\BibitemShut {NoStop}%
\bibitem [{\citenamefont {Balents}(2010)}]{QSL}%
  \BibitemOpen
  \bibfield  {author} {\bibinfo {author} {\bibfnamefont {L.}~\bibnamefont
  {Balents}},\ }\bibfield  {title} {\bibinfo {title} {Spin liquids in
  frustrated magnets},\ }\href {https://doi.org/10.1038/nature08917} {\bibfield
   {journal} {\bibinfo  {journal} {Nature}\ }\textbf {\bibinfo {volume}
  {464}},\ \bibinfo {pages} {199} (\bibinfo {year} {2010})}\BibitemShut
  {NoStop}%
\bibitem [{\citenamefont {{Sriram Shastry}}\ and\ \citenamefont
  {Sutherland}(1981)}]{Shastry-Sutherland}%
  \BibitemOpen
  \bibfield  {author} {\bibinfo {author} {\bibfnamefont {B.}~\bibnamefont
  {{Sriram Shastry}}}\ and\ \bibinfo {author} {\bibfnamefont {B.}~\bibnamefont
  {Sutherland}},\ }\bibfield  {title} {\bibinfo {title} {Exact ground state of
  a quantum mechanical antiferromagnet},\ }\href
  {https://doi.org/https://doi.org/10.1016/0378-4363(81)90838-X} {\bibfield
  {journal} {\bibinfo  {journal} {Physica B+C}\ }\textbf {\bibinfo {volume}
  {108}},\ \bibinfo {pages} {1069} (\bibinfo {year} {1981})}\BibitemShut
  {NoStop}%
\bibitem [{\citenamefont {Nakano}\ and\ \citenamefont
  {Sakai}(2018)}]{EDforPhaseBoundaryOfDSandNeel}%
  \BibitemOpen
  \bibfield  {author} {\bibinfo {author} {\bibfnamefont {H.}~\bibnamefont
  {Nakano}}\ and\ \bibinfo {author} {\bibfnamefont {T.}~\bibnamefont {Sakai}},\
  }\bibfield  {title} {\bibinfo {title} {Third boundary of the
  shastry–sutherland model by numerical diagonalization},\ }\href
  {https://doi.org/10.7566/JPSJ.87.123702} {\bibfield  {journal} {\bibinfo
  {journal} {Journal of the Physical Society of Japan}\ }\textbf {\bibinfo
  {volume} {87}},\ \bibinfo {pages} {123702} (\bibinfo {year} {2018})},\
  \Eprint {https://arxiv.org/abs/https://doi.org/10.7566/JPSJ.87.123702}
  {https://doi.org/10.7566/JPSJ.87.123702} \BibitemShut {NoStop}%
\bibitem [{\citenamefont {Koga}\ and\ \citenamefont
  {Kawakami}(2000)}]{PhaseTransitionDS-PS-Neel1}%
  \BibitemOpen
  \bibfield  {author} {\bibinfo {author} {\bibfnamefont {A.}~\bibnamefont
  {Koga}}\ and\ \bibinfo {author} {\bibfnamefont {N.}~\bibnamefont
  {Kawakami}},\ }\bibfield  {title} {\bibinfo {title} {Quantum phase
  transitions in the shastry-sutherland model for
  ${\mathrm{srcu}}_{2}({\mathrm{bo}}_{3}{)}_{2}$},\ }\href
  {https://doi.org/10.1103/PhysRevLett.84.4461} {\bibfield  {journal} {\bibinfo
   {journal} {Phys. Rev. Lett.}\ }\textbf {\bibinfo {volume} {84}},\ \bibinfo
  {pages} {4461} (\bibinfo {year} {2000})}\BibitemShut {NoStop}%
\bibitem [{\citenamefont {Zhao}\ \emph {et~al.}(2019)\citenamefont {Zhao},
  \citenamefont {Weinberg},\ and\ \citenamefont
  {Sandvik}}]{DQCPandEmergentO4inPSandNeel}%
  \BibitemOpen
  \bibfield  {author} {\bibinfo {author} {\bibfnamefont {B.}~\bibnamefont
  {Zhao}}, \bibinfo {author} {\bibfnamefont {P.}~\bibnamefont {Weinberg}},\
  and\ \bibinfo {author} {\bibfnamefont {A.~W.}\ \bibnamefont {Sandvik}},\
  }\bibfield  {title} {\bibinfo {title} {Symmetry-enhanced discontinuous phase
  transition in a two-dimensional quantum magnet},\ }\href
  {https://doi.org/10.1038/s41567-019-0484-x} {\bibfield  {journal} {\bibinfo
  {journal} {Nature Physics}\ }\textbf {\bibinfo {volume} {15}},\ \bibinfo
  {pages} {678} (\bibinfo {year} {2019})}\BibitemShut {NoStop}%
\bibitem [{\citenamefont {Corboz}\ and\ \citenamefont
  {Mila}(2013)}]{PSbyTensorNetwork}%
  \BibitemOpen
  \bibfield  {author} {\bibinfo {author} {\bibfnamefont {P.}~\bibnamefont
  {Corboz}}\ and\ \bibinfo {author} {\bibfnamefont {F.}~\bibnamefont {Mila}},\
  }\bibfield  {title} {\bibinfo {title} {Tensor network study of the
  shastry-sutherland model in zero magnetic field},\ }\href
  {https://doi.org/10.1103/PhysRevB.87.115144} {\bibfield  {journal} {\bibinfo
  {journal} {Phys. Rev. B}\ }\textbf {\bibinfo {volume} {87}},\ \bibinfo
  {pages} {115144} (\bibinfo {year} {2013})}\BibitemShut {NoStop}%
\bibitem [{\citenamefont {Weihong}\ \emph {et~al.}(1999)\citenamefont
  {Weihong}, \citenamefont {Hamer},\ and\ \citenamefont
  {Oitmaa}}]{PhaseTransitionDS-Neel1}%
  \BibitemOpen
  \bibfield  {author} {\bibinfo {author} {\bibfnamefont {Z.}~\bibnamefont
  {Weihong}}, \bibinfo {author} {\bibfnamefont {C.~J.}\ \bibnamefont {Hamer}},\
  and\ \bibinfo {author} {\bibfnamefont {J.}~\bibnamefont {Oitmaa}},\
  }\bibfield  {title} {\bibinfo {title} {Series expansions for a heisenberg
  antiferromagnetic model for ${\mathrm{srcu}}_{2}({\mathrm{bo}}_{3}{)}_{2}$},\
  }\href {https://doi.org/10.1103/PhysRevB.60.6608} {\bibfield  {journal}
  {\bibinfo  {journal} {Phys. Rev. B}\ }\textbf {\bibinfo {volume} {60}},\
  \bibinfo {pages} {6608} (\bibinfo {year} {1999})}\BibitemShut {NoStop}%
\bibitem [{\citenamefont {Takushima}\ \emph {et~al.}(2001)\citenamefont
  {Takushima}, \citenamefont {Koga},\ and\ \citenamefont
  {Kawakami}}]{PhaseTransitionDS-PS-Neel2}%
  \BibitemOpen
  \bibfield  {author} {\bibinfo {author} {\bibfnamefont {Y.}~\bibnamefont
  {Takushima}}, \bibinfo {author} {\bibfnamefont {A.}~\bibnamefont {Koga}},\
  and\ \bibinfo {author} {\bibfnamefont {N.}~\bibnamefont {Kawakami}},\
  }\bibfield  {title} {\bibinfo {title} {Competing spin-gap phases in a
  frustrated quantum spin system in two dimensions},\ }\href
  {https://doi.org/10.1143/JPSJ.70.1369} {\bibfield  {journal} {\bibinfo
  {journal} {Journal of the Physical Society of Japan}\ }\textbf {\bibinfo
  {volume} {70}},\ \bibinfo {pages} {1369} (\bibinfo {year} {2001})},\ \Eprint
  {https://arxiv.org/abs/https://doi.org/10.1143/JPSJ.70.1369}
  {https://doi.org/10.1143/JPSJ.70.1369} \BibitemShut {NoStop}%
\bibitem [{\citenamefont {{Albrecht, M.}}\ and\ \citenamefont {{Mila,
  F.}}(1996)}]{PhaseTransitionDS-Helix-Neel}%
  \BibitemOpen
  \bibfield  {author} {\bibinfo {author} {\bibnamefont {{Albrecht, M.}}}\ and\
  \bibinfo {author} {\bibnamefont {{Mila, F.}}},\ }\bibfield  {title} {\bibinfo
  {title} {First-order transition between magnetic order and valence bond order
  in a 2d frustrated heisenberg model},\ }\href
  {https://doi.org/10.1209/epl/i1996-00430-0} {\bibfield  {journal} {\bibinfo
  {journal} {Europhys. Lett.}\ }\textbf {\bibinfo {volume} {34}},\ \bibinfo
  {pages} {145} (\bibinfo {year} {1996})}\BibitemShut {NoStop}%
\bibitem [{\citenamefont {Miyahara}\ and\ \citenamefont
  {Ueda}(2003)}]{SCBOTheoryHelixAndPS}%
  \BibitemOpen
  \bibfield  {author} {\bibinfo {author} {\bibfnamefont {S.}~\bibnamefont
  {Miyahara}}\ and\ \bibinfo {author} {\bibfnamefont {K.}~\bibnamefont
  {Ueda}},\ }\bibfield  {title} {\bibinfo {title} {Theory of the orthogonal
  dimer heisenberg spin model for srcu2 (bo3)2},\ }\href
  {https://doi.org/10.1088/0953-8984/15/9/201} {\bibfield  {journal} {\bibinfo
  {journal} {Journal of Physics: Condensed Matter}\ }\textbf {\bibinfo {volume}
  {15}},\ \bibinfo {pages} {R327} (\bibinfo {year} {2003})}\BibitemShut
  {NoStop}%
\bibitem [{\citenamefont {Zheng}\ \emph {et~al.}(2001)\citenamefont {Zheng},
  \citenamefont {Oitmaa},\ and\ \citenamefont {Hamer}}]{DS-ColumnarDimer-Neel}%
  \BibitemOpen
  \bibfield  {author} {\bibinfo {author} {\bibfnamefont {W.}~\bibnamefont
  {Zheng}}, \bibinfo {author} {\bibfnamefont {J.}~\bibnamefont {Oitmaa}},\ and\
  \bibinfo {author} {\bibfnamefont {C.~J.}\ \bibnamefont {Hamer}},\ }\bibfield
  {title} {\bibinfo {title} {Phase diagram of the shastry-sutherland
  antiferromagnet},\ }\href {https://doi.org/10.1103/PhysRevB.65.014408}
  {\bibfield  {journal} {\bibinfo  {journal} {Phys. Rev. B}\ }\textbf {\bibinfo
  {volume} {65}},\ \bibinfo {pages} {014408} (\bibinfo {year}
  {2001})}\BibitemShut {NoStop}%
\bibitem [{\citenamefont {Lee}\ \emph {et~al.}(2019)\citenamefont {Lee},
  \citenamefont {You}, \citenamefont {Sachdev},\ and\ \citenamefont
  {Vishwanath}}]{PS&DQCPbyIDMRG2019}%
  \BibitemOpen
  \bibfield  {author} {\bibinfo {author} {\bibfnamefont {J.~Y.}\ \bibnamefont
  {Lee}}, \bibinfo {author} {\bibfnamefont {Y.-Z.}\ \bibnamefont {You}},
  \bibinfo {author} {\bibfnamefont {S.}~\bibnamefont {Sachdev}},\ and\ \bibinfo
  {author} {\bibfnamefont {A.}~\bibnamefont {Vishwanath}},\ }\bibfield  {title}
  {\bibinfo {title} {Signatures of a deconfined phase transition on the
  shastry-sutherland lattice: Applications to quantum critical
  ${\mathrm{srcu}}_{2}({\mathrm{bo}}_{3}{)}_{2}$},\ }\href
  {https://doi.org/10.1103/PhysRevX.9.041037} {\bibfield  {journal} {\bibinfo
  {journal} {Phys. Rev. X}\ }\textbf {\bibinfo {volume} {9}},\ \bibinfo {pages}
  {041037} (\bibinfo {year} {2019})}\BibitemShut {NoStop}%
\bibitem [{\citenamefont {L\"auchli}\ \emph {et~al.}(2002)\citenamefont
  {L\"auchli}, \citenamefont {Wessel},\ and\ \citenamefont
  {Sigrist}}]{PSIndeucedByDifferentPlaqutte}%
  \BibitemOpen
  \bibfield  {author} {\bibinfo {author} {\bibfnamefont {A.}~\bibnamefont
  {L\"auchli}}, \bibinfo {author} {\bibfnamefont {S.}~\bibnamefont {Wessel}},\
  and\ \bibinfo {author} {\bibfnamefont {M.}~\bibnamefont {Sigrist}},\
  }\bibfield  {title} {\bibinfo {title} {Phase diagram of the quadrumerized
  shastry-sutherland model},\ }\href
  {https://doi.org/10.1103/PhysRevB.66.014401} {\bibfield  {journal} {\bibinfo
  {journal} {Phys. Rev. B}\ }\textbf {\bibinfo {volume} {66}},\ \bibinfo
  {pages} {014401} (\bibinfo {year} {2002})}\BibitemShut {NoStop}%
\bibitem [{\citenamefont {Boos}\ \emph {et~al.}(2019)\citenamefont {Boos},
  \citenamefont {Crone}, \citenamefont {Niesen}, \citenamefont {Corboz},
  \citenamefont {Schmidt},\ and\ \citenamefont
  {Mila}}]{PSIndeucedByDifferentPlaqutte2}%
  \BibitemOpen
  \bibfield  {author} {\bibinfo {author} {\bibfnamefont {C.}~\bibnamefont
  {Boos}}, \bibinfo {author} {\bibfnamefont {S.~P.~G.}\ \bibnamefont {Crone}},
  \bibinfo {author} {\bibfnamefont {I.~A.}\ \bibnamefont {Niesen}}, \bibinfo
  {author} {\bibfnamefont {P.}~\bibnamefont {Corboz}}, \bibinfo {author}
  {\bibfnamefont {K.~P.}\ \bibnamefont {Schmidt}},\ and\ \bibinfo {author}
  {\bibfnamefont {F.}~\bibnamefont {Mila}},\ }\bibfield  {title} {\bibinfo
  {title} {Competition between intermediate plaquette phases in
  ${\mathrm{srcu}}_{2}$(${\mathrm{bo}}_{3}{)}_{2}$ under pressure},\ }\href
  {https://doi.org/10.1103/PhysRevB.100.140413} {\bibfield  {journal} {\bibinfo
   {journal} {Phys. Rev. B}\ }\textbf {\bibinfo {volume} {100}},\ \bibinfo
  {pages} {140413} (\bibinfo {year} {2019})}\BibitemShut {NoStop}%
\bibitem [{\citenamefont {Wang}\ \emph {et~al.}(2023)\citenamefont {Wang},
  \citenamefont {Li}, \citenamefont {Xi}, \citenamefont {Gao}, \citenamefont
  {Yan}, \citenamefont {Li},\ and\ \citenamefont {Su}}]{wang2023plaquette}%
  \BibitemOpen
  \bibfield  {author} {\bibinfo {author} {\bibfnamefont {J.}~\bibnamefont
  {Wang}}, \bibinfo {author} {\bibfnamefont {H.}~\bibnamefont {Li}}, \bibinfo
  {author} {\bibfnamefont {N.}~\bibnamefont {Xi}}, \bibinfo {author}
  {\bibfnamefont {Y.}~\bibnamefont {Gao}}, \bibinfo {author} {\bibfnamefont
  {Q.-B.}\ \bibnamefont {Yan}}, \bibinfo {author} {\bibfnamefont
  {W.}~\bibnamefont {Li}},\ and\ \bibinfo {author} {\bibfnamefont
  {G.}~\bibnamefont {Su}},\ }\href@noop {} {\bibinfo {title} {Plaquette singlet
  transition, magnetic barocaloric effect, and spin supersolidity in the
  shastry-sutherland model}} (\bibinfo {year} {2023}),\ \Eprint
  {https://arxiv.org/abs/2302.06596} {arXiv:2302.06596 [cond-mat.str-el]}
  \BibitemShut {NoStop}%
\bibitem [{\citenamefont {Yang}\ \emph {et~al.}(2022)\citenamefont {Yang},
  \citenamefont {Sandvik},\ and\ \citenamefont {Wang}}]{SLbyDMRG}%
  \BibitemOpen
  \bibfield  {author} {\bibinfo {author} {\bibfnamefont {J.}~\bibnamefont
  {Yang}}, \bibinfo {author} {\bibfnamefont {A.~W.}\ \bibnamefont {Sandvik}},\
  and\ \bibinfo {author} {\bibfnamefont {L.}~\bibnamefont {Wang}},\ }\bibfield
  {title} {\bibinfo {title} {Quantum criticality and spin liquid phase in the
  shastry-sutherland model},\ }\href
  {https://doi.org/10.1103/PhysRevB.105.L060409} {\bibfield  {journal}
  {\bibinfo  {journal} {Phys. Rev. B}\ }\textbf {\bibinfo {volume} {105}},\
  \bibinfo {pages} {L060409} (\bibinfo {year} {2022})}\BibitemShut {NoStop}%
\bibitem [{\citenamefont {Wang}\ \emph
  {et~al.}(2022{\natexlab{a}})\citenamefont {Wang}, \citenamefont {Zhang},\
  and\ \citenamefont {Sandvik}}]{SLbyED}%
  \BibitemOpen
  \bibfield  {author} {\bibinfo {author} {\bibfnamefont {L.}~\bibnamefont
  {Wang}}, \bibinfo {author} {\bibfnamefont {Y.}~\bibnamefont {Zhang}},\ and\
  \bibinfo {author} {\bibfnamefont {A.~W.}\ \bibnamefont {Sandvik}},\
  }\bibfield  {title} {\bibinfo {title} {Quantum spin liquid phase in the
  shastry-sutherland model detected by an improved level spectroscopic
  method},\ }\href {https://doi.org/10.1088/0256-307X/39/7/077502} {\bibfield
  {journal} {\bibinfo  {journal} {Chinese Physics Letters}\ }\textbf {\bibinfo
  {volume} {39}},\ \bibinfo {pages} {077502} (\bibinfo {year}
  {2022}{\natexlab{a}})}\BibitemShut {NoStop}%
\bibitem [{\citenamefont {Kele\ifmmode~\mbox{\c{s}}\else \c{s}\fi{}}\ and\
  \citenamefont {Zhao}(2022)}]{PhysRevB.105.L041115}%
  \BibitemOpen
  \bibfield  {author} {\bibinfo {author} {\bibfnamefont {A.}~\bibnamefont
  {Kele\ifmmode~\mbox{\c{s}}\else \c{s}\fi{}}}\ and\ \bibinfo {author}
  {\bibfnamefont {E.}~\bibnamefont {Zhao}},\ }\bibfield  {title} {\bibinfo
  {title} {Rise and fall of plaquette order in the shastry-sutherland magnet
  revealed by pseudofermion functional renormalization group},\ }\href
  {https://doi.org/10.1103/PhysRevB.105.L041115} {\bibfield  {journal}
  {\bibinfo  {journal} {Phys. Rev. B}\ }\textbf {\bibinfo {volume} {105}},\
  \bibinfo {pages} {L041115} (\bibinfo {year} {2022})}\BibitemShut {NoStop}%
\bibitem [{\citenamefont {Zhou}\ \emph {et~al.}(2017)\citenamefont {Zhou},
  \citenamefont {Kanoda},\ and\ \citenamefont {Ng}}]{RevModPhys.89.025003}%
  \BibitemOpen
  \bibfield  {author} {\bibinfo {author} {\bibfnamefont {Y.}~\bibnamefont
  {Zhou}}, \bibinfo {author} {\bibfnamefont {K.}~\bibnamefont {Kanoda}},\ and\
  \bibinfo {author} {\bibfnamefont {T.-K.}\ \bibnamefont {Ng}},\ }\bibfield
  {title} {\bibinfo {title} {Quantum spin liquid states},\ }\href
  {https://doi.org/10.1103/RevModPhys.89.025003} {\bibfield  {journal}
  {\bibinfo  {journal} {Rev. Mod. Phys.}\ }\textbf {\bibinfo {volume} {89}},\
  \bibinfo {pages} {025003} (\bibinfo {year} {2017})}\BibitemShut {NoStop}%
\bibitem [{\citenamefont {Sachdev}(1992)}]{spin-liquid-boson}%
  \BibitemOpen
  \bibfield  {author} {\bibinfo {author} {\bibfnamefont {S.}~\bibnamefont
  {Sachdev}},\ }\bibfield  {title} {\bibinfo {title}
  {Kagome\ifmmode\acute\else\textasciiacute\fi{}- and triangular-lattice
  heisenberg antiferromagnets: Ordering from quantum fluctuations and
  quantum-disordered ground states with unconfined bosonic spinons},\ }\href
  {https://doi.org/10.1103/PhysRevB.45.12377} {\bibfield  {journal} {\bibinfo
  {journal} {Phys. Rev. B}\ }\textbf {\bibinfo {volume} {45}},\ \bibinfo
  {pages} {12377} (\bibinfo {year} {1992})}\BibitemShut {NoStop}%
\bibitem [{\citenamefont {Arovas}\ and\ \citenamefont
  {Auerbach}(1988)}]{PhysRevB.38.316}%
  \BibitemOpen
  \bibfield  {author} {\bibinfo {author} {\bibfnamefont {D.~P.}\ \bibnamefont
  {Arovas}}\ and\ \bibinfo {author} {\bibfnamefont {A.}~\bibnamefont
  {Auerbach}},\ }\bibfield  {title} {\bibinfo {title} {Functional integral
  theories of low-dimensional quantum heisenberg models},\ }\href
  {https://doi.org/10.1103/PhysRevB.38.316} {\bibfield  {journal} {\bibinfo
  {journal} {Phys. Rev. B}\ }\textbf {\bibinfo {volume} {38}},\ \bibinfo
  {pages} {316} (\bibinfo {year} {1988})}\BibitemShut {NoStop}%
\bibitem [{\citenamefont {Read}\ and\ \citenamefont
  {Sachdev}(1991)}]{PhysRevLett.66.1773}%
  \BibitemOpen
  \bibfield  {author} {\bibinfo {author} {\bibfnamefont {N.}~\bibnamefont
  {Read}}\ and\ \bibinfo {author} {\bibfnamefont {S.}~\bibnamefont {Sachdev}},\
  }\bibfield  {title} {\bibinfo {title} {Large-n expansion for frustrated
  quantum antiferromagnets},\ }\href
  {https://doi.org/10.1103/PhysRevLett.66.1773} {\bibfield  {journal} {\bibinfo
   {journal} {Phys. Rev. Lett.}\ }\textbf {\bibinfo {volume} {66}},\ \bibinfo
  {pages} {1773} (\bibinfo {year} {1991})}\BibitemShut {NoStop}%
\bibitem [{\citenamefont {Auerbach}\ and\ \citenamefont
  {Arovas}(2011)}]{Ch14-IntroFrusMag}%
  \BibitemOpen
  \bibfield  {author} {\bibinfo {author} {\bibfnamefont {A.}~\bibnamefont
  {Auerbach}}\ and\ \bibinfo {author} {\bibfnamefont {D.~P.}\ \bibnamefont
  {Arovas}},\ }\bibinfo {title} {Schwinger boson approaches to quantum
  antiferromagnetism},\ in\ \href {https://doi.org/10.1007/978-3-642-10589-0}
  {\emph {\bibinfo {booktitle} {Introduction to Frustrated Magnetism}}},\
  \bibinfo {series and number} {Springer Series in Solid State Sciences},\
  \bibinfo {editor} {edited by\ \bibinfo {editor} {\bibfnamefont
  {C.}~\bibnamefont {Lacroix}}, \bibinfo {editor} {\bibfnamefont
  {P.}~\bibnamefont {Mendels}},\ and\ \bibinfo {editor} {\bibfnamefont
  {F.}~\bibnamefont {Mila}}}\ (\bibinfo  {publisher} {Springer Berlin,
  Heidelberg},\ \bibinfo {year} {2011})\ Chap.~\bibinfo {chapter} {14}, pp.\
  \bibinfo {pages} {365--378}\BibitemShut {NoStop}%
\bibitem [{\citenamefont {Wen}(2002)}]{PSG-fermion}%
  \BibitemOpen
  \bibfield  {author} {\bibinfo {author} {\bibfnamefont {X.-G.}\ \bibnamefont
  {Wen}},\ }\bibfield  {title} {\bibinfo {title} {Quantum orders and symmetric
  spin liquids},\ }\href {https://doi.org/10.1103/PhysRevB.65.165113}
  {\bibfield  {journal} {\bibinfo  {journal} {Phys. Rev. B}\ }\textbf {\bibinfo
  {volume} {65}},\ \bibinfo {pages} {165113} (\bibinfo {year}
  {2002})}\BibitemShut {NoStop}%
\bibitem [{\citenamefont {Wang}\ and\ \citenamefont
  {Vishwanath}(2006)}]{PSG-boson}%
  \BibitemOpen
  \bibfield  {author} {\bibinfo {author} {\bibfnamefont {F.}~\bibnamefont
  {Wang}}\ and\ \bibinfo {author} {\bibfnamefont {A.}~\bibnamefont
  {Vishwanath}},\ }\bibfield  {title} {\bibinfo {title} {Spin-liquid states on
  the triangular and kagom\'e lattices: A projective-symmetry-group analysis of
  schwinger boson states},\ }\href {https://doi.org/10.1103/PhysRevB.74.174423}
  {\bibfield  {journal} {\bibinfo  {journal} {Phys. Rev. B}\ }\textbf {\bibinfo
  {volume} {74}},\ \bibinfo {pages} {174423} (\bibinfo {year}
  {2006})}\BibitemShut {NoStop}%
\bibitem [{\citenamefont {{Tchernyshyov, O.}}\ \emph
  {et~al.}(2006)\citenamefont {{Tchernyshyov, O.}}, \citenamefont {{Moessner,
  R.}},\ and\ \citenamefont {{Sondhi, S. L.}}}]{TchernyshyovEPL06}%
  \BibitemOpen
  \bibfield  {author} {\bibinfo {author} {\bibnamefont {{Tchernyshyov, O.}}},
  \bibinfo {author} {\bibnamefont {{Moessner, R.}}},\ and\ \bibinfo {author}
  {\bibnamefont {{Sondhi, S. L.}}},\ }\bibfield  {title} {\bibinfo {title}
  {Flux expulsion and greedy bosons: Frustrated magnets at large n},\ }\href
  {https://doi.org/10.1209/epl/i2005-10389-2} {\bibfield  {journal} {\bibinfo
  {journal} {Europhys. Lett.}\ }\textbf {\bibinfo {volume} {73}},\ \bibinfo
  {pages} {278} (\bibinfo {year} {2006})}\BibitemShut {NoStop}%
\bibitem [{\citenamefont {Polyakov}(1977)}]{confinement}%
  \BibitemOpen
  \bibfield  {author} {\bibinfo {author} {\bibfnamefont {A.}~\bibnamefont
  {Polyakov}},\ }\bibfield  {title} {\bibinfo {title} {Quark confinement and
  topology of gauge theories},\ }\href
  {https://doi.org/https://doi.org/10.1016/0550-3213(77)90086-4} {\bibfield
  {journal} {\bibinfo  {journal} {Nuclear Physics B}\ }\textbf {\bibinfo
  {volume} {120}},\ \bibinfo {pages} {429} (\bibinfo {year}
  {1977})}\BibitemShut {NoStop}%
\bibitem [{\citenamefont {Yang}\ and\ \citenamefont
  {Wang}(2016)}]{piFluxOrder}%
  \BibitemOpen
  \bibfield  {author} {\bibinfo {author} {\bibfnamefont {X.}~\bibnamefont
  {Yang}}\ and\ \bibinfo {author} {\bibfnamefont {F.}~\bibnamefont {Wang}},\
  }\bibfield  {title} {\bibinfo {title} {Schwinger boson spin-liquid states on
  square lattice},\ }\href {https://doi.org/10.1103/PhysRevB.94.035160}
  {\bibfield  {journal} {\bibinfo  {journal} {Phys. Rev. B}\ }\textbf {\bibinfo
  {volume} {94}},\ \bibinfo {pages} {035160} (\bibinfo {year}
  {2016})}\BibitemShut {NoStop}%
\bibitem [{\citenamefont {Takano}\ \emph {et~al.}(2011)\citenamefont {Takano},
  \citenamefont {Tsunetsugu},\ and\ \citenamefont {Zhitomirsky}}]{SCSW1}%
  \BibitemOpen
  \bibfield  {author} {\bibinfo {author} {\bibfnamefont {J.}~\bibnamefont
  {Takano}}, \bibinfo {author} {\bibfnamefont {H.}~\bibnamefont {Tsunetsugu}},\
  and\ \bibinfo {author} {\bibfnamefont {M.~E.}\ \bibnamefont {Zhitomirsky}},\
  }\bibfield  {title} {\bibinfo {title} {Self-consistent spin wave analysis of
  the magnetization plateau in triangular antiferromagnet},\ }\href
  {https://doi.org/10.1088/1742-6596/320/1/012011} {\bibfield  {journal}
  {\bibinfo  {journal} {Journal of Physics: Conference Series}\ }\textbf
  {\bibinfo {volume} {320}},\ \bibinfo {pages} {012011} (\bibinfo {year}
  {2011})}\BibitemShut {NoStop}%
\bibitem [{\citenamefont {Mkhitaryan}\ and\ \citenamefont {Ke}(2021)}]{SCSW2}%
  \BibitemOpen
  \bibfield  {author} {\bibinfo {author} {\bibfnamefont {V.~V.}\ \bibnamefont
  {Mkhitaryan}}\ and\ \bibinfo {author} {\bibfnamefont {L.}~\bibnamefont
  {Ke}},\ }\bibfield  {title} {\bibinfo {title} {Self-consistently renormalized
  spin-wave theory of layered ferromagnets on the honeycomb lattice},\ }\href
  {https://doi.org/10.1103/PhysRevB.104.064435} {\bibfield  {journal} {\bibinfo
   {journal} {Phys. Rev. B}\ }\textbf {\bibinfo {volume} {104}},\ \bibinfo
  {pages} {064435} (\bibinfo {year} {2021})}\BibitemShut {NoStop}%
\bibitem [{\citenamefont {Nakano}\ and\ \citenamefont
  {Sakai}(2023)}]{doi:10.7566/JPSCP.38.011166}%
  \BibitemOpen
  \bibfield  {author} {\bibinfo {author} {\bibfnamefont {H.}~\bibnamefont
  {Nakano}}\ and\ \bibinfo {author} {\bibfnamefont {T.}~\bibnamefont {Sakai}},\
  }\bibinfo {title} {Large-scale numerical-diagonalization study of the
  shastry-sutherland model}\ (\bibinfo  {publisher} {Journal of the Physical
  Society of Japan},\ \bibinfo {year} {2023})\ \bibinfo {note} {0}\BibitemShut
  {NoStop}%
\bibitem [{\citenamefont {Moessner}\ \emph {et~al.}(2001)\citenamefont
  {Moessner}, \citenamefont {Sondhi},\ and\ \citenamefont
  {Fradkin}}]{PhysRevB.65.024504}%
  \BibitemOpen
  \bibfield  {author} {\bibinfo {author} {\bibfnamefont {R.}~\bibnamefont
  {Moessner}}, \bibinfo {author} {\bibfnamefont {S.~L.}\ \bibnamefont
  {Sondhi}},\ and\ \bibinfo {author} {\bibfnamefont {E.}~\bibnamefont
  {Fradkin}},\ }\bibfield  {title} {\bibinfo {title} {Short-ranged resonating
  valence bond physics, quantum dimer models, and ising gauge theories},\
  }\href {https://doi.org/10.1103/PhysRevB.65.024504} {\bibfield  {journal}
  {\bibinfo  {journal} {Phys. Rev. B}\ }\textbf {\bibinfo {volume} {65}},\
  \bibinfo {pages} {024504} (\bibinfo {year} {2001})}\BibitemShut {NoStop}%
\bibitem [{\citenamefont {Boyack}\ \emph {et~al.}(2018)\citenamefont {Boyack},
  \citenamefont {Lin}, \citenamefont {Zerf}, \citenamefont {Rayyan},\ and\
  \citenamefont {Maciejko}}]{PhysRevB.98.035137}%
  \BibitemOpen
  \bibfield  {author} {\bibinfo {author} {\bibfnamefont {R.}~\bibnamefont
  {Boyack}}, \bibinfo {author} {\bibfnamefont {C.-H.}\ \bibnamefont {Lin}},
  \bibinfo {author} {\bibfnamefont {N.}~\bibnamefont {Zerf}}, \bibinfo {author}
  {\bibfnamefont {A.}~\bibnamefont {Rayyan}},\ and\ \bibinfo {author}
  {\bibfnamefont {J.}~\bibnamefont {Maciejko}},\ }\bibfield  {title} {\bibinfo
  {title} {Transition between algebraic and ${\mathbb{z}}_{2}$ quantum spin
  liquids at large $n$},\ }\href {https://doi.org/10.1103/PhysRevB.98.035137}
  {\bibfield  {journal} {\bibinfo  {journal} {Phys. Rev. B}\ }\textbf {\bibinfo
  {volume} {98}},\ \bibinfo {pages} {035137} (\bibinfo {year}
  {2018})}\BibitemShut {NoStop}%
\bibitem [{\citenamefont {Essin}\ and\ \citenamefont {Hermele}(2013)}]{Z2}%
  \BibitemOpen
  \bibfield  {author} {\bibinfo {author} {\bibfnamefont {A.~M.}\ \bibnamefont
  {Essin}}\ and\ \bibinfo {author} {\bibfnamefont {M.}~\bibnamefont
  {Hermele}},\ }\bibfield  {title} {\bibinfo {title} {Classifying
  fractionalization: Symmetry classification of gapped ${\mathbb{z}}_{2}$ spin
  liquids in two dimensions},\ }\href
  {https://doi.org/10.1103/PhysRevB.87.104406} {\bibfield  {journal} {\bibinfo
  {journal} {Phys. Rev. B}\ }\textbf {\bibinfo {volume} {87}},\ \bibinfo
  {pages} {104406} (\bibinfo {year} {2013})}\BibitemShut {NoStop}%
\bibitem [{\citenamefont {Holstein}\ and\ \citenamefont
  {Primakoff}(1940)}]{HP}%
  \BibitemOpen
  \bibfield  {author} {\bibinfo {author} {\bibfnamefont {T.}~\bibnamefont
  {Holstein}}\ and\ \bibinfo {author} {\bibfnamefont {H.}~\bibnamefont
  {Primakoff}},\ }\bibfield  {title} {\bibinfo {title} {Field dependence of the
  intrinsic domain magnetization of a ferromagnet},\ }\href
  {https://doi.org/10.1103/PhysRev.58.1098} {\bibfield  {journal} {\bibinfo
  {journal} {Phys. Rev.}\ }\textbf {\bibinfo {volume} {58}},\ \bibinfo {pages}
  {1098} (\bibinfo {year} {1940})}\BibitemShut {NoStop}%
\bibitem [{\citenamefont {Sachdev}(2012)}]{Sachdev2012}%
  \BibitemOpen
  \bibfield  {author} {\bibinfo {author} {\bibfnamefont {S.}~\bibnamefont
  {Sachdev}},\ }\bibinfo {title} {Quantum phase transitions of antiferromagnets
  and the cuprate superconductors},\ in\ \href
  {https://doi.org/10.1007/978-3-642-10449-7_1} {\emph {\bibinfo {booktitle}
  {Modern Theories of Many-Particle Systems in Condensed Matter Physics}}},\
  \bibinfo {editor} {edited by\ \bibinfo {editor} {\bibfnamefont {D.~C.}\
  \bibnamefont {Cabra}}, \bibinfo {editor} {\bibfnamefont {A.}~\bibnamefont
  {Honecker}},\ and\ \bibinfo {editor} {\bibfnamefont {P.}~\bibnamefont
  {Pujol}}}\ (\bibinfo  {publisher} {Springer Berlin Heidelberg},\ \bibinfo
  {address} {Berlin, Heidelberg},\ \bibinfo {year} {2012})\ pp.\ \bibinfo
  {pages} {1--51}\BibitemShut {NoStop}%
\bibitem [{\citenamefont {Wang}\ \emph
  {et~al.}(2022{\natexlab{b}})\citenamefont {Wang}, \citenamefont {Zhang},\
  and\ \citenamefont {Sandvik}}]{ED}%
  \BibitemOpen
  \bibfield  {author} {\bibinfo {author} {\bibfnamefont {L.}~\bibnamefont
  {Wang}}, \bibinfo {author} {\bibfnamefont {Y.}~\bibnamefont {Zhang}},\ and\
  \bibinfo {author} {\bibfnamefont {A.~W.}\ \bibnamefont {Sandvik}},\
  }\bibfield  {title} {\bibinfo {title} {Quantum spin liquid phase in the
  shastry–sutherland model detected by an improved level spectroscopic
  method},\ }\href {https://doi.org/10.1088/0256-307X/39/7/077502} {\bibfield
  {journal} {\bibinfo  {journal} {Chinese Physics Letters}\ }\textbf {\bibinfo
  {volume} {39}},\ \bibinfo {eid} {077502} (\bibinfo {year}
  {2022}{\natexlab{b}})}\BibitemShut {NoStop}%
\end{thebibliography}%
\end{document}